\documentclass[11pt]{article}
\usepackage{amsmath,amssymb,bm,epsf,epsfig,graphicx}
\input epsf.sty
\usepackage[utf8]{inputenc}
\usepackage[english]{babel}
\usepackage{amsmath}
\usepackage{latexsym}
\usepackage{amsfonts}
\usepackage{mathtools}
\usepackage{amssymb}
\usepackage{scalerel}
\usepackage{extpfeil}
\usepackage[left=3cm,right=3cm,top=3.1cm,bottom=3.1cm]{geometry}
\usepackage{cite}
\usepackage{tensor}
\usepackage{tikz}
\usepackage{tikz-cd}
\usetikzlibrary{arrows.meta, bending}
\tikzset{C/.style={circle, minimum size=8mm,
		node contents={},
		append after command={\pgfextra{%
				\draw[-{Straight Barb[flex']}](\tikzlastnode.150) arc (450:110:2.8mm);}
	}}
}

\newcommand{\bs}[1]{\boldsymbol{#1}}

\usepackage{hyperref}
\numberwithin{equation}{section}

\def\p{\partial}

\def \be {\begin{eqnarray}}
\def \ee {\end{eqnarray}}
\def \bdm {\begin{displaymath}}
\def \edm {\end{displaymath}}

\def\0{\nonumber}

\begin{document}

\vspace*{.1cm}

\centerline{\LARGE \bf Classical algebraic structures } \vspace{.2cm}
\centerline{\LARGE \bf in string theory effective actions}
\vspace*{0.8cm}

\begin{center}

{\large Harold Erbin$^{(a)}$\footnote{Email: harold.erbin at gmail.com}, Carlo Maccaferri$^{(a)}$\footnote{Email: maccafer at gmail.com} , Martin Schnabl$^{(b,c)}$\footnote{Email: schnabl.martin at gmail.com}  and Jakub Vo\v{s}mera$^{(b,d)}$\footnote{Email: vosmera at gmail.com} }
\vskip 0.8 cm
$^{(a)}${\it Dipartimento di Fisica, Universit\`a di Torino, \\INFN  Sezione di Torino and Arnold-Regge Center\\
Via Pietro Giuria 1, I-10125 Torino, Italy}
\vskip .25 cm
$^{(b)}${\it Institute of Physics of the Czech Academy of Sciences, \\
Na Slovance 2, 182 21 Prague 8, Czech Republic}
\vskip .25 cm
$^{(c)}${\it  Institute of Mathematics of the Czech Academy of Sciences,\\ \v{Z}itn\'{a}~25, 115~67~Prague~1, Czech Republic}
\vskip .25 cm
$^{(d)}${\it  Institute of Particle and Nuclear Physics, Charles University,\\ V Hole\v{s}ovi\v{c}k\'{a}ch 2,  Prague 8, Czech~Republic}

\end{center}

\vspace*{5.0ex}

\centerline{\bf Abstract}
\bigskip
We study generic properties of string theory effective actions obtained by classically integrating out massive excitations from string field theories based on cyclic homotopy algebras of  $A_\infty$ or $L_\infty$ type.
 We construct   observables in the UV theory and we discuss their fate after integration-out. Furthermore, we discuss how to compose two subsequent integrations of degrees of freedom (horizontal composition) and how to integrate out degrees of freedom after deforming the UV theory with a new consistent interaction (vertical decomposition).
We then apply our general results to the open bosonic string using Witten's open string field theory.  There we  show how the  horizontal  composition can be used to systematically integrate out the Nakanishi-Lautrup field from the set of massless excitations, ending with a non-abelian  $A_\infty$-gauge theory for just the open string gluon. 
Moreover we show how the vertical decomposition can be used to construct effective open-closed couplings by deforming Witten OSFT with  a tadpole given by the Ellwood invariant. Also, we discuss how the effective theory controls the possibility of removing the tadpole in the microscopic theory, giving a  new framework for studying D-branes deformations induced by changes in the closed string background.\baselineskip=16pt
\newpage
\setcounter{tocdepth}{2}
\tableofcontents

\section{Introduction}\label{intro}
In recent years complete constructions of super string field theories have become available \cite{Moosavian:2019ydz, Kunitomo:2019glq, Erler:2017onq, Konopka:2016grr, Erler:2016ybs, Sen:2015uaa,  Kunitomo:2015usa} and there has been interest in explicitly computing the effective action of a given microscopic string field theory, after integrating out the massive degrees of freedom \cite{Erbin:2019spp,Maccaferri:2019ogq, Maccaferri:2018vwo, Asada:2017ykq}, as originally done in \cite{BS}.
The construction of effective actions from string field theory is instrumental for both giving a useful low-dimensional handle on the space of classical solutions \cite{Erler:2019fye, Vosmera:2019mzw,  Erler:2019xof, Kudrna:2018mxa, Cho:2018nfn, Larocca:2017pbo, Kojita:2016jwe,  Kudrna:2016ack,  Maccaferri:2015cha,  Erler:2014eqa, Maccaferri:2014cpa,Kudrna:2014rya,Erler:2013wda,  Kudrna:2012re,Erler:2012qn,  Murata:2011ep, Hata:2011ke, Kiermaier:2010cf, Bonora:2010hi, Schnabl:2005gv} as well as for a better grounded approach to superstring perturbation theory \cite{Sen:2019jpm, deLacroix:2018tml, Sen:2017szq, Sen:2016qap,Sen:2016uzq, Sen:2016gqt, Pius:2016jsl, Sen-restoration, Sen:2015hha, Sen:2014dqa, Sen:2014pia}.
See \cite{Erler:2019loq,Erler:2019vhl,deLacroix:2017lif} for recent reviews on SFT.

Since the pioneering works of Kajiura\cite{Kajiura:2001ng, Kajiura:2003ax} and more recently of Sen \cite{Sen:2016qap} it has been recognized that the symmetry structure of the theory in the ``ultraviolet'' (i.e. the initial microscopic theory) is reflected in the ``infrared'' (i.e. the effective theory for the light fields at low energy). 
In particular if the original theory has a gauge invariance encoded in a $A_\infty$ or $L_\infty$ structure, then an isomorphic homotopy structure is retained after the RG flow. 

In this paper we elucidate several aspects of the  structures which are transferred from the UV to the IR under the flow induced by {\it classically} integrating out a set of degrees of freedom from the original action. 
By classically integrating-out we  mean that we solve the equations of motion for a set of degrees of freedom in function of the remaining ones and then plug the solution back in the original action. This gives the tree-level effective action, which corresponds to keep the leading contribution in the saddle-point expansion of the path integral of the massive fields.  Loop corrections, following the general structure of \cite{Sen:2016qap}, can be also  considered  but we will not do so here \footnote{Some aspects in this regard are discussed in \cite{Masuda:2020tfa}.}.

The (gauge-fixed) solution which expresses the massive fields in terms of the massless fields is constructed perturbatively in the number of massless fields insertions.  We consider the standard Siegel gauge $b_0=0$ for open strings and $b_0+\bar b_0=0$ for closed strings. Thanks to  the structure of the BRST charge the equation of motion for the Siegel-gauge part of the massive fields is always solvable in terms of massless fields and Siegel gauge propagators $\frac{b_0}{L_0}(1-P_0)$, where $P_0$ is the projector on the kernel of $L_0$.  On the other hand, the equations of motion for the non-Siegel  part of the massive fields remain as  gauge constraints which cannot be derived anymore from the gauge fixed-action. However, in the process of our analysis,  we realize that these gauge constraints are in fact automatically accounted for by the effective equations of motion for the massless fields.  So nothing is lost in fixing Siegel gauge for the massive fields.\footnote{This consistency property was already observed up to the first few orders in perturbation theory in WZW-like heterotic string field theory \cite{Erbin:2019spp}.}

This direct perturbative approach towards the effective action we have just outlined  becomes quickly cumbersome, just like any Feynman diagram expansion. However the underlying homotopy  structure of the original action allows to package the perturbation theory in the convenient language of co-algebras, co-derivations and co-homomorphisms in the (symmetrized, in case of $L_\infty$) tensor algebra \cite{Gaberdiel:1997ia, Erler:2013xta, Erler:2015uba}. Equipped with this  convenient language we are able to find closed-form expressions for the solution of the massive fields in terms of the massless fields and, more importantly, a closed form expression for the all-order tree-level effective action and its corresponding effective vertices. 
The process of projecting out a set of fields in the tensor algebra can be rephrased as a strong-deformation-retract (SDR) \cite{Kajiura:2003ax} and the final form of the effective vertices is in fact directly implied by the homological perturbation lemma \cite{crainic2004perturbation, Konopka:2015tta, Matsunaga:2019fnc} which describes how the SDR for the initially free theory is deformed by switching on interactions. This nicely parallels what we have obtained by directly solving the equations of motion for the massive fields by automatically encoding in the co-algebra language all the tree-level Feynman diagrams.  In this approach it is evident that the full equations of motion of string field theory (including the out-of-gauge equations for the massive fields) are obeyed whenever the massless fields solve the equations of motion of the effective action.

The co-algebra language turns out to be very efficient also to discuss a new general class of observables (i.e. gauge invariant quantities) beside the action itself. Just like the action is associated with the set of  multi-string vertices encoded into odd cyclic  (nilpotent) coderivations  ${\bf m}=\sum_{n=1}^\infty {\bf m}_n$,  these new observables are associated to a set of odd cyclic (not necessarily nilpotent) coderivations ${\bf e}=\sum_{n=0}^\infty{\bf e}_n$ which commute (as coderivations) with the $A_\infty$  products ${\bf m}$.
These observables get modified in the IR upon integrating out the massive fields but they remain invariant under the gauge transformations of the effective action. Interestingly the explicit form of the effective observables coincide with the original observables of the UV theory where the original massive fields are substituted by their on-shell expression in terms of the massless fields, just  as it happens  for the effective action itself.

We then continue our analysis by studying what happens when, after having integrated out some degrees of freedom, we decide to  integrate out more fields. The two processes can be performed one after the other and at each stage the homotopy-algebraic structure is obviously preserved. But in fact the double-step integration can be performed in a single step by considering an Hodge-Kodaira decomposition of the BRST charge in which the propagator is the sum of the two subsequent propagators and the corresponding projector is the product of the two subsequent projectors. This results in a very compact way to handle the resulting doubled perturbation theory in a single set of diagrams where external legs are the final projected fields and the internal propagators are the sum of the two propagators. We call this process {\it horizontal composition}. 

On a complementary line,  it is often useful to consider deformations of the original UV action that preserve the homotopy structure and therefore the gauge invariance. This is for example what happens by adding to the action an observable of the kind discussed above whose defining odd coderivation $\bf e$ is also nilpotent, so that $({\bf m} +\mu {\bf e})^2=0$.  It is interesting  to explore the structure of the effective action after the deformation. The full result of integrating out can be obtained from  the homological perturbation lemma by deforming the free theory with ${\bf m}+\mu{\bf e}$, however the resulting perturbation theory is not very clearly organized in this form. In fact, we would be more physically interested in computing the effective action starting directly from the interacting theory with ${\bf m}$ and treating $\mu {\bf e}$ as a deformation, by running the homological perturbation lemma on the initial interacting SDR (and not on the free one).  But in fact it turns out that the two effective actions are just the same and this is guaranteed by the possibility of (de)composing interacting SDR's by simply decomposing the corresponding interactions. We  call this process  {\it vertical decomposition}. Interestingly the  final effective action will contain  infinite non-linear terms in the coderivation  $\bf e$ accounting for the fact that the coderivation of the induced effective observable is not nilpotent anymore, in general, and therefore it is not enough to add it to the undeformed effective vertices to retain the $A_\infty$-relations in the infrared. 

The second part of the paper is focused on examples of the above general structures in the context of Witten bosonic open string field theory.
We first describe the process of horizontal composition needed to consistently treat the auxiliary level-zero field given by $c_0|0,k\rangle$, which plays the role of a Nakanishi-Lautrup (NL) field for the open string gluon. This field would be set to zero in Siegel gauge, however  at level zero this would leave out-of-gauge equations which would not be  accounted for by the remaining massless fields, contrary to what happens  for the massive fields. Therefore this field has to be integrated out, rather than being set to zero and this is thus an  instance where horizontal composition becomes handy. The resulting effective action is expressed via off-shell amplitudes for the gauge field where in addition to the usual worldsheet propagator $\frac{b_0}{L_0}(1-P_0)$ there is also an ``algebraic'' propagator of the form $\frac12 c_0b_{-1}b_1P_0$. This propagator (which already appeared in \cite{Erbin:2019spp} in  heterotic string field theory, with the understood superstring corrections) has been recently discussed by Sen in the context of string  perturbation theory in D-instantons background \cite{Sen:2020cef} and our analysis offers a  complementary (and equivalent in our tree-level treatment) viewpoint.

At last we consider the Ellwood Invariant \cite{Hashimoto:2001sm, Gaiotto:2001ji, Ellwood:2008jh}
as an example of our generic class of observables. This observable is constructed using a single nilpotent coderivation consisting of a zero-string product given by the insertion of an on-shell closed string state at the midpoint of the identity string field. This defines an open string state which behaves as a tadpole in the action.  We show how the vertical decomposition allows to  account for all the tree-level open-closed effective couplings whose structure can be systematically extracted out at every order in perturbation theory.  The fate of the tadpole in the full theory depends on the possibility of removing the tadpole in the effective theory. Indeed we show that the obstructions to the vacuum shift in the full theory are just the equations of motion for the vacuum shift in the effective theory. If these equations are solvable then a new shifted vacuum will show up, with no tadpole anymore and a deformed spectrum of physical states, corresponding to the original D-brane having adapted to the new bulk CFT. However it is also possible that the vacuum shift is obstructed at some order which physically corresponds to the fact that the closed string deformation is incompatible with the boundary conditions of the starting D-brane system. 

The paper is organized as follows. In section \ref{sec:eff} we present the detailed construction of the tree-level effective action for a string field theory with $A_\infty$ gauge symmetry. We start pedagogically in the product-notation and we gradually upgrade the language to tensor co-algebras where we can write down explicit all-order statements. 
We  discuss observables in the UV and in the IR in subsection \ref{sec:obs}. In subsection \ref{subsec:HorComp} we describe the concept of horizontal composition which is useful to integrate out further degrees of freedom and in subsection \ref{subsec:VertComp} we describe vertical decomposition which is useful to calculate the deformation of the effective action under a deformation of the microscopic action. In section \ref{sec:witten} we apply our constructions to Witten OSFT. In subsection \ref{subsec:EffEx1} we show how to integrate-out the NL field, maintaining the $A_\infty$ gauge symmetry and ending up with off-shell amplitudes whose propagator is the sum of the usual Siegel gauge propagator and a new algebraic propagator. In subsection \ref{Ellwood} we give the structure of the effective action deformed by the Ellwood invariant and we discuss some of the physics associated to the closed string tadpole and the corresponding change in the bulk CFT. We conclude in section \ref{concl} with final comments  and an outlook for the future. In appendix \ref{A:hpl} we review the necessary mathematics of the strong deformation retract and the homological perturbation lemma which we use thoroughly during the paper and finally in appendix \ref{A:Linfty} we extend the construction of the effective action to string field theories based on $L_\infty$-algebras. The technical new result here is given by the construction of the uplift of the propagator to an appropriate operator in  the symmetrized tensor co-algebra which is consistent with the co-algebraic Hodge-Kodaira decomposition and therefore reproduces the correct perturbation theory.

Some of the results presented in this paper also appear in the thesis \cite{jakub-thesis} written by one of the authors.

{\bf Note added.} During the writing of this paper we have  learnt that \cite{yuji} also obtained and discussed the effective open-closed couplings in Witten OSFT. Our papers will appear on the same day.
\section{Effective physics for an $A_\infty$ theory}\label{sec:eff}

The goal of this section will be to outline the framework for computing tree-level effective actions for general string field theories based on a cyclic $A_\infty$ structure. Parallel considerations can be applied to a cyclic $L_\infty$ structure, and the details are presented in Appendix \ref{A:Linfty}.

\subsection{Product notation}
\label{subsec:EffProd}

Let us first lay out the basic principles of constructing tree-level SFT effective actions using the intuitive language of products on the 
 string Hilbert space $\mathcal{H}$. This is a graded vector space, where the grading will be provided by the degree $d(A)=|A|+1$ (with $|A|$ denoting the ghost-number of $A\in\mathcal{H}$). The vertices of the full SFT actions which we will be considering are given by degree-odd multilinear products
\begin{align}
m_k:\mathcal{H}^{\otimes k}\longrightarrow \mathcal{H}\,,
\end{align}
which satisfy the \emph{$A_\infty$-relations} \cite{Gaberdiel:1997ia}
\begin{align}
\sum_{l=1}^{k} m_l m_{k+1-l} =0\,.\label{eq:Ainfrel}
\end{align}
For $k=1,2,3,\ldots$, these can be explicitly written out as
\begingroup\allowdisplaybreaks
\begin{subequations}
	\begin{align}
	0&=m_1(m_1(A_1))\,,\label{eq:nil}\\
	0&=m_1(m_2(A_1,A_2))+m_2(m_1(A_1),A_2)+\nonumber\\
	&\hspace{4cm}+(-1)^{d(A_1)}m_2(A_1,m_1(A_2))\,,\label{eq:der}\\
	0&=m_1(m_3(A_1,A_2,A_3))+m_2(m_2(A_1,A_2),A_3)+\nonumber\\
	&\hspace{4cm}+(-1)^{d(A_1)}m_2(A_1,m_2(A_2,A_3))+\nonumber\\
	&\hspace{1cm}+m_3(m_1(A_1),A_2,A_3)+(-1)^{d(A_1)}m_3(A_1,m_1(A_2),A_3)+\nonumber\\
	&\hspace{4cm}+(-1)^{d(A_1)+d(A_2)}m_3(A_1,A_2,m_1(A_3))\,,\label{eq:ass}\\
	&\hspace{0.2cm}\vdots\nonumber
	\end{align}
\end{subequations}
\endgroup
for $A_1,A_2,A_3,\ldots\in\mathcal{H}$. The property \eqref{eq:nil} says that the operation $m_1$ is nilpotent, the Leibniz-like property \eqref{eq:der} tells us that $m_1$ is a derivation of the 2-product $m_2$, while the property \eqref{eq:ass} says that the failure of $m_1$ to be a derivation of $m_3$ is exactly balanced by the failure of the associativity of $m_2$ (that is, $m_2$ is associative up to a homotopy).
The 1-string product $m_1$ is usually given by the BRST charge $Q$
\be
m_1=Q.
\ee 
The only remaining ingredient needed to write down the action is the symplectic form $\omega:\mathcal{H}^{\otimes 2}\longrightarrow \mathcal{H}$. This is a graded anti-symmetric bilinear map with respect to which the products $m_k$ are cyclic
\begin{subequations}
	\begin{align}
	\omega(A_1,A_2) &= -(-1)^{d(A_1)d(A)_2}\omega(A_2,A_1)\,,\\
	\omega(A_1,m_k(A_2,\ldots,A_{k+1}))&=-(-1)^{d(A_1)}\omega(m_k(A_1,\ldots,A_k),A_{k+1})\,,
	\end{align}
\end{subequations}
for $A_i\in\mathcal{H}$. In practice, $\omega$ is usually given in terms of the BPZ inner product on $\mathcal{H}$. Fixing a degree-even element $\Psi\in\mathcal{H}$ to denote the dynamical string field, the full SFT action can then be written as
\begin{align}
S(\Psi) = \sum_{k=1}^\infty\frac{1}{k+1}\omega(\Psi,m_k(\Psi^{\otimes k}))\,.\label{eq:SAinf}
\end{align}
We will often find it useful to get rid of the fractional coefficients $1/(k+1)$ in the action \eqref{eq:SAinf} by introducing an arbitrary smooth interpolation $\Psi(t)$ for $0\leq t\leq 1$ such that $\Psi(0)=0$, $\Psi(1)=\Psi$.
Using cyclicity of $m_k$ with respect to $\omega$, we can then rewrite the action \eqref{eq:SAinf} as
\begin{align}
S(\Psi) = \int_0^1 dt\, \sum_{k=1}^\infty\omega(\dot{\Psi}(t),m_k(\Psi(t)^{\otimes k}))\,,\label{eq:ActInterp}
\end{align}
where the $t$-dependence is purely topological. Varying the action with respect to $\Psi$ and using cyclicity, we obtain the equation of motion 
\begin{align}
\text{EOM}(\Psi)=\sum_{k=1}^\infty m_k(\Psi^{\otimes k})\equiv Q\Psi+\mathcal{J}(\Psi)\,,\label{eq:MC}
\end{align}
where the interacting part of the equation of motion $\mathcal{J}(\Psi)$ has been defined.
In the mathematical context, this is usually called the {Maurer-Cartan equation} and any $\Psi^\ast\in\mathcal{H}$ which satisfies $\text{EOM}(\Psi^\ast)=0$ (i.e.\ a classical solution) is called a {Maurer-Cartan} element. It is also a simple exercise to show that the action \eqref{eq:SAinf} is invariant under the linearized gauge transformation
\begin{align}
\delta_\Lambda\Psi
&=\sum_{k=1}^\infty\sum_{l=0}^{k-1}m_k(\Psi^{\otimes l},\Lambda,\Psi^{\otimes( k-l-1)})\,,\label{eq:gtAinf}
\end{align}
where $\Lambda\in\mathcal{H}$ is a degree-odd gauge parameter. Indeed, using cyclicity of $m_k$ and graded anti-symmetry of $\omega$ a number of times, we have
\begin{subequations}
	\label{eq:AinfGauge}
	\begin{align}
	\delta_\Lambda S  &= \sum_{n=1}^\infty \omega(\delta_\Lambda\Psi, m_n(\Psi^{\otimes n}))\\
	&=\sum_{n=1}^\infty \sum_{k=1}^\infty\omega(\Lambda, m_km_n(\Psi^{\otimes k+n-1}))\\[+2mm]
	&=0\,,
	\end{align}
\end{subequations}
where the last line holds by the $A_\infty$ relations \eqref{eq:Ainfrel}. 

Especially for the illustrative purposes of fixing  Siegel gauge for the massive fields at the beginning of our presentation, we should keep in mind the concrete examples of the bosonic cubic OSFT (where the products truncate at $m_2$) or the  $A_\infty$ open superstring field theory constructed in \cite{Erler:2013xta, Konopka:2016grr}. At some point we will, however, realize that there is an abstract way of fixing the gauge purely in terms of the propagator for the massive modes, which does not require referring to concrete operators such as $b_0,c_0$ (which might be, in principle, theory-specific).

\subsubsection{Splitting the string field}

Let us consider a projector $P$ acting on $\mathcal{H}$ (and denote $\bar{P}=1-P$) which is BPZ-even
\begin{align}
\omega(PA_1,A_2) = \omega(A_1,PA_1)\,.\label{eq:BPZproj}
\end{align}
For the purposes of our initial exposition in Siegel gauge, we will also require that $\text{ker}\,L_0 \subset \text{im}\, P$. Introducing then an operator $(b_0/L_0)\bar{P}\equiv h$, we can write a Hodge-Kodaira decomposition
\begin{align}
h Q + Q h = 1-P\,.\label{eq:HK}
\end{align}
Here we note that $h$ is always well-defined because since we assume that $\text{ker}\,L_0\subset\text{im}P$, we have (denoting by $P_0$ the projector onto $\text{ker}\, L_0$) $\bar{P}=\bar{P}_0\bar{P}$ and $(b_0/L_0)\bar{P}_0$ is well-defined by construction. As a consequence of \eqref{eq:HK} and the super-Jacobi identity, we have $[P,Q]=[P,h]=0$. 
Note that since we have $(b_0)^2=0=P\bar{P}=\bar{P}P$, we recover the conditions $h^2 =Ph = hP=0$. 
We then decompose the string field as 
\be
\Psi = \psi+\mathcal{R},\label{splitstring}
\ee where $\psi = P\Psi$ and $\mathcal{R}=\bar{P}\Psi$. Using the BPZ property \eqref{eq:BPZproj} of the projector $P$ and varying the action separately with respect to $\psi$ and $\mathcal{R}$, the equations of motion for $\psi$ and $\mathcal{R}$ read
\begin{subequations}
	\begin{align}
	\text{EOM}_\psi(\psi,\mathcal{R}) &= P\,\text{EOM}(\psi+\mathcal{R})= Q\psi + P\mathcal{J}(\psi+\mathcal{R})\,,\label{eq:EOMpsi}\\
	\text{EOM}_\mathcal{R}(\psi,\mathcal{R}) &= \bar{P}\,\text{EOM}(\psi+\mathcal{R})= Q\mathcal{R} + \bar{P}\mathcal{J}(\psi+\mathcal{R})\,.\label{eq:EOMR}
	\end{align}
\end{subequations}

\subsubsection{Fixing Siegel gauge for $\mathcal{R}$}

We now want to use the equation of motion \eqref{eq:EOMR} to integrate $\mathcal{R}$ out using $h$ as a propagator and extract the effective dynamics of $\psi$. In order to do this, we will need to fix a gauge for $\mathcal{R}$. For pedagogical reasons, we shall first do so by explicitly applying the Siegel gauge condition $b_0\mathcal{R}=0$. 
Let us therefore assume that, as in the case of the open (super)string, we can decompose
\begin{align}
Q = c_0 L_0 +b_0 M^++\widehat{Q}\,,
\label{eq:Qdecomp}
\end{align}
where $M$ and $\widehat{Q}$ do not contain any zero modes (see e.g.\ \cite{Asano:2016rxi} for concrete expressions for $M^+$ and $\widehat{Q}$ for both bosonic string and superstring).

Defining the Siegel-gauge projector $P_s=b_0 c_0$ together with $\bar{P}_s=1-P_s=c_0b_0$ and assuming $[P,P_s]=0$ (which is clearly the case for instance for $P=P_0$), we then decompose $\mathcal{R}=R+\tilde{R}$ where $R=P_s\mathcal{R}$ and $\tilde{R}=\bar{P}_s\mathcal{R}$. Gauge-fixing of the $\mathcal{R}$ component of the string field can then be effected by requiring $\tilde{R}=0$. The equation of motion $\text{EOM}_\mathcal{R}(\psi,\mathcal{R})$ therefore decomposes into two components
\begin{subequations}
	\begin{align}
	\text{EOM}_R(\psi,R) &= \bar{P}_sQ R + \bar{P}_s\bar{P}\mathcal{J}(\psi+R)\,,\\
	\text{EOM}_{\tilde{R}}(\psi,R) &= P_sQ R +P_s\bar{P}\mathcal{J}(\psi+R)\,.\label{eq:oosAinf}
	\end{align}
\end{subequations}
The first gives the equation of motion for $R$ which is to be solved for $R(\psi)$. The second gives the gauge constraint (out-of-Siegel equation) which generally needs to be kept alongside the in-Siegel equation of motion. 

\subsubsection{Solving for \texorpdfstring{$R(\psi)$}{R(psi)}}

Note that using \eqref{eq:Qdecomp}, we can rewrite
\begin{align}
\text{EOM}_R(\psi,R)  = c_0 L_0 R + c_0 b_0\bar{P}\mathcal{J}(\psi+R)\,.
\end{align}
That is, since $b_0 R=0$, the solution $R(\psi)$ needs to satisfy
\begin{align}
R(\psi) = -h\mathcal{J}(\Psi)\big|_{\Psi=\psi+R(\psi)}\,.\label{eq:RecRel}
\end{align}
Denoting $\mathcal{G}(A) = -h\mathcal{J}(A)$
and assuming the initial condition $R(0)=0$, we therefore obtain the solution
\begin{align}
R(\psi) = \mathcal{G}(\psi+\mathcal{G}(\psi+\mathcal{G}(\psi+\ldots))).\,
\end{align}
Up to cubic order in $\psi$,  $R(\psi)$ can be expanded as
\begin{align}
R(\psi) &= -hm_2(\psi,\psi)-hm_3(\psi,\psi,\psi)+\nonumber\\
&\hspace{1cm}+hm_2(hm_2(\psi,\psi),\psi)+hm_2(\psi,hm_2(\psi,\psi))+\ldots
\end{align}
so that substituting back into the splitting of the string field \eqref{splitstring}, we obtain
\begin{subequations}
	\begin{align}
	\Psi(\psi) 
	&\equiv \psi+R(\psi)\\
	&= \psi -hm_2(\psi,\psi)-hm_3(\psi,\psi,\psi)+\nonumber\\
	&\hspace{1cm}+hm_2(hm_2(\psi,\psi),\psi)+hm_2(\psi,hm_2(\psi,\psi))+\ldots
	\label{eq:Psipsi}
	\end{align}
\end{subequations}
Note that the terms inside $\Psi(\psi)$ containing $k$ powers of $\psi$ can be given a Feynman-diagrammatic interpretation as consisting of all possible rooted trees with $k$ leaves and at least 3-valent nodes. This means that the number of terms arising at order $\psi^{\otimes k}$ is given by the $k^{\mathrm{th}}$ super-Catalan number.

\subsubsection{Checking the out-of-Siegel constraint}

Let us now show that the out-of-Siegel constraint \eqref{eq:oosAinf} is, in fact, automatically satisfied whenever $\psi$ solves the equation of motion \eqref{eq:EOMpsi}.
We can first act $Q$ on \eqref{eq:RecRel} and then use the Hodge-Kodaira decomposition \eqref{eq:HK} to show that
\begin{align}
QR(\psi) = -\bar{P}\mathcal{J}(\Psi(\psi))+hQ\mathcal{J}(\Psi(\psi))\,.\label{eq:HKidentity}
\end{align}
Substituting this into \eqref{eq:oosAinf}, we obtain
\begingroup\allowdisplaybreaks
\begin{align}
\text{EOM}_{\tilde{R}}(\psi,R(\psi)) 
&=hQ\mathcal{J}(\Psi(\psi))\,.\label{eq:Aoos}
\end{align}
\endgroup
Using the $A_\infty$ relations we may now show that
\begin{align}
Q\mathcal{J}(\Psi) 
&=  -\sum_{k=2}^\infty\sum_{l=0}^{k-1} m_k(\Psi^{\otimes l},Q\Psi,\Psi^{\otimes (k-1-l)})+\nonumber\\
&\hspace{3cm}-\sum_{k=3}^\infty \sum_{m=2}^{k-1}m_m m_{k+1-m}(\Psi^{\otimes k})\,.\label{eq:Aref1}
\end{align}
Expressing $Q\Psi$ as 
\begin{align}
Q\Psi(\psi)
&= \text{EOM}(\Psi)-\mathcal{J}(\Psi)\,\label{eq:Aref2}
\end{align}
and substituting \eqref{eq:Aref2} into the r.h.s.\ of \eqref{eq:Aref1} by straightforward manipulation of the products, we eventually obtain 
\begingroup\allowdisplaybreaks
\begin{align}
Q\mathcal{J}(\Psi) 
&= 
-\sum_{k=2}^\infty\sum_{l=0}^{k-1} m_k(\Psi^{\otimes l},\text{EOM}(\Psi(\psi)),\Psi^{\otimes (k-1-l)})\,.\label{eq:5117}
\end{align}
\endgroup
Substituting this back into \eqref{eq:Aoos}, as well as assuming that the equation of motion for $\psi$ is solved (that is, taking $\psi=\psi^\ast$ such that $\text{EOM}_\psi(\psi^\ast,R(\psi^\ast))=0$) and noting that $\text{EOM}_R(\psi,R(\psi))=0$, we obtain
\begin{align}
\text{EOM}_{\tilde{R}}(\psi^\ast,R(\psi^\ast))  &= 
\mathcal{F}[\text{EOM}_{\tilde{R}}(\psi^\ast,R(\psi^\ast))]\,,
\end{align}
where we have defined the linear operator
\begin{align}
\mathcal{F}[X]=-\sum_{k=2}^\infty\sum_{l=0}^{k-1} hm_k(\Psi(\psi^\ast)^{\otimes l},X,\Psi(\psi^\ast)^{\otimes k-1-l})\,.
\end{align}
Therefore, assuming that the operator $1-\mathcal{F}$ is invertible\footnote{This should be the case at least for small $\psi^\ast$ because, since $R(0)=0$, then also $\mathcal{F}$ should be small for small $\psi^\ast$}, it follows that
\begin{align}
\text{EOM}_{\tilde{R}}(\psi^\ast,R(\psi^\ast))=0\,.
\end{align}

\subsubsection{Abstract gauge-fixing}

There is a more abstract (but nevertheless equivalent) way of fixing the gauge for $\mathcal{R}$ and solving \eqref{eq:EOMR}, namely by requiring that $hR=0$ (see \cite{Kajiura:2001ng,Kajiura:2003ax}). Assuming this condition, it is then possible to derive the key recursion relation \eqref{eq:RecRel} by simply hitting \eqref{eq:EOMR} with $h$ and using the Hodge-Kodaira decomposition \eqref{eq:HK}. Consistency of \eqref{eq:RecRel} then also requires that we have $h^2 =Ph=0$ (as can be seen by acting on \eqref{eq:RecRel} with $h$ and $P$ and requiring the gauge condition $hR=0$, as well as that $PR=0$). Finally, $h^2=0$ in conjunction with the Hodge-Kodaira decomposition implies $[h,P]=0$ which in turn gives the remaining condition $hP=0$.

The associated out-of-gauge constraints can then be indirectly seen to hold by noting that the total SFT equation of motion is automatically satisfied whenever the equation of motion for $\psi$ is solved (namely that $\text{EOM}(\Psi(\psi^\ast))=0$ whenever we consider $\psi=\psi^\ast$ such that $\text{EOM}_\psi(\psi^\ast,R(\psi^\ast))=0$). Indeed, using the identity \eqref{eq:HKidentity} (which is purely a consequence of acting with $Q$ on \eqref{eq:RecRel} and applying the Hodge-Kodaira decomposition \eqref{eq:HK}), we may show that
\begin{align}
\text{EOM}(\Psi(\psi)) 
&=\text{EOM}_\psi(\psi,R(\psi)) -hQ\mathcal{J}(\Psi(\psi))\,.
\end{align}
Using the result \eqref{eq:5117} (which is derived independently of having fixed Siegel gauge) and substituting $\psi=\psi^\ast$, we  obtain 
\begin{align}
\text{EOM}(\Psi(\psi^\ast)) 
&=\mathcal{F}[\text{EOM}(\Psi(\psi^\ast))]\,.
\end{align}	
Assuming again invertibility of $1-\mathcal{F}$, it follows that $\text{EOM}(\Psi(\psi^\ast))=0$. This is how the gauge constraints are trivialized for the abstract gauge fixing $hR=0$.

We can conclude that the conditions $hP=Ph=h^2=0$, as well as the Hodge-Kodaira decomposition \eqref{eq:HK} seem to be the key ingredients for the whole construction of tree-level effective action to work (ref.\ \cite{Kajiura:2003ax} arrives at the same conclusion). Noting that we can split $P=I\Pi$, where \be\Pi:\mathcal{H}\longrightarrow P\mathcal{H}\ee is the canonical projection and \be I:P\mathcal{H}\longrightarrow \mathcal{H}\ee is the canonical inclusion (so that we also have $ \Pi h = hI=0$), it neatly follows that the algebraic properties we have encountered so far can be summarized by the \emph{strong deformation retract} (or SDR; see Appendix \ref{A:hpl} for a working review)
\begin{align}
\mathrel{\raisebox{+14.5pt}{\rotatebox{-110}{  \begin{tikzcd}[sep=12mm,
			arrow style=tikz,
			arrows=semithick,
			diagrams={>={Straight Barb}}
			]
			\ar[to path={ node[pos=.5,C] }]{}
			\end{tikzcd}
}}}\hspace{-2.7mm}
(-h)\,(\mathcal{H},Q)
\hspace{-3.2mm}\raisebox{-0.2pt}{$\begin{array}{cc} {\text{\scriptsize $\Pi$}}\\[-3.0mm] \mathrel{\begin{tikzpicture}[node distance=1cm]
		\node (A) at (0, 0) {};
		\node (B) at (1.5, 0) {};
		\draw[->, to path={-> (\tikztotarget)}]
		(A) edge (B);
		\end{tikzpicture}}\\[-4mm] {\begin{tikzpicture}[node distance=1cm]
		\node (A) at (1.5, 0) {};
		\node (B) at (0, 0) {};
		\draw[->, to path={-> (\tikztotarget)}]
		(A) edge (B);
		\end{tikzpicture}}\\[-3.5mm]
	\text{\scriptsize $I$} \end{array}$} \hspace{-3.4mm}
(P\mathcal{H},\Pi Q I)\,,\label{eq:SDRH}
\end{align}
\vspace{-7mm}

\noindent where the propagator $h$ is (minus) the {\it contracting homotopy operator}.

\subsubsection{Effective action and the minimal model theorem}

Substituting the solution \eqref{eq:Psipsi} into \eqref{eq:EOMpsi}, we obtain that the equation of motion for $\psi$ can be rewritten as
\begin{align}
\text{eom}(\psi) 
&= \tilde{m}_1(\psi) + \tilde{m}_2(\psi,\psi)+\tilde{m}_3(\psi,\psi,\psi)+\ldots
\label{eq:eompsi}
\end{align}
where we have introduced new multi-linear products $\tilde{m}_k:\mathcal{H}^{\otimes k}\longrightarrow \mathcal{H}$
\begingroup\allowdisplaybreaks
\begin{subequations}
	\label{eq:effprod}
	\begin{align}
	\tilde{m}_1(A_1) &= Pm_1(A_1)\,,\\
	\tilde{m}_2(A_1,A_2) &= Pm_2(A_1,A_2)\,,\\
	\tilde{m}_3(A_1,A_2,A_3) &= Pm_3(A_1,A_2,A_3)+\nonumber\\
	&\hspace{0.5cm}-Pm_2(hm_2(A_1,A_2),A_3)-Pm_2(A_1,hm_2(A_2,A_3))\,,\\
	&\hspace{0.2cm}\vdots\nonumber
	\end{align}
\end{subequations}
\endgroup
for $A_1,A_2,A_3\in P\mathcal{H}$. Below we will prove, using the techniques of tensor coalgebra, that the products $\tilde{m}_k$ satisfy $A_\infty$ relations. Also, assuming that the contracting homotopy operator $(-h) $ is BPZ self-conjugate (which is clearly true for the Siegel-gauge propagator $h=(b_0/L_0)\bar{P}$)
\begin{align}
\omega(A_1,h A_2) = (-1)^{d(A_1)}\omega(h A_1,A_2)\,,
\end{align}
we will show that the products $\tilde{{m}}_k$ are cyclic with respect to $\tilde{\omega}\equiv \omega|_{P\mathcal{H}}$ (in accordance with \cite{Kajiura:2001ng,Kajiura:2003ax}). Nevertheless, it is a rewarding exercise  to verify cyclicity and $A_\infty$ relations explicitly at least for $\tilde{m}_1$, $\tilde{m}_2$ and $\tilde{m}_3$ and we encourage the  reader who might not have familiarity with this to do so. These results, together with the above result that the out-of-gauge constraints are automatically solved when $\psi$ satisfies $\text{EOM}_\psi(\psi,R(\psi))$, imply that the dynamics of $\psi$ is completely captured by the effective action $\tilde{S}(\psi) = S(\psi+R(\psi))$ where
\begin{align}
\tilde{S}(\psi) = \sum_{k=1}^\infty \frac{1}{k+1}\tilde{\omega}(\psi,\tilde{m}_k(\psi^{\otimes k}))\label{eq:Stilde}\,.
\end{align}
Furthermore, the action \eqref{eq:Stilde} manifests the gauge invariance 
\begin{align}
\tilde{\delta}_\lambda \psi &= \tilde{m}_1(\lambda) + \tilde{m}_2(\lambda,\psi)+\tilde{m}_2(\psi,\lambda)+\nonumber\\
&\hspace{2cm}+\tilde{m}_3(\lambda,\psi,\psi)+\tilde{m}_3(\psi,\lambda,\psi)+\tilde{m}_3(\psi,\psi,\lambda)+\ldots\,,
\end{align}
associated with the effective products $\tilde{m}_k$ (where $\lambda\in P\mathcal{H}$ is a gauge parameter). We shall see below how this is related to the gauge transformation of the full SFT.

All of these results have appeared in some form in \cite{Kajiura:2001ng,Kajiura:2003ax}, where the author mostly specializes on integrating out all fields outside of the BRST cohomology. As a result, he obtains effective products with $\tilde{m}_1 =0$. The effective $A_\infty$ structure $(P\mathcal{H},\{\tilde{m}_k\}_{k=2}^\infty)$ therefore provides the \emph{minimal model}\footnote{An $A_\infty$ algebra $(\mathcal{H},\{m_k\}_{k=1}^\infty)$ is called minimal if $m_1=0$.} for the original UV $A_\infty$ structure $(\mathcal{H},\{m_k\}_{k=1}^\infty)$. Existence of such minimal model is guaranteed by the \emph{minimal model theorem} \cite{kadeishvili1982algebraic}. 

Notice, however, that it is a priori not always clear how the products $\tilde{m}_k$ of the minimal model can be explicitly constructed in practice for any given SFT: while it is straightforward how to expand the effective products $\tilde{m}_k$ in terms of the propagator $h$ and the UV products $m_k$, it is not immediately obvious what is the explicit expression for $h$ which would implement integrating out all modes outside of the cohomology of $Q$. For instance, considering the cubic OSFT, the Siegel-gauge propagator $(b_0/L_0)\bar{P}_0$ only integrates out the modes outside of $\text{ker}\,L_0$ while there are known examples of states in $\text{ker}\, L_0$ (such as $\p c$) which are clearly not BRST closed.\footnote{At this point it is important to remember that we have only established that it is consistent to fix Siegel gauge for the massive modes which are projected away by $P_0$ (and which were already integrated out). It does not seem to be possible to fix Siegel gauge also for the modes $\psi\in\text{ker}\,L_0$ in such a way  that the corresponding out-of-Siegel equations would be trivialized by the in-Siegel equations of motion, as it was the case above for the massive modes \cite{Erbin:2019spp}.} Below in subsection \ref{subsec:HorComp} we will present a method of integrating such modes out by adding a correction to the Siegel-gauge propagator, thus providing an explicit example  of an effective SFT action given by a minimal $A_\infty$ algebra (by restricting to zero momentum), which is related to the original UV SFT $A_\infty$ algebra by means of an explicit $A_\infty$ quasi-isomorphism.

\subsection{Tensor coalgebra language}
\label{subsec:EffTens}

We will now unleash the full power of tensor coalgebras \cite{Gaberdiel:1997ia} (see especially \cite{Erler:2015uba} for a self-contained introduction of the necessary concepts, whose knowledge we will  assume here) and homological perturbation theory (see Appendix \ref{A:hpl} for a working review) to derive the $A_\infty$ effective action in a closed and compact form.  As the story will develop, we will recognize that the mechanism behind constructing the effective action is clearly governed by the homological perturbation lemma (for strong deformation retracts), as introduced in Appendix \ref{A:hpl}: see also \cite{Konopka:2015tta,Erler:2016rxg}. 

\subsubsection{$A_\infty$ SFT in tensor coalgebra language}

Let us start by lifting the various maps and products on $\mathcal{H}$ defined in subsection \ref{subsec:EffProd} to the tensor-product space
\begin{align}
T\mathcal{H} = \mathcal{H}^{\otimes 0}\oplus\mathcal{H}^{\otimes 1}\oplus\mathcal{H}^{\otimes 2}\oplus\ldots\,,
\end{align}
where $\mathcal{H}^{\otimes 0}$ consists of scalars multiplying the identity element $1_{T\mathcal{H}}$ of the tensor-product space $T\mathcal{H}$ (that is $1_{T\mathcal{H}}\otimes V=V\otimes 1_{T\mathcal{H}}=V$ for all $V\in T\mathcal{H}$).
We will denote by \be\pi_k:T\mathcal{H}\longrightarrow \mathcal{H}^{\otimes k}\ee the projection onto the $k$-string component $\mathcal{H}^{\otimes k}$ of $T\mathcal{H}$. The space $T\mathcal{H}$ can be equipped with a co-associative deconcatenation co-product $\Delta_{T\mathcal{H}}:T\mathcal{H}\longrightarrow T\mathcal{H}\otimes'T\mathcal{H}$ which acts as
\begin{align}
\Delta_{T\mathcal{H}}(A_1\otimes \ldots \otimes A_k) = \sum_{l=0}^{k} (A_1\otimes\ldots \otimes A_l)\otimes' (A_{l+1}\otimes\ldots \otimes A_{k})\,.\label{eq:coprT}
\end{align}
In the cases where the summation index $l$ in \eqref{eq:coprT} attains the values $l=0$ or $l=k$, the summand should be understood as being equal to $1_{T\mathcal{H}}\otimes'(A_1\otimes\ldots\otimes A_k)$ and $(A_1\otimes\ldots\otimes A_k)\otimes' 1_{T\mathcal{H}}$, respectively. The pair $(T\mathcal{H},\Delta_{T\mathcal{H}})$ then constitutes a \emph{tensor coalgebra}.
We will also denote by $\nabla_{T\mathcal{H}}$ the corresponding associative concatenation product $\nabla_{T\mathcal{H}} :T\mathcal{H}\otimes' T\mathcal{H}\longrightarrow T\mathcal{H}$,
which simply acts by replacing $\otimes '$ with $\otimes$.

\noindent
Considering first the $k$-string products $m_k$, let us define the maps \be\mathbf{m}_k:T\mathcal{H}\longrightarrow T\mathcal{H}\ee
by requiring that on $\mathcal{H}^{\otimes N}$,  they act as (denoting by $1_{\mathcal{H}}$ the identity map on $\mathcal{H}$, i.e.\ $1_{\mathcal{H}}(A)=A$ for all $A\in\mathcal{H}$)
\begin{align}
\mathbf{m}_k\pi_N=\sum_{n=0}^{N-k}\big[({1}_\mathcal{H})^{\otimes N-k-n}\otimes m_k\otimes ({1}_\mathcal{H})^{\otimes n}\big]\pi_N\label{eq:corresp}
\end{align}
for $N\geq k$ and that they vanish on $\mathcal{H}^{\otimes N}$ for $N<k$. The maps $\mathbf{m}_k$ then act as \emph{coderivations}, that is \be\Delta_{T\mathcal{H}} \mathbf{m}_k = (\mathbf{m}_k\otimes '\mathbf{1}_{T\mathcal{H}}+\mathbf{1}_{T\mathcal{H}}\otimes '\mathbf{m}_k)\Delta_{T\mathcal{H}},\ee where $\mathbf{1}_{T\mathcal{H}}$ denotes the identity \emph{cohomomorphism} on $T\mathcal{H}$ (i.e.\ we have $\mathbf{1}_{T\mathcal{H}}(V)=V$ for all $V\in T\mathcal{H}$, as well as $\Delta_{T\mathcal{H}} \mathbf{1}_{T\mathcal{H}} = (\mathbf{1}_{T\mathcal{H}}\otimes'\mathbf{1}_{T\mathcal{H}})\Delta_{T\mathcal{H}}$). Defining the total coderivation \be\mathbf{m}=\sum_{k=1}^\infty \mathbf{m}_k,\ee the $A_\infty$ relations \eqref{eq:Ainfrel} can be succinctly expressed as
\begin{align}
[\mathbf{m},\mathbf{m}]=0\,.\label{eq:nilpm}
\end{align}
Introducing the bra-notation $\langle \omega|:\mathcal{H}^{\otimes 2}\longrightarrow \mathbb{C}$ for the symplectic form $\omega$ by writing $\omega(A_1,A_2)=\langle \omega|A_1\otimes A_2$, cyclicity of the coderivation $\mathbf{m}$ is simply expressed as
\begin{align}
\langle \omega |\pi_2 \mathbf{m}=0\,.
\end{align}
The UV SFT action \eqref{eq:ActInterp} is then rewritten as 
\begin{align}
S(\Psi) = \int_0^1 dt\, \langle \omega|\pi_1 \bs{\p}_t \frac{1}{1-\Psi(t)}\otimes \pi_1 \mathbf{m}\frac{1}{1-\Psi(t)}\,,\label{eq:ActCoder}
\end{align}
where $\bs{\p}_t$ is the coderivation corresponding to the operator $\p_t$ understood as a 1-string product on $\mathcal{H}$ and 
\begin{align}
V_{\Psi(t)}\equiv\frac{1}{1-\Psi(t)} = 1_{T\mathcal{H}}+\Psi(t)+\Psi(t)\otimes \Psi(t)+\ldots\in T\mathcal{H}\,,
\end{align}
is the \emph{group-like element} corresponding to $\Psi(t)\in\mathcal{H}$, which satisfies \be\Delta_{T\mathcal{H}} V_{\Psi(t)} = V_{\Psi(t)} \otimes ' V_{\Psi(t)} .\ee Introducing an infinitesimal variation
\begin{align}
\delta\Psi(t) = \pi_1 \bs{\delta}(t) \frac{1}{1-\Psi(t)}  \label{eq:varPsi}
\end{align}
for some even coderivation $\bs{\delta}(t)$, it is then straightforward to use cyclicity\footnote{Given any two coderivations $\mathbf{d}_1$, $\mathbf{d}_2$ and a cyclic coderivation $\mathbf{s}$ on $T\mathcal{H}$, it is possible to establish \cite{Erler:2015rra,Erler:2015uoa} the identity 
	\begin{align}
	&\langle \omega |\pi_1\mathbf{s}\mathbf{d}_1\frac{1}{1-A}\otimes \pi_1\mathbf{d}_2 \frac{1}{1-A}=-(-1)^{d(\mathbf{s})d(\mathbf{d}_1)}\langle \omega |\pi_1\mathbf{d}_1\frac{1}{1-A}\otimes \pi_1\mathbf{s}\mathbf{d}_2 \frac{1}{1-A}\label{eq:canceld}
	\end{align}
	for any $A\in\mathcal{H}$.
} of $\mathbf{m}$ to show that $\delta S(\Psi)$ receives contributions only from the boundary terms, namely
\begin{align}
\delta S(\Psi) = \langle \omega |\delta\Psi(1)\otimes \pi_1 \mathbf{m}\frac{1}{1-\Psi(1)}-\langle \omega |\delta\Psi(0)\otimes \pi_1 \mathbf{m}\frac{1}{1-\Psi(0)}\,.\label{eq:bdyVar}
\end{align}
Setting first $\bs{\delta}(0)=\bs{\delta}(1)=0$, so that $\delta\Psi(0)=\delta\Psi(1)=0$, then \eqref{eq:bdyVar} clearly gives $\delta S(\Psi)=0$ which confirms that the $t$-dependence in \eqref{eq:ActCoder} is purely topological. Second, setting $\bs{\delta}(0)=0$ and keeping $\bs{\delta}(1)$ (and therefore $\delta\Psi$) arbitrary, gives us the equation of motion for $\Psi$
\begin{align}
\text{EOM}(\Psi) = \pi_1 \mathbf{m}\frac{1}{1-\Psi} = m_1(\Psi)+m_2(\Psi,\Psi)+\ldots\,.
\end{align}
Finally, setting $\bs{\delta}(t)=[\mathbf{m},\mathbf{\Lambda}(t)]$ for a degree-odd cyclic coderivation $\mathbf{\Lambda}(t)$ such that $\mathbf{\Lambda}(0)=0$, $\mathbf{\Lambda}(1)=\mathbf{\Lambda}$, the action \eqref{eq:ActCoder} can also be shown (using cyclicity of both $\mathbf{m}$, $\mathbf{\Lambda}$, as well as \eqref{eq:nilpm}) to remain invariant under the infinitesimal gauge transformation\footnote{In general, it is possible to show that an infinitesimal transformation
	\begin{align}
	\delta \Psi = \pi_1 \mathbf{S} \frac{1}{1-\Psi}
	\end{align}
	generated by a cyclic degree-even coderivation $\mathbf{S}$ is a \emph{symmetry} of the action whenever we have $[\mathbf{m},\mathbf{S}]=0$. This condition is clearly satisfied when $\mathbf{S}=[\mathbf{m},\mathbf{\Lambda}]$, that is when $\mathbf{S}$ generates an infinitesimal gauge transformation (by invoking the $A_\infty$ relations and the super-Jacobi identity).
} 
\begin{align}
\delta\Psi = \pi_1[\mathbf{m},\mathbf{\Lambda}] \frac{1}{1-\Psi}\,.\label{eq:gauge}
\end{align}
Notice that unless we choose the coderivation $\mathbf{\Lambda}$ so that it corresponds purely to a 0-string product (that is $\pi_1\mathbf{\Lambda}\pi_k= 0$ for all $k>0$), the gauge transformation will contain trivial pieces which vanish on-shell. Indeed, in general we may consider $\mathbf{\Lambda} = \sum_{k=0}^\infty \mathbf{\Lambda}_k$ where $\mathbf{\Lambda}_k$ are cyclic coderivations corresponding to $k$-string products $\Lambda_k:\mathcal{H}^{\otimes k}\longrightarrow \mathcal{H}$. Expanding \eqref{eq:gauge} in terms of the products $m_k$, $\Lambda_k$, we would obtain
\begin{subequations}
	\begin{align}
	\delta\Psi&=\sum_{k=1}^\infty\sum_{r=0}^{\infty}[m_k,\Lambda_{r}](\Psi^{\otimes {k+r-1}})\\
	&=\sum_{k=1}^\infty\sum_{r=0}^\infty\sum_{l=0}^{k-1}\bigg\{m_k(\Psi^{\otimes l},\Lambda_{r}(\Psi^{\otimes r}),\Psi^{\otimes (k-l-1)})+\nonumber\\
	&\hspace{6cm}+\Lambda_r(\Psi^{\otimes l},m_{k}(\Psi^{\otimes k}),\Psi^{\otimes (r-l-1)})\bigg\}\,,\label{eq:245b}
	\end{align}
\end{subequations}
where the first term in \eqref{eq:245b} gives the usual gauge transformation (generally with a $\Psi$-dependent gauge parameter $\sum_{r=0}^\infty \Lambda_r(\Psi^{\otimes r})$), while the second term in \eqref{eq:245b} (which is present only if $\Lambda_r\neq 0$ for some $r>0$) constitutes a trivial transformation which vanishes on-shell.

\subsubsection{Unperturbed SDR}

Let us further define a projector \be\mathbf{P}:T\mathcal{H}\longrightarrow T\mathcal{H}\ee acting on the tensor-product space $T\mathcal{H}$ by requiring
\begin{align}
\mathbf{P}\pi_k = P^{\otimes k}\pi_k\,.
\end{align}
The map $\mathbf{P}$  clearly acts as a cohomomorphism, namely $\Delta_{T\mathcal{H}} \mathbf{P} = (\mathbf{P}\otimes'\mathbf{P})\Delta_{T\mathcal{H}}$. Given this definition, we can therefore write \be\mathbf{P}(T\mathcal{H})= (P\mathcal{H})^{\otimes 1}\oplus (P\mathcal{H})^{\otimes 2}\oplus \ldots\equiv TP\mathcal{H}\subset T\mathcal{H},\ee where $TP\mathcal{H}$ can again be equipped with a coassociative deconcatenation coproduct $\Delta_{TP\mathcal{H}}:TP\mathcal{H}\longrightarrow TP\mathcal{H}\otimes' TP\mathcal{H}$ (which is induced from $T\mathcal{H}$) so that the pair $(TP\mathcal{H},\Delta_{TP\mathcal{H}})$ constitutes a tensor coalgebra.
Defining also $\mathbf{Q}$ to be the coderivation corresponding to the 1-string product $Q$, we have $[\mathbf{Q},\mathbf{P}]=0$ (because $[Q,P]=0$ as a consequence of the decomposition \eqref{eq:HK} and the super-Jacobi identity). We also define the map $\mathbf{h}: T\mathcal{H}\longrightarrow T\mathcal{H}$
by requiring
\begin{align}
\mathbf{h}\pi_k &= \sum_{l=0}^{k-1}\big[(1_\mathcal{H})^{\otimes l}\otimes h\otimes P^{\otimes(k-1-l)}\big]\pi_k\,.\label{eq:huplift}
\end{align}
This definition can be motivated by the fact that $\mathbf{h}$ then satisfies the tensor coalgebra version of the Hodge-Kodaira decomposition
\begin{align}
\mathbf{Q}\mathbf{h}+\mathbf{h}\mathbf{Q}=\mathbf{1}_{T\mathcal{H}}-\mathbf{P}\equiv 
\bar{\mathbf{P}}\,,\label{eq:HKtens}
\end{align}
as it is easy to check explicitly. However, note that such $\mathbf{h}$ does not quite behave as a coderivation: instead we turn out to have
\begin{align}
\Delta_{T\mathcal{H}} \mathbf{h} = (\mathbf{h}\otimes '\mathbf{P}+\mathbf{1}_{T\mathcal{H}}\otimes' \mathbf{h})\Delta_{T\mathcal{H}}\,.
\end{align}
The annihilation conditions $h^2 =Ph=hP=0 $ clearly imply that $\mathbf{h}^2 =\mathbf{Ph}=\mathbf{hP}=0 $. 
Also note that we can formally separate $$\mathbf{P}=\mathbf{I\Pi},$$ where
$$\mathbf{\Pi}:T\mathcal{H}\longrightarrow TP\mathcal{H},$$ $$\mathbf{I}:TP\mathcal{H}\longrightarrow T\mathcal{H}$$
are the canonical projection and inclusion, respectively, mapping between $T\mathcal{H}$ and $TP\mathcal{H}$. On the other hand, we clearly have $$\mathbf{\Pi}\mathbf{I}=\mathbf{1}_{TP\mathcal{H}}.$$ Both $\mathbf{\Pi}$ and $\mathbf{I}$ again act as cohomomorphisms, that is we have $$\Delta_{T\mathcal{H}} \mathbf{I} = (\mathbf{I}\otimes'\mathbf{I})\Delta_{TP\mathcal{H}}$$ and $$\Delta_{TP\mathcal{H}}\mathbf{\Pi}=(\mathbf{\Pi}\otimes'\mathbf{\Pi})\Delta_{T\mathcal{H}}.$$
Since we also have the annihilation conditions $\mathbf{\Pi h} = \mathbf{hI}=0=\mathbf{h}^2$, as well as the Hodge-Kodaira decomposition \eqref{eq:HKtens}, we have therefore established the following tensor coalgebra version of the SDR \eqref{eq:SDRH}
\begin{align}
\mathrel{\raisebox{+14.5pt}{\rotatebox{-110}{  \begin{tikzcd}[sep=12mm,
			arrow style=tikz,
			arrows=semithick,
			diagrams={>={Straight Barb}}
			]
			\ar[to path={ node[pos=.5,C] }]{}
			\end{tikzcd}
}}}\hspace{-2.7mm}
(-\mathbf{h})\,(T\mathcal{H},\mathbf{Q})
\hspace{-3.2mm}\raisebox{-0.2pt}{$\begin{array}{cc} {\text{\scriptsize $\mathbf{\Pi}$}}\\[-3.0mm] \mathrel{\begin{tikzpicture}[node distance=1cm]
		\node (A) at (0, 0) {};
		\node (B) at (1.5, 0) {};
		\draw[->, to path={-> (\tikztotarget)}]
		(A) edge (B);
		\end{tikzpicture}}\\[-4mm] {\begin{tikzpicture}[node distance=1cm]
		\node (A) at (1.5, 0) {};
		\node (B) at (0, 0) {};
		\draw[->, to path={-> (\tikztotarget)}]
		(A) edge (B);
		\end{tikzpicture}}\\[-3.5mm]
	\text{\scriptsize $\mathbf{I}$} \end{array}$} \hspace{-3.4mm}
(TP\mathcal{H},\mathbf{\Pi Q I})\,.\label{eq:SDRtensor}
\end{align}
\vspace{-7mm}

\noindent
Apart from defining the co-derivation $\mathbf{m}=\sum_{k=1}^\infty\mathbf{m}_k$, we also define $\delta\mathbf{m}=\sum_{k=2}^\infty\mathbf{m}_k$ so that we can view the full interacting set of products $\mathbf{m}$ as a perturbation of the free-theory product $\mathbf{Q}$, namely $\mathbf{m}=\mathbf{Q}+\delta\mathbf{m}$.

\subsubsection{Perturbed inclusion \texorpdfstring{$\tilde{\mathbf{I}}$}{Itilde}}

Let us start with the full SFT action \eqref{eq:ActCoder}. Splitting the string field using the BPZ-even projector $P$ as in the previous subsection,  the equations of motion for $\psi$ and $\mathcal{R}$ (\ref{eq:EOMpsi},\ref{eq:EOMR}) can be expressed as
\begin{subequations}
	\begin{align}
	\text{EOM}_\psi(\Psi)&=\pi_1\mathbf{P}\mathbf{m}\frac{1}{1-\Psi} \,,\label{eq:EOMpsiCoalg}\\
	\text{EOM}_{\mathcal{R}}(\Psi)&=\pi_1\bar{\mathbf{P}}\mathbf{m}\frac{1}{1-\Psi} \,.
	\end{align}
\end{subequations}
Isolating the interacting part of the full equation of motion
\begin{align}
\mathcal{J}(\Psi) = \pi_1\delta\mathbf{m}\frac{1}{1-\Psi}\,,
\end{align}
we observe that having fixed the gauge $h R=0$ for $R$, the recursive relation \eqref{eq:RecRel} for $R(\psi)$ can be recast as
\begin{align}
R(\psi)=-\pi_1\mathbf{h}\delta\mathbf{m}\frac{1}{1-\Psi(\psi)}\,.
\end{align}
This therefore allows us to write the following equation for $\Psi(\psi)$
\begin{align}
\Psi(\psi)&= \psi-\pi_1\mathbf{h}\delta\mathbf{m}\frac{1}{1-\Psi(\psi)}\,.\label{eq:previous}
\end{align}
Expressing $\Psi(\psi)$ in terms of $\pi_1$ acting on the corresponding group-like element, moving the second term on the r.h.s.\ of \eqref{eq:previous} to the l.h.s., and finally, assuming that the map $\mathbf{1}_{T\mathcal{H}}+\mathbf{h}\delta\mathbf{m}$ is invertible, we can further rewrite \eqref{eq:previous} as
\begin{align}
\pi_1(\mathbf{1}_{T\mathcal{H}}+\mathbf{h}\delta\mathbf{m})\bigg(\frac{1}{1-\Psi(\psi)} - \frac{1}{\mathbf{1}_{T\mathcal{H}}+\mathbf{h}\delta\mathbf{m}}\mathbf{I}\frac{1}{1-\psi}\bigg)=0\,.\label{eq:EqCoh}
\end{align}
At this point, it is useful to note that the map $\tilde{\mathbf{I}}:TP\mathcal{H}\longrightarrow T\mathcal{H}$ defined by
\begin{align}
\tilde{\mathbf{I}} = \frac{1}{\mathbf{1}_{T\mathcal{H}}+\mathbf{h}\delta\mathbf{m}}\mathbf{I}\,\label{eq:defIt}
\end{align}
is in fact a cohomomorphism (by virtue of the annihilation conditions $\mathbf{hI}=\mathbf{\Pi h}=\mathbf{h}^2=0$; see e.g.\ Appendix A of \cite{Konopka:2015tta} for a proof). 
Using the fact that cohomomorphisms map  group-like elements to group-like elements, we can write the unique solution to \eqref{eq:EqCoh} satisfying $\Psi(0)=0$ as
\begin{align}
\Psi(\psi) = \pi_1\tilde{\mathbf{I}}\frac{1}{1-\psi}\,.\label{eq:Psi(psi)}
\end{align}
As we will see below, the image of $\tilde{\mathbf{I}}$ does not span the whole of $T\mathcal{H}$ so that the cohomomorphism $\tilde{\mathbf{I}}$ is only invertible on its image. Unpackaging the tensor coalgebra notation in terms of ordinary products on $\mathcal{H}$, we can write (thanks to \eqref{eq:huplift})
\begingroup\allowdisplaybreaks
\begin{subequations}
	\begin{align}
	\Psi(\psi) &=\pi_1 \tilde{\mathbf{I}}\frac{1}{1-\psi}\\
	&= \psi-\mathbf{hm}_2\mathbf{I}(\psi\otimes \psi)-(\mathbf{hm}_3-\mathbf{hm}_2\mathbf{hm}_2)\mathbf{I}(\psi\otimes \psi\otimes \psi)+\ldots\\
	&= \psi -hm_2(\psi,\psi)-hm_3(\psi,\psi,\psi)+\nonumber\\
	&\hspace{2cm}+hm_2(hm_2(\psi,\psi),\psi)+hm_2(\psi,hm_2(\psi,\psi))+\ldots
	\end{align}
\end{subequations}
\endgroup
which agrees with our previous result \eqref{eq:Psipsi}. 

\subsubsection{Perturbed projection \texorpdfstring{$\tilde{\mathbf{\Pi}}$}{Pitilde}}

Let us proceed with defining the cohomomorphism
\begin{align}
\tilde{\mathbf{\Pi}}=\mathbf{\Pi}\frac{1}{\mathbf{1}_{T\mathcal{H}}+\delta\mathbf{m}\mathbf{h}}\,.\label{eq:defPit}
\end{align}
To check that $\tilde{\mathbf{\Pi}}$ is indeed a cohomomorphism, we proceed in a parallel manner to the proof for $\tilde{\mathbf{I}}$.\footnote{This does not seem to work for general homotopy equivalence data, but only if we work with an SDR where we have the annihilation conditions $\mathbf{hI}=\mathbf{Ph}=\mathbf{h}^2=0$.} We then clearly have
\begin{subequations}
	\label{eq:5146}
	\begin{align}
	\tilde{\mathbf{\Pi}}\tilde{\mathbf{I}} &=\mathbf{\Pi}\frac{1}{\mathbf{1}_{T\mathcal{H}}+\delta\mathbf{m}\mathbf{h}} \frac{1}{\mathbf{1}_{T\mathcal{H}}+\mathbf{h}\delta\mathbf{m}}\mathbf{I}\\
	&=\mathbf{\Pi}\mathbf{I}\\[+1.5mm]
	&=\mathbf{1}_{TP\mathcal{H}}\,,
	\end{align}
\end{subequations}
where we have used that $\mathbf{h}^2 =0$. Also note that similarly $\mathbf{hI}=0$ implies $\tilde{\mathbf{\Pi}}\mathbf{I}=\mathbf{1}_{TP\mathcal{H}}$ and that $\mathbf{Ph}=0$ implies ${\mathbf{\Pi}}\tilde{\mathbf{I}}=\mathbf{1}_{TP\mathcal{H}}$.
We also define the cohomomorphism $\tilde{\mathbf{P}}:T\mathcal{H}\longrightarrow T\mathcal{H}$ by
\begin{subequations}
	\begin{align}
	\tilde{\mathbf{P}} &= \tilde{\mathbf{I}}\tilde{\mathbf{\Pi}}\\
	&=\frac{1}{\mathbf{1}_{T\mathcal{H}}+\mathbf{h}\delta\mathbf{m}}\mathbf{P}\frac{1}{\mathbf{1}_{T\mathcal{H}}+\delta\mathbf{m}\mathbf{h}}\,.
	\end{align}
\end{subequations}
Note that adding the interactions $\delta\mathbf{m}$ creates a new (perturbed) embedding of $TP\mathcal{H}$ inside $T\mathcal{H}$ given by the image of $\tilde{\mathbf{P}}$ (which is the same as the image of $\tilde{\mathbf{I}}$): given an element $\psi\in P\mathcal{H}$, it may be uniquely associated to an element $\Psi\in \text{im}\,\tilde{\mathbf{P}}\subset T\mathcal{H}$ (but not in the whole of $T\mathcal{H}$). Put in another way, the cohomomorphisms $\tilde{\mathbf{\Pi}}$ and $\tilde{\mathbf{I}}$ are invertible only if we restrict the domain of $\tilde{\mathbf{\Pi}}$ and the target of $\tilde{\mathbf{I}}$ on $\text{im}\,\tilde{\mathbf{P}}\subset T\mathcal{H}$.

\subsubsection{Effective products}

We can now substitute the solution for $\Psi(\psi)$ into the equation of motion \eqref{eq:EOMpsiCoalg} for $\psi$ which yields
\begingroup\allowdisplaybreaks
\begin{subequations}
	\begin{align}
	\text{eom}(\psi) 
	&= \pi_1 \mathbf{\Pi}\mathbf{m}\tilde{\mathbf{I}}\frac{1}{1-\psi}\\
	&\equiv\pi_1 \tilde{\mathbf{m}} \frac{1}{1-\psi}\,.\label{eq:eom}
	\end{align}
\end{subequations}
\endgroup
Here we have introduced a new map
\begin{subequations}
	\label{eq:defmt}
	\begin{align}
	\tilde{\mathbf{m}} 
	&\equiv\mathbf{\Pi}\mathbf{m}\tilde{\mathbf{I}}\\[+2mm]
	&=\mathbf{\Pi}\mathbf{Q}\mathbf{I}+\mathbf{\Pi}\delta\mathbf{m}\frac{1}{\mathbf{1}_{T\mathcal{H}}+\mathbf{h}\delta\mathbf{m}}\mathbf{I}\,\label{eff-products}
	\end{align}
\end{subequations}
and where in the last equality, we have used the fact that $\mathbf{P}\mathbf{h}=0$ (see also \eqref{eq:Altm} below for some alternative ways of expressing $\tilde{\mathbf{m}}$).
Let us show that $\tilde{\mathbf{m}}$ is a coderivation on $TP\mathcal{H}$: we first have
\begin{align}
\Delta_{TP\mathcal{H}}\tilde{\mathbf{m}}
&=\big(\mathbf{\Pi}\mathbf{m}\tilde{\mathbf{I}}\otimes' \mathbf{\Pi}\tilde{\mathbf{I}}+\mathbf{\Pi}\tilde{\mathbf{I}}\otimes'\mathbf{\Pi}\mathbf{m}\tilde{\mathbf{I}}\big)\Delta_{TP\mathcal{H}}\,,\label{eq:prev}
\end{align}	
where we note that $\mathbf{\Pi}\mathbf{h}=0$ implies $\mathbf{\Pi}\tilde{\mathbf{I}}= \mathbf{1}_{TP\mathcal{H}}$,
so that the map $\tilde{\mathbf{m}}=\mathbf{\Pi}\mathbf{m}\tilde{\mathbf{I}}$ is indeed a coderivation on $TP\mathcal{H}$. It is also straightforward to unpackage the coalgebra notation and see that the definition \eqref{eq:defmt} of $\tilde{\mathbf{m}}$ gives $k$-products $\pi_1\tilde{\mathbf{m}}\pi_k$ which precisely agree with the effective products \eqref{eq:effprod} computed in the previous subsection.

\subsubsection{Effective theory as a homotopy transfer}

By now it should be easy to observe that comparing the definitions \eqref{eq:defIt}, \eqref{eq:defPit}, \eqref{eq:defmt} with the output \eqref{eq:HPLdefs} of the homological perturbation lemma applied on the SDR \eqref{eq:SDRtensor} (where we perturb $\mathbf{Q}\to\mathbf{m}=\mathbf{Q}+\delta\mathbf{m}$) we have established the perturbed SDR
\begin{align}
\mathrel{\raisebox{+14.5pt}{\rotatebox{-110}{  \begin{tikzcd}[sep=12mm,
			arrow style=tikz,
			arrows=semithick,
			diagrams={>={Straight Barb}}
			]
			\ar[to path={ node[pos=.5,C] }]{}
			\end{tikzcd}
}}}\hspace{-2.7mm}
(-\tilde{\mathbf{h}})\,(T\mathcal{H},\mathbf{m})
\hspace{-3.2mm}\raisebox{-0.2pt}{$\begin{array}{cc} {\text{\scriptsize $\tilde{\mathbf{\Pi}}$}}\\[-3.0mm] \mathrel{\begin{tikzpicture}[node distance=1cm]
		\node (A) at (0, 0) {};
		\node (B) at (1.5, 0) {};
		\draw[->, to path={-> (\tikztotarget)}]
		(A) edge (B);
		\end{tikzpicture}}\\[-4mm] {\begin{tikzpicture}[node distance=1cm]
		\node (A) at (1.5, 0) {};
		\node (B) at (0, 0) {};
		\draw[->, to path={-> (\tikztotarget)}]
		(A) edge (B);
		\end{tikzpicture}}\\[-3.0mm]
	\text{\scriptsize $\tilde{\mathbf{I}}$} \end{array}$} \hspace{-3.4mm}
(TP\mathcal{H},\tilde{\mathbf{m}})\,,\label{eq:ContTilde}
\end{align}
\vspace{-7mm}

\noindent
provided that we also introduce (minus) the perturbed contracting homotopy\footnote{This can be shown to satisfy the expected property
	\begin{align}
	\Delta_{T\mathcal{H}}\tilde{\mathbf{h}}
	&=(\mathbf{1}_{T\mathcal{H}}\otimes'\tilde{\mathbf{h}}+\tilde{\mathbf{h}}\otimes'\tilde{\mathbf{P}})\Delta_{T\mathcal{H}}\,.
	\end{align}}
\begin{align}
\tilde{\mathbf{h}} 
&=\frac{1}{\mathbf{1}_{T\mathcal{H}}+\mathbf{h}\delta\mathbf{m}}\mathbf{h}\,.
\end{align}
The homological perturbation lemma therefore immediately tells us that $\tilde{\mathbf{m}}$ must be a nilpotent differential, namely $\tilde{\mathbf{m}}^2=0$. The products $\tilde{m}_k=\pi_1\tilde{\mathbf{m}}\pi_k$ encoded by the coderivation $\tilde{\mathbf{m}}$ therefore indeed satisfy $A_\infty$ relations, as we claimed in the previous subsection. Another consequence of the perturbed SDR \eqref{eq:ContTilde} is the corresponding chain-map relation
\begin{align}
\mathbf{m}\tilde{\mathbf{I}} = \tilde{\mathbf{I}}\tilde{\mathbf{m}}\,.\label{eq:mIIm}
\end{align}
This implies that the cohomomorphism $\tilde{\mathbf{I}}$ is in fact an $A_\infty$-morphism. In particular, in the cases where $\pi_1\tilde{\mathbf{m}}\pi_1=0$, this construction provides the minimal model for $(T\mathcal{H},\mathbf{m})$. Similarly we have the chain-map relation
\begin{align}
\tilde{\mathbf{\Pi}}\mathbf{m} = \tilde{\mathbf{m}}\tilde{\mathbf{\Pi}}\,.\label{eq:PmmP}
\end{align}
Also it is important to note that we can in fact express $\tilde{\mathbf{m}}$ in multiple ways as
\begin{align}
\tilde{\mathbf{m}} = \mathbf{\Pi}\mathbf{m}\tilde{\mathbf{I}}=\tilde{\mathbf{\Pi}}\mathbf{m}\mathbf{I}=\tilde{\mathbf{\Pi}}\mathbf{m}\tilde{\mathbf{I}}\,,\label{eq:Altm}
\end{align}
because \eqref{eq:mIIm} implies $\tilde{\mathbf{\Pi}}\mathbf{m}\mathbf{I}=\tilde{\mathbf{m}}\tilde{\mathbf{\Pi}}\mathbf{I}=\tilde{\mathbf{m}}$ and $\tilde{\mathbf{\Pi}}{\mathbf{m}}\tilde{\mathbf{I}} =\tilde{\mathbf{\Pi}}\tilde{\mathbf{I}}\tilde{\mathbf{m}}
=\tilde{\mathbf{m}}$, while \eqref{eq:PmmP} implies ${\mathbf{\Pi}}\mathbf{m}\tilde{\mathbf{I}}={\mathbf{\Pi}}\mathbf{I}\tilde{\mathbf{m}}=\tilde{\mathbf{m}}$. We therefore learn that the effective IR SFT interactions are given by a \emph{homotopy transfer} applied to the full UV SFT interactions. Finally, we note in passing that the perturbed Hodge-Kodaira decomposition 
\begin{align}
\tilde{\mathbf{h}}\mathbf{m}+\mathbf{m}\tilde{\mathbf{h}} = \mathbf{1}_{T\mathcal{H}}-\tilde{\mathbf{P}}\label{eq:pertHK}
\end{align}
together with the super-Jacobi identity imply that $[\tilde{\mathbf{P}},\mathbf{m}]=0$.

\subsubsection{Out-of-gauge constraints and classical solutions}

Let us now substitute the string field $\Psi(\psi)$ after integrating out the unwanted degrees of freedom (as expressed in terms of the cohomomorphism $\tilde{\mathbf{I}}$ in \eqref{eq:Psi(psi)}) into the full SFT equation of motion $\text{EOM}(\Psi)$. Using the chain-map property \eqref{eq:mIIm}, we can first write
\begin{align}
\text{EOM}(\Psi(\psi))&=\pi_1 \tilde{\mathbf{I}}\tilde{\mathbf{m}}\frac{1}{1-\psi}\,.
\end{align}
Realizing then that any coderivation $\mathbf{d}$ satisfies the identity (for any $A\in \mathcal{H}$)
\begin{align}
\mathbf{d} \frac{1}{1-A} = \frac{1}{1-A}\otimes \bigg(\pi_1 \mathbf{d}\frac{1}{1-A}\bigg)\otimes \frac{1}{1-A}
\end{align}
and recalling the form \eqref{eq:eom} of the equation of motion for $\psi$, we can finally express
\begingroup\allowdisplaybreaks
\begin{align}
\text{EOM}(\Psi(\psi)) &= \pi_1 \tilde{\mathbf{I}}\bigg\{\frac{1}{1-\psi}\otimes\text{eom}(\psi)\otimes \frac{1}{1-\psi}\bigg\}\,.	\label{eq:compeom}
\end{align}
\endgroup
We therefore obtain that $\text{eom}(\psi)=0$ implies $\text{EOM}(\Psi(\psi))=0$.
This means that once the effective equation of motion $\text{eom}(\psi)$ is satisfied, there are no additional constraints on the dynamics of $\Psi$. In other words, the out-of-gauge constraints $\text{EOM}_{\tilde{R}}$ are automatically satisfied on any solution of $\text{eom}$. Any classical solution $\psi^\ast\in TP\mathcal{H}$ of the effective theory therefore automatically provides a classical solution $\Psi^\ast\in \text{im}\,\tilde{\mathbf{P}}\subset T\mathcal{H}$ of the full theory which is given by 
\begin{align}
\Psi^\ast = \pi_1\tilde{\mathbf{I}}\frac{1}{1-\psi^\ast}\,.\label{eq:clCorresp}
\end{align}
Hence, the above-described effective framework enables us to look for certain solutions of the full SFT equation of motion $\text{EOM}(\Psi)$ by only working with a smaller (possibly finite) number of degrees of freedom $\psi$ which we can anticipate to be dominantly excited by such solutions. Then, after solving the effective equation of motion $\text{eom}(\psi)$ for $\psi^\ast$, we can always construct\footnote{Modulo possible issues with convergence of $\Psi^\ast$ after applying the cohomomorphism $\tilde{\mathbf{I}}$ on $\psi^\ast$.} a solution $\Psi^\ast$ to the full SFT equation of motion by using \eqref{eq:clCorresp}. 

We can summarize our discussion up to this point by saying that the homological perturbation lemma automatically takes care of integrating out degrees of freedom from an interacting $A_\infty$ SFT whenever the modes $\psi$ we wish to keep are given by a BPZ-even projector ${P}$, and, the remaining modes $R$ can be integrated out by a propagator ${h}$, where ${h}$ and ${P}$ are such that we may write an SDR of the form \eqref{eq:SDRH}. In other words, the lemma provides a way of packaging the Feynman diagram expansion of tree-level effective interactions in any $A_\infty$ SFT. The propagator (a.k.a.\ --minus-- the contracting homotopy operator) also implicitly imposes the gauge-fixing condition $hR=0$ in such a way that the out-of-gauge constraints are automatically satisfied upon using the equation of motion for the remaining modes $\psi$.

\subsubsection{Obstructions to marginal deformations and the massless equation of motion}

We shall now give a more tangible example of how the solutions of the tree-level effective equation of motion $\text{eom}(\psi)$ provide classical solutions in the full SFT. In particular, we set $P=P_0$ (where $P_0$ projects onto $\text{ker}\,L_0$) and look for continuously parametrized families of classical solutions of $\text{eom}(\psi)$
\begin{align}
\psi(\lambda)=\sum_{k=1}^\infty \lambda^k \psi_k\in P_0\mathcal{H}\,.\label{eq:start}
\end{align}
Then, expanding the equation of motion \eqref{eq:eom} order by order in $\lambda$ using the explicit expressions \eqref{eq:effprod}, we obtain equations
\begin{subequations}
	\label{eq:obstrEff}
	\begin{align}
	\text{eom}_1&= P_0 m_1(\psi_1)\,,\\
	\text{eom}_2&= P_0 m_1(\psi_2)+ P_0m_2(\psi_1,\psi_1)\,,\\
	\text{eom}_3&= P_0 m_1(\psi_3)+ P_0m_2(\psi_1,\psi_2)+ P_0m_2(\psi_2,\psi_1)+\nonumber\\
	&\hspace{1cm}+P_0m_3(\psi_1,\psi_1,\psi_1)-P_0m_2(h_0m_2(\psi_1,\psi_1),\psi_1)+\nonumber\\
	&\hspace{4.45cm}-P_0m_2(\psi_1,h_0m_2(\psi_1,\psi_1))\,,\\
	&\hspace{0.2cm}\vdots\nonumber
	\end{align}
\end{subequations}
where we have introduced the propagator $h_0 = (b_0/L_0)\bar{P}_0$ for the massive modes. The corresponding classical solution
\begin{align}
\Psi(\psi(\lambda)) = \pi_1 \tilde{\mathbf{I}} \frac{1}{1-\psi(\lambda)}
\end{align}
of the full SFT equation of motion can then be expanded order by order in $\lambda$ as
\begin{align}
\Psi(\psi(\lambda))=\sum_{k=1}^\infty \lambda^k \Psi_k\,,
\end{align}
where we have
\begingroup\allowdisplaybreaks
\begin{subequations}
	\label{eq:MargDefEff}
	\begin{align}
	\Psi_1 &=\psi_1 \,,\\
	\Psi_2 &=\psi_2 -h_0m_2(\psi_1 ,\psi_1 )\,,\\
	\Psi_3 &=\psi_3 -h_0m_2(\psi_1 ,\psi_2 )-h_0m_2(\psi_2 ,\psi_1 )+\nonumber\\
	&\hspace{1cm}-h_0m_3(\psi_1 ,\psi_1 ,\psi_1 )+h_0m_2(h_0m_2(\psi_1 ,\psi_1 ),\psi_1 )+\nonumber\\
	&\hspace{4.45cm}+h_0m_2(\psi_1 ,h_0m_2(\psi_1 ,\psi_1 ))\,.\\
	&\hspace{0.2cm}\vdots\nonumber
	\end{align}
\end{subequations}
\endgroup
It is therefore manifest (see e.g.\ \cite{Mattiello:2019gxc,Vosmera:2019mzw,Erbin:2019spp} and the references therein) that $\text{eom}_k$ should be interpreted precisely as the obstructions to exactness of the marginal deformation
$\Psi(\psi(\lambda))$ arising at order $\lambda^k$. The individual terms of $\Psi(\psi(\lambda))$ (as given by \eqref{eq:MargDefEff}) then exactly agree with order-by-order expansion of classical solution of the full SFT equation of motion which corresponds to a marginal deformation of the original perturbative vacuum.
We can therefore conclude that exactly marginal deformations of the given open-string background are in one-to-one correspondence (via the cohomomorphism $\tilde{\mathbf{I}}$) with those classical solutions $\psi(\lambda)$ to the $P_0$-effective equation of motion, which are continuously connected to the effective perturbative vacuum $\psi_\mathrm{v}=0$. Put in other words, the string fields $\psi(\lambda)$ traversing local minima of the $P_0$-effective potential such that $\psi(0)=0$, are in one-to-one correspondence with exactly marginal deformations of the full SFT for the background at hand. This is how the \emph{moduli spaces} of consistent open string backgrounds make their appearance in string field theory.

         \subsubsection{Cyclicity} \label{subsub:cyc}

Let us now proceed with showing that the effective products $\tilde{m}_k=\pi_1\tilde{\mathbf{m}}\pi_k$ are cyclic with respect to the following symplectic form $\langle \tilde{\omega}|:P\mathcal{H}^{\otimes 2}\longrightarrow \mathbb{C}$ on $P\mathcal{H}$
\begin{align}
\langle \tilde{\omega}|\pi_2   \equiv \langle \omega|  \pi_2{\mathbf{I}}\,,\label{eq:561}
\end{align}
whenever $\mathbf{m}$ is cyclic with respect to $\langle \omega|$.
Crucially, in order for this proof to work, we will need to assume that $P$ and $h$ are BPZ-selfconjugate, that is
\begin{subequations}
	\label{eq:BPZprop}
	\begin{align}
	\omega(\Psi_1,h\Psi_2)&=(-1)^{d(\Psi_1)}\omega(h\Psi_1,\Psi_2)\,,	\label{eq:BPZproph}\\
	\omega(\Psi_1,P\Psi_2)&=\omega(P\Psi_1,\Psi_2)\,,\label{eq:BPZpropP}
	\end{align}
\end{subequations}
for any $\Psi_1,\Psi_2\in \mathcal{H}$. The definition \eqref{eq:561} is clearly motivated by the expression \eqref{eq:Stilde} for the effective action we derived above (note that the same definition is used in \cite{Kajiura:2003ax}). In other words, the symplectic form $\tilde{\omega}$ is defined so that the cohomomorphism $\mathbf{I}$ is cyclic.\footnote{A cohomomorphism $\mathbf{F}:T\mathcal{H}\longrightarrow T\mathcal{H}'$ is said to be cyclic with respect to the symplectic forms $\omega$ and $\omega'$ on $T\mathcal{H}$ and $T\mathcal{H}'$ whenever we have
	\begin{align}
	\langle \omega'| \pi_2\mathbf{F} = \langle \omega |\pi_2\,.
	\end{align}
} On the other hand, neither the cohomomorphism $\mathbf{\Pi}$, nor the cohomomorphism $\mathbf{P}$ are cyclic because we clearly have 
\begin{align}
\omega(P\Psi_1,P\Psi_2) \neq \omega(\Psi_1,\Psi_2)
\end{align}
for general $\Psi_1,\Psi_2\in\mathcal{H}$.
Note that the definition of $\tilde{\omega}$ clearly makes it graded anti-symmetric. 

Given these assumptions, we will first show that also the perturbed inclusion cohomomorphism $\tilde{\mathbf{I}}$ is cyclic, namely
\begin{align}
\langle \tilde{\omega}|\pi_2  = \langle \omega| \pi_2 \tilde{\mathbf{I}}\,.\label{eq:cycItilde}
\end{align}
To show this, we note that the product $\nabla_{T\mathcal{H}}$ and the coproduct $\Delta_{T\mathcal{H}}$ satisfy the identity (see \cite{Erler:2015uba} for a more detailed discussion)
\begin{align}
\pi_{k+l} = \nabla_{T\mathcal{H}}(\pi_k\otimes '\pi_l)\Delta_{T\mathcal{H}}\,.\label{eq:split}
\end{align}
Applying the splitting property \eqref{eq:split} on $\pi_2$ and using the fact that $\tilde{\mathbf{I}}$ is a cohomomorphism, it is then possible to write
\begin{align}
\langle\omega|\pi_2\tilde{\mathbf{I}} = \langle\omega|\nabla_{T\mathcal{H}}(\pi_1\tilde{\mathbf{I}} \otimes' \pi_1\tilde{\mathbf{I}})\Delta_{TP\mathcal{H}}\,.
\end{align}
As a consequence of the BPZ properties \eqref{eq:BPZprop} and the annihilation conditions $hP=Ph=h^2 =0$, we clearly have
\begin{subequations}
	\label{eq:BPZ2}
	\begin{align}
	\omega(I\psi_1,h\Psi_2)&=0\,,\\
	\omega(h\Psi_1,h\Psi_2)&=0\,,
	\end{align}
\end{subequations}
for any $\psi_1\in P\mathcal{H}$, $\Psi_1,\Psi_2\in\mathcal{H}$.  Noting then that we have $\pi_1 \mathbf{I}= I\pi_1$, $\pi_1 \mathbf{h}= h\pi_1$ and expanding the perturbed inclusion $\tilde{\mathbf{I}}$ in terms of $\mathbf{I},\mathbf{h},\delta\mathbf{m}$, we therefore learn that
\begin{subequations}
	\begin{align}
	\langle\omega|\nabla_{T\mathcal{H}}(\pi_1\tilde{\mathbf{I}} \otimes' \pi_1\tilde{\mathbf{I}})\Delta_{TP\mathcal{H}}&=\langle\omega|\nabla_{T\mathcal{H}}(\pi_1{\mathbf{I}} \otimes' \pi_1{\mathbf{I}})\Delta_{TP\mathcal{H}}\\
	&=\langle \omega|  \pi_2{\mathbf{I}}\\
	&=\langle \tilde{\omega}|  \pi_2\,,
	\end{align}
\end{subequations}
which concludes the proof of \eqref{eq:cycItilde}.

By exploiting cyclicity of the cohomomorphism $\tilde{\mathbf{I}}$, it is now straightforward to show that the coderivation $\tilde{\mathbf{m}}$ is cyclic. This is because we in addition have the property $[\mathbf{m},\tilde{\mathbf{P}}]=0$. Indeed we can first use cyclicity of $\tilde{\mathbf{I}}$ to write
\begin{align}
\langle \tilde{\omega}|\pi_2 \tilde{\mathbf{m}} = \langle \omega|\pi_2 \tilde{\mathbf{I}}\tilde{\mathbf{\Pi}}\mathbf{m}\tilde{\mathbf{I}}\,.
\end{align}
Realizing that $\tilde{\mathbf{I}}\tilde{\mathbf{\Pi}}=\tilde{\mathbf{P}}$ and that $[\mathbf{m},\tilde{\mathbf{P}}]=0$, we can therefore write
\begin{subequations}
	\begin{align}
	\langle\tilde{\omega}|\pi_2\tilde{\mathbf{m}}
	&=\langle{\omega}|\pi_2{\mathbf{m}}\tilde{\mathbf{P}}\tilde{\mathbf{I}}\\[0.3mm]
	&=0\,,
	\end{align}
\end{subequations}
which gives us the required result. This shows that the effective products $\tilde{m}_k$ are cyclic with respect to the symplectic form $\tilde{\omega}$. It is instructive to work out the first couple of orders explicitly. While the results for $\tilde{m}_1$ and $\tilde{m}_2$ are arguably trivial, for $\tilde{m}_3$ we can write (for any $A_1,A_2,A_3,A_4\in P\mathcal{H}$)
\begingroup\allowdisplaybreaks
\begin{subequations}
	\begin{align}
	\tilde{\omega}(A_1,\tilde{m}_3(A_2,A_3,A_4))&=\omega(A_1,Pm_3(A_2,A_3,A_4))+\nonumber\\
	&\hspace{2cm}-\omega(A_1,Pm_2(hm_2(A_2,A_3),A_4))\nonumber\\
	&\hspace{2cm}-\omega(A_1,Pm_2(A_2,hm_2(A_3,A_4)))\\
	&=-(-1)^{d(A_1)}\omega(Pm_3(A_1,A_2,A_3),A_4)+\nonumber\\
	&\hspace{2cm}+(-1)^{d(A_1)}\omega(Pm_2(A_1,hm_2(A_2,A_3)),A_4)\nonumber\\
	&\hspace{2cm}+(-1)^{d(A_1)}\omega(m_2(A_1,A_2),hm_2(A_3,A_4))\\
	&=-(-1)^{d(A_1)}\omega(Pm_3(A_1,A_2,A_3),A_4)+\nonumber\\
	&\hspace{2cm}+(-1)^{d(A_1)}\omega(Pm_2(A_1,hm_2(A_2,A_3)),A_4)\nonumber\\
	&\hspace{2cm}-(-1)^{d(A_2)}\omega(hm_2(A_1,A_2),m_2(A_3,A_4))\\
	&=-(-1)^{d(A_1)}\omega(Pm_3(A_1,A_2,A_3),A_4)+\nonumber\\
	&\hspace{2cm}+(-1)^{d(A_1)}\omega(Pm_2(A_1,hm_2(A_2,A_3)),A_4)\nonumber\\
	&\hspace{2cm}+(-1)^{d(A_1)}\omega(Pm_2(hm_2(A_1,A_2),A_3),A_4)\\
	&= -(-1)^{d(A_1)}\tilde{\omega}(\tilde{m}_3(A_1,A_2,A_3),A_4)\,,
	\end{align}
\end{subequations}
\endgroup
where in the first step we have substituted from \eqref{eq:effprod}, in the second step we have
used the BPZ property \eqref{eq:BPZpropP} of $P$ and cyclicity of $m_2$, $m_3$, in the third step we have used the BPZ property \eqref{eq:BPZproph}, while in the fourth step we have again made use of cyclicity of $m_2$, thus finally showing that $\tilde{m}_3$ is cyclic with respect to $\tilde{\omega}$. Similarly for $\tilde{m}_4$, $\tilde{m}_5$ and so on.

Finally, let us consider a general coderivation $\mathbf{d}: T\mathcal{H}\longrightarrow T\mathcal{H}$ which is cyclic with respect to\ $\omega$, that is $\langle {\omega}|\pi_2 {\mathbf{d}}=0$, but, which may \emph{not} satisfy that $[\mathbf{d},\tilde{\mathbf{P}}]=0$.
We will now show that then the coderivation $\tilde{\mathbf{d}}=\tilde{\mathbf{\Pi}}\mathbf{d}\tilde{\mathbf{I}}$ is still cyclic with respect to\ $\tilde{\omega}$, namely $\langle\tilde{\omega}|\pi_2 \tilde{\mathbf{d}}=0$.
Following some straightforward manipulations, we first write
\begingroup\allowdisplaybreaks
\begin{subequations}
	\begin{align}
	\langle\tilde{\omega}|\pi_2 \tilde{\mathbf{d}} 
	&= 	\langle{\omega}|\pi_2\tilde{\mathbf{I}}\tilde{\mathbf{\Pi}} {\mathbf{d}} \tilde{\mathbf{I}}\\
	&= 	\langle{\omega}|\pi_2\big(\mathbf{1}_{T\mathcal{H}}-\mathbf{m}\tilde{\mathbf{h}}-\tilde{\mathbf{h}}\mathbf{m}\big) {\mathbf{d}} \tilde{\mathbf{I}}\\
	&= -\langle{\omega}|\pi_2 \tilde{\mathbf{h}}\mathbf{m} {\mathbf{d}} \tilde{\mathbf{I}}\,,
	\end{align}
\end{subequations}
\endgroup
where we have first used cyclicity of $\tilde{\mathbf{I}}$, then we have realized that $\tilde{\mathbf{I}}\tilde{\mathbf{\Pi}}$, substituted the perturbed Hodge-Kodaira decomposition \eqref{eq:pertHK} and finally made use of cyclicity of both $\mathbf{m}$ and $\mathbf{d}$. Using now the property \eqref{eq:split} in the form $\pi_2 = \nabla_{T\mathcal{H}}(\pi_1\otimes'\pi_1)\Delta_{T\mathcal{H}}$, as well as the perturbed annihilation condition $\tilde{\mathbf{h}}\tilde{\mathbf{I}}=0$ and the chain-map property $\mathbf{m}\tilde{\mathbf{I}}=\tilde{\mathbf{I}}\tilde{\mathbf{m}}$, we eventually find
\begin{align}
\langle{\omega}|\pi_2 \tilde{\mathbf{h}}\mathbf{m} {\mathbf{d}} \tilde{\mathbf{I}}
&= 	\langle{\omega}| \nabla_{T\mathcal{H}}(\pi_1\tilde{\mathbf{h}}\mathbf{m} {\mathbf{d}}\tilde{\mathbf{I}}\otimes' \pi_1\tilde{\mathbf{P}}\tilde{\mathbf{I}}+(-1)^{d(\mathbf{d})} \pi_1\tilde{\mathbf{h}}{\mathbf{d}}\tilde{\mathbf{I}}\otimes' \pi_1\tilde{\mathbf{P}}\mathbf{m}\tilde{\mathbf{I}}+\nonumber\\
&\hspace{2.5cm}-
\pi_1\tilde{\mathbf{P}}\tilde{\mathbf{I}}\tilde{\mathbf{m}}\otimes'\pi_1\tilde{\mathbf{h}} {\mathbf{d}}\tilde{\mathbf{I}}+\pi_1\tilde{\mathbf{P}}\tilde{\mathbf{I}}\otimes'\pi_1\tilde{\mathbf{h}}\mathbf{m}{\mathbf{d}}\tilde{\mathbf{I}}
)\Delta_{TP\mathcal{H}}\,.
\end{align}
Finally, we note that when $\tilde{\mathbf{h}}$ acts on anything, the $\pi_1$ projection of the result will always have an overall factor of $h$ in front. Similarly, when $\tilde{\mathbf{P}}$ acts on anything, the $\pi_1$ projection of the result will be always have an overall factor of either $P$ or $h$ in front. But at the same time, we have the BPZ properties \eqref{eq:BPZ2} so that this leads us to conclude that we indeed have
\begin{align}
\langle\tilde{\omega}|\pi_2 \tilde{\mathbf{d}} =0\,.\label{eq:cyct}
\end{align}

\subsubsection{Effective action}

Realizing that the coderivation $\tilde{\mathbf{m}}$ encoding the effective multi-string products $\tilde{m}_k = \pi_1 \tilde{\mathbf{m}}\pi_k$ is cyclic with respect to\ $\tilde{\omega}$, we can conclude that the equation of motion \eqref{eq:eom} must be reproduced by the action
\begin{align}
\tilde{S}(\psi) = \int_0^1 dt \,\langle \tilde{\omega} |\pi_1 \bs{\p}_t \frac{1}{1-\psi(t)}\otimes \pi_1 \tilde{\mathbf{m}} \frac{1}{1-\psi(t)}=\sum_{k=1}^{\infty}\frac1{k+1}\omega\left(\psi,\tilde {m}_k\left(\psi^{\otimes k}\right)\right)\,,\label{eq:EffActC}
\end{align}
where we have introduced an interpolation $\psi(t)\in \Pi\mathcal{H}$ for $0\leq t \leq 1$ with $\psi(0)=0$ and $\psi(1)=\psi$.
Since we have shown that the equation of motion $\text{eom}(\psi)$ automatically implies the full equation of motion $\text{EOM}(\Psi(\psi))$, the action \eqref{eq:EffActC} fully captures the dynamics of $\psi$ and can be therefore called the effective action for $\psi$. Let us now show that \eqref{eq:EffActC} can be also derived by directly substituting the group-like element
\begin{align}
\frac{1}{1-\Psi(\psi)} = \tilde{\mathbf{I}}\frac{1}{1-\psi}
\end{align}
into the full SFT action \eqref{eq:ActCoder}. To this end, let us choose a particular interpolation $\Psi(t)\in\mathcal{H}$, namely such that
\begin{align}
\frac{1}{1-\Psi(t)} = \tilde{\mathbf{I}}\frac{1}{1-\psi(t)}\,.\label{eq:interp}
\end{align}
Note that this is a valid choice, because $\tilde{\mathbf{I}}$ maps group-like elements on $TP\mathcal{H}$ to group-like elements on $\text{im}\,\tilde{\mathbf{P}}\subset T\mathcal{H}$ and we also have $\Psi(0) =  \pi_1\tilde{\mathbf{I}}1_{TP\mathcal{H}}=0$. Substituting \eqref{eq:interp} into the action \eqref{eq:ActCoder}, we first obtain
\begin{align}
S(\Psi(\psi)) 
&= \int_0^1 dt\, \langle \omega |\pi_1 \tilde{\mathbf{I}} \bs{\p}_t \frac{1}{1-\psi(t)}\otimes \pi_1 \tilde{\mathbf{I}}\tilde{\mathbf{m}}\frac{1}{1-\psi(t)}
\end{align}
where we have used the chain-map relation \eqref{eq:mIIm}, as well as the fact that $[\tilde{\mathbf{I}},\bs{\p}_t]=0$. We then have to realize that for any two coderivations $\mathbf{d}_1$, $\mathbf{d}_2$ on $T\mathcal{H}$ and any cohomomorphism $\mathbf{F}:T\mathcal{H}\longrightarrow T\mathcal{H}'$ which is cyclic with respect to the symplectic forms $\langle \omega|$, $\langle \omega'|$, we have the identity (see \cite{Erler:2015rra,Erler:2015uoa} for a proof)
\begin{align}
\langle \omega'|\pi_1\mathbf{F}\mathbf{d}_1 \frac{1}{1-A}\otimes \pi_1 \mathbf{F}\mathbf{d}_2 \frac{1}{1-A} = \langle \omega |\pi_1 \mathbf{d}_1\frac{1}{1-A}\otimes\pi_1 \mathbf{d}_2\frac{1}{1-A}\label{eq:cancelF}
\end{align}
for any $A\in\mathcal{H}$. Using cyclicity of $\tilde{\mathbf{I}}$ and \eqref{eq:cancelF}, we can finally write 
\begingroup\allowdisplaybreaks
\begin{subequations}
	\label{eq:compEffAct}
	\begin{align}
	S(\Psi(\psi)) &= \int_0^1 dt\, \langle \omega |\pi_1  \bs{\p}_t \frac{1}{1-\psi(t)}\otimes \pi_1 \tilde{\mathbf{m}}\frac{1}{1-\psi(t)}\\[2mm]
	&=\tilde{S}(\psi)\,.
	\end{align}
\end{subequations}
We have therefore managed to reproduce the result \eqref{eq:EffActC} for the effective IR SFT action by directly substituting $\Psi(\psi)$ into the full UV SFT action \eqref{eq:ActCoder}.

\subsubsection{Gauge transformation}

We will now look at the interplay of the homotopy transfer with the gauge transformation in both UV and IR. Consider first a gauge transformation of the effective SFT
\begin{align}
\delta \psi =\pi_1 [\tilde{\mathbf{m}},\boldsymbol{\lambda}] \frac{1}{1-\psi}\,,\label{eq:IRgauge}
\end{align}
where the degree-odd coderivation $\bs{\lambda}$ plays the role of a gauge parameter. One should then be interested into the corresponding gauge transformation of the full UV SFT induced on the image of $\tilde{\mathbf{P}}$ from \eqref{eq:IRgauge} by acting with the perturbed inclusion $\tilde{\mathbf{I}}$ on \eqref{eq:IRgauge}. We clearly have
\begin{align}
\delta \Psi(\psi) &=\pi_1\tilde{\mathbf{I}}[\tilde{\mathbf{m}},\boldsymbol{\lambda}]\tilde{\mathbf{\Pi}} \frac{1}{1-\Psi(\psi)}\,.
\end{align}
Recalling that the chain map properties \eqref{eq:mIIm} and \eqref{eq:PmmP} give us that $\tilde{\mathbf{I}}\tilde{\mathbf{m}}\tilde{\mathbf{\Pi}} = \tilde{\mathbf{P}}\mathbf{m}=\mathbf{m}\tilde{\mathbf{P}}$, we note that we can rewrite the gauge transformation induced on the UV SFT in terms of
$\tilde{\mathbf{I}}[\tilde{\mathbf{m}},\boldsymbol{\lambda}]\tilde{\mathbf{\Pi}} =[\mathbf{m},\mathbf{\Lambda}]$ where we have denoted 
\begin{align}
\mathbf{\Lambda} \equiv \tilde{\mathbf{I}}\boldsymbol{\lambda}\tilde{\mathbf{\Pi}}\,,
\end{align}
which clearly satisfies $\tilde{\mathbf{P}}\mathbf{\Lambda}=\mathbf{\Lambda}\tilde{\mathbf{P}}=\mathbf{\Lambda}$.
It can be easily shown that the parameter $\mathbf{\Lambda}$ is a coderivation only when restricted onto $\text{im}\,\tilde{\mathbf{P}}$, and, that $\mathbf{\Lambda}$ is cyclic provided that also $\boldsymbol{\lambda}$ is cyclic. Altogether we learn that the effective gauge transformation with parameter $\bs{\lambda}$ induces a gauge transformation on $\text{im}\,\tilde{\mathbf{P}}\subset T\mathcal{H}$ with parameter $\mathbf{\Lambda} = \tilde{\mathbf{I}}\boldsymbol{\lambda}\tilde{\mathbf{\Pi}}$. 

Observe that even if we choose $\boldsymbol{\lambda}$ to correspond only to a 0-string product (that is $\boldsymbol{\lambda}=\boldsymbol{\lambda}_0$ and $\pi_1\boldsymbol{\lambda}\pi_k=0$ for $k>0$) then we generally have $\pi_1\mathbf{\Lambda}\pi_k\neq 0$ for $k>0$. For instance, unpackaging the coalgebra notation, we obtain
\begin{subequations}
	\begin{align}
	\pi_1\mathbf{\Lambda}\pi_1
	&=	\pi_1\frac{1}{\mathbf{1}_{T\mathcal{H}}+\mathbf{h}\delta\mathbf{m}}{\mathbf{I}}\boldsymbol{\lambda}_0{\mathbf{\Pi}}\frac{1}{\mathbf{1}_{T\mathcal{H}}+\delta\mathbf{m}\mathbf{h}}\pi_1\\
	&=	\pi_1\frac{1}{\mathbf{1}_{T\mathcal{H}}+\mathbf{h}\delta\mathbf{m}}(I\lambda_0\otimes P+P\otimes I\lambda_0)\\[+2mm]
	&=	-hm_2(I\lambda_0\otimes P)-hm_2(P\otimes I\lambda_0)\,,
	\end{align}
\end{subequations}
that is, the induced coderivation $\mathbf{\Lambda}$ contains a 1-string product $\Lambda_1$ with
\begin{align}
\Lambda_1(\Psi) &=-hm_2(\lambda_0,P\Psi)-hm_2(P\Psi,\lambda_0)
\end{align}
(as well as the 0-string product $\pi_1 \mathbf{\Lambda}\pi_0  =\lambda_0$ and higher products $\pi_1\mathbf{\Lambda}\pi_k$ for $k>1$). As a result, the gauge transformation induced on the full SFT will contain trivial pieces which vanish on-shell.

Going in the other direction, let us consider a gauge transformation of the full theory 
\begin{align}
\delta\Psi = [\mathbf{m},\mathbf{\Lambda}]\frac{1}{1-\Psi}\,,
\end{align}
where $\mathbf{\Lambda}$ is a cyclic coderivation. This clearly generates the transformation of the effective theory
\begin{align}
\delta\psi = \pi_1\tilde{\mathbf{\Pi}}[\mathbf{m},\mathbf{\Lambda}]\tilde{\mathbf{I}}\frac{1}{1-\psi}\,,
\end{align}
on $TP\mathcal{H}$. It can be straightforwardly shown that $\tilde{\mathbf{\Pi}}[\mathbf{m},\mathbf{\Lambda}]\tilde{\mathbf{I}}$ is always a coderivation on $TP\mathcal{H}$. Note that using the fact that $[\mathbf{m},\tilde{\mathbf{P}}]=0$, we also have
$\tilde{\mathbf{\Pi}}[\mathbf{m},\mathbf{\Lambda}]\tilde{\mathbf{I}}=[\tilde{\mathbf{m}},{\boldsymbol{\lambda}}]$ where we have denoted
\begin{align}
\boldsymbol{\lambda}=\tilde{\mathbf{\Pi}}\mathbf{\Lambda}\tilde{\mathbf{I}}\,.
\end{align}
Given that $\mathbf{\Lambda}$ is a cyclic coderivation, it follows from \eqref{eq:cyct} that $\boldsymbol{\lambda}$ is also a cyclic coderivation. We can therefore conclude that an infinitesimal gauge symmetry $\mathbf{\Lambda}$ of the full theory on $T\mathcal{H}$ always induces an infinitesimal gauge symmetry on $TP\mathcal{H}$, which is generated by $\boldsymbol{\lambda}=\tilde{\mathbf{\Pi}}\mathbf{\Lambda}\tilde{\mathbf{I}}$.

\subsection{Observables}\label{sec:obs}

Having discussed at some length the effective actions for $A_\infty$ SFTs, as well as their associated gauge symmetries, it is only fitting that we now turn to considering the fate of observables (gauge-invariant quantities) after integrating out some portion of degrees of freedom. We will start by briefly discussing a possible framework for observables within the context of $A_\infty$ SFTs (more details are to be presented in \cite{Schnabl:2020xx}). We will then show that applying the homotopy transfer onto an observable falling into this class always yields an observable for the effective theory.

\subsubsection{General discussion}

Let $(\mathcal{H},\{m_k\}_{k\geq 1},\omega)$ be a cyclic $A_\infty$ algebra defining an $A_\infty$ SFT given by the action \eqref{eq:SAinf} for a degree-even string field $\Psi$. Let us start by considering a quantity
\begin{align}
\mathcal{E}(\Psi) = \sum_{k=0}^\infty \frac{1}{k+1}\omega(\Psi,e_k(\Psi^{\otimes k}))\,,\label{eq:generalForm}
\end{align}
where $e_k$ are cyclic degree-odd products.
It is easy to see that in a manner completely parallel to the action, this may be rewritten in the tensor coalgebra notation as
\begin{align}
\mathcal{E}(\Psi) =\int_0^1 dt\,\langle\omega| \pi_1 \bs{\p}_t \frac{1}{1-\Psi(t)}\otimes \pi_1 \mathbf{e} \frac{1}{1-\Psi(t)}\,,\label{eq:generalFormTensor}
\end{align}
where $\mathbf{e}=\sum_k\mathbf{e}_k$ and $\mathbf{e}_k$ are the cyclic coderivations corresponding via \eqref{eq:corresp} to the cyclic products $e_k$. As usual, $\Psi(t)$ for $0\leq t\leq 1$ interpolates between $\Psi(0)=0$ and $\Psi(1)=\Psi$. We would now like to isolate conditions on $\mathbf{e}$ such that $\mathcal{E}(\Psi)$ is gauge-invariant (possibly up to trivial pieces which vanish on-shell). To this end, let us introduce an infinitesimal variation $\delta\Psi(t)$ given by \eqref{eq:varPsi} in terms of a degree-even coderivation $\bs{\delta}(t)$. As in the case of the action, it is possible to use cyclicity of $\mathbf{e}$ (by applying \eqref{eq:canceld}) to show that $\delta\mathcal{E}(\Psi)$ receives contributions only from the boundary terms, namely
\begin{align}
\delta\mathcal{E}(\Psi) = \langle \omega |\delta\Psi(1)\otimes \pi_1 \mathbf{e}\frac{1}{1-\Psi(1)}-\langle \omega |\delta\Psi(0)\otimes \pi_1 \mathbf{e}\frac{1}{1-\Psi(0)}\,.
\end{align}
Setting first $\delta\Psi(0)=\delta\Psi(1)=0$ again serves to confirm that the $t$-dependence in \eqref{eq:generalFormTensor} is topological. On the other hand, considering the gauge transformation $\bs{\delta}(t)=[\mathbf{m},\mathbf{\Lambda}(t)]$ where $\mathbf{\Lambda}(t)$ with $\mathbf{\Lambda}(0)=0$, $\mathbf{\Lambda}(1)=\mathbf{\Lambda}$ is a degree-odd cyclic coderivation, we obtain (denoting by $\ldots$ pieces that vanish on-shell),
\begin{align}
\delta\mathcal{E}(\Psi) = \langle \omega |\pi_1\mathbf{\Lambda} \frac{1}{1-\Psi}\otimes \pi_1 [\mathbf{m},\mathbf{e}]\frac{1}{1-\Psi}+\ldots\,,
\end{align}
where we have used cyclicity of $\mathbf{m}$ through the property \eqref{eq:canceld}.
Hence, the condition on $\mathcal{E}(\Psi)$ to be an observable (on-shell gauge-invariant) reads
\begin{align}
[\mathbf{m},\mathbf{e}] \frac{1}{1-\Psi^\ast}=0\label{eq:onshell}
\end{align}
for any classical solution $\Psi^\ast$. In particular, in order for \eqref{eq:onshell} to be satisfied, it is therefore sufficient to require that $[\mathbf{m},\mathbf{e}]=0$. An example of an observable, which is present for any $A_\infty$ SFT, is clearly the action, because setting $\mathbf{e}=\mathbf{m}$ gives $[\mathbf{e},\mathbf{m}]=[\mathbf{m},\mathbf{m}]=0$. Another example of an observable is the Ellwood invariant \cite{Hashimoto:2001sm, Gaiotto:2001ji, Ellwood:2008jh} in cubic OSFT, which is given by $\mathbf{e}=\mathbf{e}_0$, where $\mathbf{e}_0$ is the coderivation corresponding to a 0-string product $e_0$ whose output is a midpoint insertion of an on-shell primary $(h,\bar{h})=(0,0)$ closed-string state. This has been recently generalized \cite{jak-talk} to the case of the ``Munich'' $A_\infty$ open superstring field theory \cite{Erler:2013xta}, where the corresponding coderivation $\mathbf{e}=\mathbf{E}$ turns out to consist of $k$-string products $E_k$ for all $k\geqslant 0$ (details will be reported in \cite{MV,Schnabl:2020xx}). In both the cubic OSFT case and the Munich case, the coderivation $\mathbf{e}$ is nilpotent with $[\mathbf{e},\mathbf{m}]=0$, so that introducing the perturbed coderivation $\mathbf{M}(\mu)=\mathbf{m}+\mu\mathbf{e}$, we obtain
\begin{align}
\mathbf{M}(\mu)^2 = \mathbf{m}^2 +\mu[\mathbf{m},\mathbf{e}]+\mu^2\mathbf{e}^2=0\,.
\end{align}
Hence, any nilpotent coderivation $\mathbf{e}$ giving rise to an observable with $[\mathbf{e},\mathbf{m}]=0$ via \eqref{eq:generalFormTensor} can be used to deform the products of the theory so that they continue to satisfy $A_\infty$ relations.\footnote{In general, these will be the \emph{weak} $A_\infty$ relations, that is, we will have $M_0(\mu)\neq 0$.} See below subsections \ref{subsec:VertComp} and \ref{Ellwood} for more details.

Also note that any observable for which we can write $\mathbf{e}=[\mathbf{m},\mathbf{s}]$ for some arbitrary degree-even cyclic coderivation $\mathbf{s}$, is automatically trivial: while it is true that we then have $[\mathbf{m},\mathbf{e}]=0$ by super-Jacobi identity, we can use cyclicity of both $\mathbf{s}$ and $\mathbf{m}$ to show that $\mathcal{E}(\Psi)$ is equal to the variation of the action induced by $\mathbf{s}$ so that on-shell it necessarily vanishes.

\subsubsection{Homotopy transfer of observables}

Let us now assume that a degree-odd cyclic coderivation $\mathbf{e}$ provides via \eqref{eq:generalFormTensor} an observable for the full theory on $T\mathcal{H}$ (so that \eqref{eq:onshell} holds for any classical solution $\Psi^\ast$ of the full SFT). Introducing now a new degree-odd coderivation $\tilde{\mathbf{e}} = \tilde{\mathbf{\Pi}}\mathbf{e}\tilde{\mathbf{I}}$ on $TP\mathcal{H}$ (which is cyclic by virtue of the discussion preceding \eqref{eq:cyct}), it is straightforward to show (for any classical solution $\psi^\ast$ of the effective theory)
\begin{subequations}
	\begin{align}
	[\tilde{\mathbf{m}},\tilde{\mathbf{e}}]\frac{1}{1-\psi^\ast}&=(\tilde{\mathbf{\Pi}}{\mathbf{m}}\tilde{\mathbf{P}}{\mathbf{e}}\tilde{\mathbf{I}}+\tilde{\mathbf{\Pi}}{\mathbf{e}}\tilde{\mathbf{P}}{\mathbf{m}}\tilde{\mathbf{I}})\frac{1}{1-\psi^\ast}\\
	&=\tilde{\mathbf{\Pi}}[{\mathbf{m}},{\mathbf{e}}]\tilde{\mathbf{I}}\frac{1}{1-\psi^\ast}\,,\label{eq:Pime}
	\end{align}
\end{subequations}
where we have used that $[\mathbf{m},\tilde{\mathbf{P}}]=0$ as well as that $\tilde{\mathbf{P}}\tilde{\mathbf{I}}=\tilde{\mathbf{I}}$ and $\tilde{\mathbf{\Pi}}\tilde{\mathbf{P}}=\tilde{\mathbf{\Pi}}$. Recalling then \eqref{eq:clCorresp}, we note that since $\Psi(\psi^\ast)=\tilde{\mathbf{I}}\frac{1}{1-\psi^\ast}$ is a classical solution of the full SFT, then \eqref{eq:Pime} needs to vanish by virtue of \eqref{eq:onshell}. This shows that $\tilde{\mathbf{e}}$ satisfies the condition \eqref{eq:onshell} as well, so that the quantity
\begin{align}
\tilde{\mathcal{E}}(\psi) &=
\int_0^1 dt\,\langle \tilde{\omega} | \pi_1 \bs{\p}_t \frac{1}{1-\psi(t)}\otimes \pi_1 \tilde{\mathbf{e}}\frac{1}{1-\psi(t)}\label{eq:EffObs}
\end{align} 
is an on-shell gauge invariant of the effective theory. Note that explicitly we may expand
\begin{subequations}
	\label{eq:etexpl}
	\begin{align}
	\tilde{e}_0 &= Pe_0\,,\\
	\tilde{e}_1(\psi) &= Pe_1(\psi)-\tilde{m}_2(he_0,\psi)-\tilde{m}_2(\psi,he_0)\,,\\
	\tilde{e}_2(\psi,\psi)&= Pe_2(\psi,\psi)-Pe_1(hm_2(\psi,\psi))+\nonumber\\
	&\hspace{0.5cm}-\tilde{m}_2(he_1(\psi),\psi)-\tilde{m}_2(\psi,he_1(\psi))+\nonumber\\
	&\hspace{0.5cm}-\tilde{m}_3(he_0,\psi,\psi)-\tilde{m}_3(\psi,he_0,\psi)-\tilde{m}_3(\psi,\psi,he_0)\\
	&\hspace{0.2cm}\vdots
	\end{align}
\end{subequations}
which in turn gives the expansion of $\tilde{\mathcal{E}}(\psi)$. Note that when we have $e_k=0$ for $k>0$ (which is for instance the case for the Ellwood invariant in cubic OSFT), the formulae \eqref{eq:etexpl} reduce to
\begin{align}
\tilde{e}_k(\psi^{\otimes k}) = -\sum_{l=0}^{k-1}\tilde{m}_k(\psi^{\otimes l},he_0,\psi^{\otimes k-1-l})\,,\label{eq:teexpl}
\end{align}
which is valid for $k>0$.

In general we can in fact show that it is possible to write
\begin{align}
\tilde{\mathcal{E}}(\psi) = \mathcal{E}(\Psi(\psi))\,,\label{eq:Esubst}
\end{align}
namely that it is possible to express the effective observable $\tilde{\mathcal{E}}(\psi)$ by substituting the string field $\Psi(\psi)$ after having integrated out the unwanted degrees of freedom into the UV SFT observable $\mathcal{E}(\Psi)$. Indeed, using cyclicity of $\tilde{\mathbf{I}}$, we can first express
\begin{align}
\tilde{\mathcal{E}}(\psi)
&=\int_0^1 dt\,\langle{\omega} | \pi_1 \tilde{\mathbf{I}}\bs{\p}_t \frac{1}{1-\psi(t)}\otimes \pi_1 \tilde{\mathbf{P}}{\mathbf{e}}\tilde{\mathbf{I}}\frac{1}{1-\psi(t)}\,,
\end{align}
where we have also used that $\tilde{\mathbf{I}}\tilde{\mathbf{e}} = \tilde{\mathbf{P}}\mathbf{e}\tilde{\mathbf{I}}$. We can now substitute for $\tilde{\mathbf{P}}$ from the perturbed Hodge-Kodaira decomposition \eqref{eq:pertHK}: note that the $\tilde{\mathbf{h}}\mathbf{m}$ part actually does not contribute because $\pi_1\tilde{\mathbf{h}}$ acting on anything always has $h$ as a prefactor and therefore (using the BPZ property \eqref{eq:BPZprop} and the fact that we have the annihilation conditions $hP=Ph=h^2=0$) necessarily gives zero because $\pi_1\tilde{\mathbf{I}}$ acting on anything always contains either $h$ or $P$ as a prefactor. Using that $[\bs{\p}_t,\tilde{\mathbf{I}}]=0$, we therefore obtain
\begin{align}
\tilde{\mathcal{E}}(\psi)&=\mathcal{E}(\Psi(\psi))-\int_0^1 dt\,\langle{\omega} | \pi_1 \bs{\p}_t \tilde{\mathbf{I}}\frac{1}{1-\psi(t)}\otimes \pi_1 \mathbf{m}\tilde{\mathbf{h}}{\mathbf{e}}\tilde{\mathbf{I}}\frac{1}{1-\psi(t)}\,.\label{eq:tildeEalmost}
\end{align}
We then notice that since $\tilde{\mathbf{h}}\tilde{\mathbf{I}}=0$, we can replace $\tilde{\mathbf{h}}\mathbf{e}$ in \eqref{eq:tildeEalmost} with $[\tilde{\mathbf{h}},\mathbf{e}]$. This then acts as a coderivation on the group-like element $(1-\Psi(\psi(t)))^{-1}=\tilde{\mathbf{I}}(1-\psi(t))^{-1}$ because we have
\begin{subequations}
	\begin{align}
	\Delta_{T\mathcal{H}} [\tilde{\mathbf{h}},\mathbf{e}] \tilde{\mathbf{I}} \frac{1}{1-\psi(t)}
	&=\big([\tilde{\mathbf{h}},\mathbf{e}]\otimes' \tilde{\mathbf{P}}
	+{\mathbf{1}}_{T\mathcal{H}}\otimes'[\tilde{\mathbf{h}},\mathbf{e}]+\nonumber\\
	&\hspace{4cm}+\tilde{\mathbf{h}}\otimes'[\tilde{\mathbf{P}},\mathbf{e}]\big)(\tilde{\mathbf{I}}\otimes'\tilde{\mathbf{I}})\Delta_{TP\mathcal{H}} \frac{1}{1-\psi(t)}\\
	&=\big({\mathbf{1}}_{T\mathcal{H}}\otimes'[\tilde{\mathbf{h}},\mathbf{e}]	+[\mathbf{e},\tilde{\mathbf{h}}]\otimes' \mathbf{1}_{T\mathcal{H}}
	\big)\Delta_{T\mathcal{H}}\tilde{\mathbf{I}} \frac{1}{1-\psi(t)}\,,
	\end{align}
\end{subequations}
where in the last step we have used $\tilde{\mathbf{P}}\tilde{\mathbf{I}}=\tilde{\mathbf{I}}$, as well as that $\tilde{\mathbf{h}}\tilde{\mathbf{I}}=0$. Given this preparation, one may then invoke cyclicity of $\mathbf{m}$ (by applying \eqref{eq:canceld}) and subsequently the chain-map property \eqref{eq:mIIm} as well as $[\bs{\p}_t,\mathbf{m}]=[\bs{\p}_t,\tilde{\mathbf{m}}]=[\bs{\p}_t,\tilde{\mathbf{I}}]=0$ to eventually obtain
\begin{align}
\tilde{\mathcal{E}}(\psi)&=\mathcal{E}(\Psi(\psi))+\int_0^1 dt\,\langle{\omega} | \pi_1 \tilde{\mathbf{I}}\tilde{\mathbf{m}}\bs{\p}_t \frac{1}{1-\psi(t)}\otimes \pi_1 \tilde{\mathbf{h}}{\mathbf{e}}\tilde{\mathbf{I}}\frac{1}{1-\psi(t)}\,,\label{eq:tildeEstill}
\end{align}
where now the second term in \eqref{eq:tildeEstill} clearly vanishes by applying \eqref{eq:BPZprop} because $\pi_1\tilde{\mathbf{h}}$ acting on anything always gives $h$ as a prefactor while $\pi_1\tilde{\mathbf{I}}$ gives either $h$ or $P$. This finally proves the equality \eqref{eq:Esubst}.

On the other hand, assuming that, in addition to \eqref{eq:onshell},  we have $\mathbf{e}^2 =0$, then
\begin{subequations}
	\begin{align}
	\tilde{\mathbf{e}}^2 &= \tilde{\mathbf{\Pi}}\mathbf{e}\tilde{\mathbf{I}} \tilde{\mathbf{\Pi}}\mathbf{e}\tilde{\mathbf{I}}\\
	&= \tilde{\mathbf{\Pi}}\mathbf{e}\tilde{\mathbf{P}} \mathbf{e}\tilde{\mathbf{I}}\,,
	\end{align}
\end{subequations}
which is in general non-zero (unless, for instance, we have $[\tilde{\mathbf{P}},\mathbf{e}]=0$). This tells us that if we use $\mathbf{e}$ to perturb the multi-string products of the parent UV action as $\mathbf{M}(\mu)=\mathbf{m}+\mu\mathbf{e}$ (in the case that $\mathbf{e}^2=0$), then the corresponding perturbation of the effective action cannot be effected simply by adding $\tilde{\mathbf{e}}$ to $\tilde{\mathbf{m}}$. Indeed, as we will see below in subsection \ref{Ellwood}, we will need to consider effective couplings containing arbitrary powers of $\mathbf{e}$ in order to implement the corresponding perturbation on the level of open SFT effective action.

\subsection{Horizontal composition}\label{subsec:HorComp}

In certain situations one needs to perform two consecutive procedures of integrating out unwanted degrees of freedom: see subsection \ref{subsec:EffEx1} for a concrete example in the context of Witten's cubic OSFT. We will now see how this can be dealt with in one step using a composite propagator.

\subsubsection{Composite propagator}\label{subsubsec:compprop}

Let us assume that we first need to integrate out degrees of freedom which are singled out using a projector $\bar{\mathbf{P}}^{(1)}\equiv \mathbf{1}_{T\mathcal{H}}-\mathbf{P}^{(1)}$, where the projector $\mathbf{P}^{(1)}:T\mathcal{H}\longrightarrow T\mathcal{H}$ is associated to a propagator $\mathbf{h}^{(1)}:T\mathcal{H}\longrightarrow T\mathcal{H}$ via the Hodge-Kodaira decomposition
\begin{align}
\mathbf{Q}\mathbf{h}^{(1)} +\mathbf{h}^{(1)}\mathbf{Q} = \mathbf{1}_{T\mathcal{H}}-\mathbf{P}^{(1)}\,.\label{eq:HK1}
\end{align}
Decomposing the projector as $\mathbf{P}^{(1)} = \mathbf{I}^{(1)} \mathbf{\Pi}^{(1)}$ into a canonical projection $\mathbf{\Pi}^{(1)}:T\mathcal{H}\longrightarrow TP^{(1)} \mathcal{H}$ and the canonical inclusion $\mathbf{I}^{(1)}:TP^{(1)} \mathcal{H}\longrightarrow T\mathcal{H}$ (so that we recover the retract relation $\mathbf{\Pi}^{(1)}\mathbf{I}^{(1)}=\mathbf{1}_{TP^{(1)}\mathcal{H}}$), we will further assume (as directed by the discussion in subsection \ref{subsec:EffProd}) the annihilation conditions $\mathbf{h}^{(1)}\mathbf{I}^{(1)} = (\mathbf{h}^{(1)})^2 = \mathbf{\Pi}^{(1)}\mathbf{h}^{(1)} =0$ (so that, in particular, \eqref{eq:HK1} yields $[\mathbf{P}^{(1)},\mathbf{Q}]=[\mathbf{P}^{(1)},\mathbf{h}^{(1)}]=0$). Altogether, we can therefore write the SDR
\begin{align}
\mathrel{\raisebox{+14.5pt}{\rotatebox{-110}{  \begin{tikzcd}[sep=12mm,
			arrow style=tikz,
			arrows=semithick,
			diagrams={>={Straight Barb}}
			]
			\ar[to path={ node[pos=.5,C] }]{}
			\end{tikzcd}
}}}\hspace{-2.7mm}
(-\mathbf{h}^{(1)})\,(T\mathcal{H},\mathbf{Q})
\hspace{-3.2mm}\raisebox{-1pt}{$\begin{array}{cc} {\text{\scriptsize $\mathbf{\Pi}^{(1)}$}}\\[-3.0mm] \mathrel{\begin{tikzpicture}[node distance=1cm]
		\node (A) at (0, 0) {};
		\node (B) at (1.5, 0) {};
		\draw[->, to path={-> (\tikztotarget)}]
		(A) edge (B);
		\end{tikzpicture}}\\[-4mm] {\begin{tikzpicture}[node distance=1cm]
		\node (A) at (1.5, 0) {};
		\node (B) at (0, 0) {};
		\draw[->, to path={-> (\tikztotarget)}]
		(A) edge (B);
		\end{tikzpicture}}\\[-3.0mm]
	\text{\scriptsize $\mathbf{I}^{(1)}$} \end{array}$} \hspace{-3.4mm}
(TP^{(1)}\mathcal{H},\mathbf{Q}^{(1)})\,,\label{eq:SDR1}
\end{align}
\vspace{-7mm}

\noindent
where we have introduced the coderivation $\mathbf{Q}^{(1)}=\mathbf{\Pi}^{(1)} \mathbf{Q}\mathbf{I}^{(1)}$. The chain-map relations $\mathbf{Q}\mathbf{I}^{(1)} = \mathbf{I}^{(1)}\mathbf{Q}^{(1)}$ and $\mathbf{\Pi}^{(1)}\mathbf{Q}=\mathbf{Q}^{(1)}\mathbf{\Pi}^{(1)}$ clearly follow using the fact that $[\mathbf{P}^{(1)},\mathbf{Q}]=0$.

Let us consider that furthermore, we want to integrate out some degrees of freedom from $\mathrm{im}\,\mathbf{P}^{(1)}$. These are specified by a projector $\bar{\mathbf{P}}^{(2)}\equiv\mathbf{1}_{TP^{(1)}\mathcal{H}}-\mathbf{P}^{(2)}$, where the projector $\mathbf{P}^{(2)}$ should be understood as a map $\mathbf{P}^{(2)}:TP^{(1)}\mathcal{H}\longrightarrow TP^{(1)}\mathcal{H}$ (so that it is implicit that $\mathrm{im}\,\mathbf{P}^{(2)} \subset\mathrm{im}\,\mathbf{P}^{(1)}$). Assuming that the degrees of freedom outside of  $\mathrm{im}\,\mathbf{P}^{(2)}$ can be integrated out using a propagator $\mathbf{h}^{(2)}:TP^{(1)}\mathcal{H}\longrightarrow TP^{(1)}\mathcal{H}$ satisfying the Hodge-Kodaira decomposition
\begin{align}
\mathbf{Q}^{(1)}\mathbf{h}^{(2)} +\mathbf{h}^{(2)}\mathbf{Q}^{(1)} = \mathbf{1}_{TP^{(1)}\mathcal{H}}-\mathbf{P}^{(2)}\,,\label{eq:HK2}
\end{align}
(so that, in particular, we have $[\mathbf{P}^{(2)},\mathbf{Q}^{(1)}]=[\mathbf{P}^{(2)},\mathbf{h}^{(2)}]=0$) where the associated canonical projection $\mathbf{\Pi}^{(2)}:TP^{(1)}\mathcal{H}\longrightarrow TP^{(2)} P^{(1)}\mathcal{H}$ and inclusion $\mathbf{I}^{(2)}:TP^{(2)} P^{(1)}\mathcal{H}\longrightarrow TP^{(1)}\mathcal{H}$ satisfy $\mathbf{I}^{(2)}\mathbf{\Pi}^{(2)} = \mathbf{P}^{(2)}$, as well as the retract relation $\mathbf{\Pi}^{(2)}\mathbf{I}^{(2)} =\mathbf{1}_{TP^{(2)}P^{(1)}\mathcal{H}}$ and the annihilation relations $\mathbf{h}^{(2)}\mathbf{I}^{(2)} = (\mathbf{h}^{(2)})^2=\mathbf{\Pi}^{(2)}\mathbf{h}^{(2)}=0$, we can therefore write the SDR
\begin{align}
\mathrel{\raisebox{+14.5pt}{\rotatebox{-110}{  \begin{tikzcd}[sep=12mm,
			arrow style=tikz,
			arrows=semithick,
			diagrams={>={Straight Barb}}
			]
			\ar[to path={ node[pos=.5,C] }]{}
			\end{tikzcd}
}}}\hspace{-2.7mm}
(-\mathbf{h}^{(2)})\,(TP^{(1)}\mathcal{H},\mathbf{Q}^{(1)})
\hspace{-3.2mm}\raisebox{-1pt}{$\begin{array}{cc} {\text{\scriptsize $\mathbf{\Pi}^{(2)}$}}\\[-3.0mm] \mathrel{\begin{tikzpicture}[node distance=1cm]
		\node (A) at (0, 0) {};
		\node (B) at (1.5, 0) {};
		\draw[->, to path={-> (\tikztotarget)}]
		(A) edge (B);
		\end{tikzpicture}}\\[-4mm] {\begin{tikzpicture}[node distance=1cm]
		\node (A) at (1.5, 0) {};
		\node (B) at (0, 0) {};
		\draw[->, to path={-> (\tikztotarget)}]
		(A) edge (B);
		\end{tikzpicture}}\\[-3.0mm]
	\text{\scriptsize $\mathbf{I}^{(2)}$} \end{array}$} \hspace{-3.4mm}
(TP^{(2)} P^{(1)}\mathcal{H},\mathbf{Q}^{(2)})\,,\label{eq:SDR2}
\end{align}
\vspace{-7mm}

\noindent where we introduce the coderivation $\mathbf{Q}^{(2)}=\mathbf{\Pi}^{(2)}\mathbf{Q}^{(1)}\mathbf{I}^{(2)} = \mathbf{\Pi}^{(2)}\mathbf{\Pi}^{(1)}\mathbf{Q}\mathbf{I}^{(1)}\mathbf{I}^{(2)}$. The corresponding chain-map relations again follow using the fact that $[\mathbf{P}^{(2)},\mathbf{Q}^{(1)}]=0$.

We will now show that defining a composite propagator $\mathbf{h}^{(12)}:T\mathcal{H}\longrightarrow T\mathcal{H}$ by
\begin{align}
\mathbf{h}^{(12)}\equiv \mathbf{h}^{(1)}\circ \mathbf{h}^{(2)}\equiv \mathbf{h}^{(1)} + \mathbf{I}^{(1)}\mathbf{h}^{(2)}\mathbf{\Pi}^{(1)}\,,\label{eq:comph}
\end{align}
as well as defining composite projection $\mathbf{\Pi}^{(12)}:T\mathcal{H}\longrightarrow TP^{(2)} P^{(1)}\mathcal{H}$ and inclusion $\mathbf{I}^{(12)}:TP^{(2)} P^{(1)}\mathcal{H} \longrightarrow T\mathcal{H}$ as
\begin{subequations}
	\begin{align}
	\mathbf{\Pi}^{(12)}&=\mathbf{\Pi}^{(2)}\mathbf{\Pi}^{(1)}\,,\\
	\mathbf{I}^{(12)}&=\mathbf{I}^{(1)}\mathbf{I}^{(2)}\,,
	\end{align}
\end{subequations}
we can write the composite SDR
\begin{align}
\mathrel{\raisebox{+14.5pt}{\rotatebox{-110}{  \begin{tikzcd}[sep=12mm,
			arrow style=tikz,
			arrows=semithick,
			diagrams={>={Straight Barb}}
			]
			\ar[to path={ node[pos=.5,C] }]{}
			\end{tikzcd}
}}}\hspace{-2.7mm}
(-\mathbf{h}^{(12)})\,(T\mathcal{H},\mathbf{Q})
\hspace{-3.2mm}\raisebox{-1pt}{$\begin{array}{cc} {\text{\scriptsize $\mathbf{\Pi}^{(12)}$}}\\[-3.0mm] \mathrel{\begin{tikzpicture}[node distance=1cm]
		\node (A) at (0, 0) {};
		\node (B) at (1.5, 0) {};
		\draw[->, to path={-> (\tikztotarget)}]
		(A) edge (B);
		\end{tikzpicture}}\\[-4mm] {\begin{tikzpicture}[node distance=1cm]
		\node (A) at (1.5, 0) {};
		\node (B) at (0, 0) {};
		\draw[->, to path={-> (\tikztotarget)}]
		(A) edge (B);
		\end{tikzpicture}}\\[-3.0mm]
	\text{\scriptsize $\mathbf{I}^{(12)}$} \end{array}$} \hspace{-3.4mm}
(TP^{(2)} P^{(1)}\mathcal{H},\mathbf{Q}^{(2)})\,,\label{eq:SDR12}
\end{align}
\vspace{-7mm}

\noindent where we can clearly write the coderivation $\mathbf{Q}^{(2)}$ as $\mathbf{Q}^{(2)} = \mathbf{\Pi}^{(12)} \mathbf{Q}\mathbf{I}^{(12)}$. To do this, we will need to establish the homotopy-equivalence relations
\begin{subequations}
\begin{align}
\mathbf{Q}\mathbf{h}^{(12)}+\mathbf{h}^{(12)}\mathbf{Q} &=\mathbf{1}_{T\mathcal{H}}-\mathbf{I}^{(12)}\mathbf{\Pi}^{(12)}\,,\\
\mathbf{Q}\mathbf{I}^{(12)}&=\mathbf{I}^{(12)}\mathbf{Q}^{(2)}\,,\\ \mathbf{\Pi}^{(12)}\mathbf{Q}&=\mathbf{Q}^{(2)}\mathbf{\Pi}^{(12)}\,,
\end{align}
\end{subequations}
as well as the retract relation
\begin{align}
\mathbf{\Pi}^{(12)} \mathbf{I}^{(12)} &=\mathbf{1}_{TP^{(2)}P^{(1)}\mathcal{H}}
\end{align}
and the annihilation conditions
\begin{align}
\mathbf{\Pi}^{(12)}\mathbf{h}^{(12)} =(\mathbf{h}^{(12)})^2= \mathbf{h}^{(12)}\mathbf{I}^{(12)}=0 \,.
\end{align}
As we will see explicitly in the following subsection, the composite propagator $\mathbf{h}^{(12)}$ provides us with a possibility of going directly from $T\mathcal{H}$ to $TP^{(2)}P^{(1)}\mathcal{H}$ by integrating out the modes outside of the image of $\mathbf{P}^{(12)}\equiv \mathbf{I}^{(12)}\mathbf{\Pi}^{(12)}$ by ``cutting out the middleman'' which takes on the form of $TP^{(1)}\mathcal{H}$.

First, the retract relation follows because
\begin{subequations}
	\begin{align}
	\mathbf{\Pi}^{(12)} \mathbf{I}^{(12)} &= \mathbf{\Pi}^{(2)}\mathbf{\Pi}^{(1)}\mathbf{I}^{(1)}\mathbf{I}^{(2)}\\
	&= \mathbf{\Pi}^{(2)}\mathbf{1}_{TP^{(1)}\mathcal{H}}\mathbf{I}^{(2)}\\
	&=\mathbf{1}_{TP^{(2)}P^{(1)}\mathcal{H}}\,,
	\end{align}
\end{subequations}
where we have used the retract relations for the SDRs \eqref{eq:SDR1} and \eqref{eq:SDR2}. Second, we can use the definition \eqref{eq:comph} to expand
\begin{align}
\mathbf{Q}\mathbf{h}^{(12)}+\mathbf{h}^{(12)}\mathbf{Q} = \mathbf{Q}\mathbf{h}^{(1)}+\mathbf{h}^{(1)}\mathbf{Q}+\mathbf{Q}\mathbf{I}^{(1)}\mathbf{h}^{(2)}\mathbf{\Pi}^{(1)}+\mathbf{I}^{(1)}\mathbf{h}^{(2)}\mathbf{\Pi}^{(1)}\mathbf{Q}\,,\label{eq:step1}
\end{align}
where we can substitute from the Hodge-Kodaira decomposition \eqref{eq:HK1} and use the chain-map relations $\mathbf{Q}\mathbf{I}^{(1)} = \mathbf{I}^{(1)}\mathbf{Q}^{(1)}$, $\mathbf{\Pi}^{(1)}\mathbf{Q} = \mathbf{Q}^{(1)}\mathbf{\Pi}^{(1)}$ to rewrite \eqref{eq:step1} as
\begin{align}
\mathbf{Q}\mathbf{h}^{(12)}+\mathbf{h}^{(12)}\mathbf{Q} = \mathbf{1}_{T\mathcal{H}}-\mathbf{I}^{(1)}\mathbf{\Pi}^{(1)}+\mathbf{I}^{(1)}\big(\mathbf{Q}\mathbf{h}^{(2)}+\mathbf{h}^{(2)}\mathbf{Q}\big)\mathbf{\Pi}^{(1)}\,.
\end{align}
Finally, substituting from the elementary Hodge-Kodaira decomposition \eqref{eq:HK2}, we therefore obtain the required composite Hodge-Kodaira decomposition for the SDR \eqref{eq:SDR12}
\begin{subequations}
	\label{eq:HK12}
	\begin{align}
	\mathbf{Q}\mathbf{h}^{(12)}+\mathbf{h}^{(12)}\mathbf{Q} &=	\mathbf{1}_{T\mathcal{H}}-\mathbf{I}^{(1)}\mathbf{\Pi}^{(1)} +\mathbf{I}^{(1)}\mathbf{1}_{TP^{(1)}\mathcal{H}}\mathbf{\Pi}^{(1)}-\mathbf{I}^{(1)}\mathbf{I}^{(2)}\mathbf{\Pi}^{(2)}\mathbf{\Pi}^{(1)}\\
	&=\mathbf{1}_{T\mathcal{H}}-\mathbf{I}^{(12)}\mathbf{\Pi}^{(12)}\,.
	\end{align}
\end{subequations}
Recalling the definition $\mathbf{P}^{(12)}\equiv\mathbf{I}^{(12)}\mathbf{\Pi}^{(12)}$ of the composite projector, the decomposition \eqref{eq:HK2} implies $[\mathbf{P}^{(12)},\mathbf{Q}]=[\mathbf{P}^{(12)},\mathbf{h}^{(12)}]=0$ which, in particular, yields the chain-map relations $\mathbf{Q}\mathbf{I}^{(12)}=\mathbf{I}^{(12)}\mathbf{Q}^{(2)}$ and $\mathbf{\Pi}^{(12)}\mathbf{Q}=\mathbf{Q}^{(2)}\mathbf{\Pi}^{(12)}$. Finally, it is straightforward to verify the composite annihilation conditions. Indeed, we can expand
\begin{align}
(\mathbf{h}^{(12)})^2 = (\mathbf{h}^{(1)})^2 +\mathbf{h}^{(1)}\mathbf{I}^{(1)}\mathbf{h}^{(2)}\mathbf{\Pi}^{(1)}+\mathbf{I}^{(1)}\mathbf{h}^{(2)}\mathbf{\Pi}^{(1)}\mathbf{h}^{(1)}+\mathbf{I}^{(1)}\mathbf{h}^{(2)}\mathbf{\Pi}^{(1)}\mathbf{I}^{(1)}\mathbf{h}^{(2)}\mathbf{\Pi}^{(1)}\,,\label{eq:h122}
\end{align}
where we can use the elementary annihilation conditions $(\mathbf{h}^{(1)})^2=0$, $\mathbf{h}^{(1)}\mathbf{I}^{(1)}=0$ and $\mathbf{\Pi}^{(1)}\mathbf{h}^{(1)}=0$ to get rid of the first three terms in \eqref{eq:h122}, while the last term can be seen to vanish by realizing that $\mathbf{\Pi}^{(1)}\mathbf{I}^{(1)}=\mathbf{1}_{TP^{(1)}\mathcal{H}}$ and $(\mathbf{h}^{(2)})^2=0$. This therefore gives us the composite annihilation condition $(\mathbf{h}^{(12)})^2=0$. We can also expand
\begin{align}
\mathbf{\Pi}^{(12)}\mathbf{h}^{(12)} =\mathbf{\Pi}^{(2)}\mathbf{\Pi}^{(1)}\mathbf{h}^{(1)}+\mathbf{\Pi}^{(2)}\mathbf{\Pi}^{(1)}\mathbf{I}^{(1)}\mathbf{h}^{(2)}\mathbf{\Pi}^{(1)}\,,
\end{align}
where the first term vanishes by the annihilation condition $\mathbf{\Pi}^{(1)}\mathbf{h}^{(1)}=0$ and the second term vanishes by realizing that $\mathbf{\Pi}^{(1)}\mathbf{I}^{(1)}=\mathbf{1}_{TP^{(1)}\mathcal{H}}$ and using the annihilation condition $\mathbf{\Pi}^{(2)}\mathbf{h}^{(2)}=0$. This then establishes the composite annihilation condition $\mathbf{\Pi}^{(12)}\mathbf{h}^{(12)}=0$. Similarly for $\mathbf{h}^{(12)}\mathbf{I}^{(12)}=0$. 

Finally, it is easy to see that the composite maps $\mathbf{\Pi}^{(12)}$, $\mathbf{I}^{(12)}$ retain the required coalgebraic properties, namely that $\mathbf{\Pi}^{(12)}$, $\mathbf{I}^{(12)}$ are again cohomomorphisms. On the other hand, we now clearly have
\begin{align}
\Delta_{T\mathcal{H}} \mathbf{h}^{(12)} \neq (\mathbf{h}^{(12)}\otimes '\mathbf{P}^{(12)}+\mathbf{1}_{T\mathcal{H}}\otimes'\mathbf{h}^{(12)})\Delta_{T\mathcal{H}}\,.
\end{align}
Thus, instead of following the rule \eqref{eq:huplift}, the action of the composite propagator on $T\mathcal{H}$ is now given by
\begin{align}
\mathbf{h}^{(12)}\pi_k &= \sum_{l=0}^{k-1}(1_\mathcal{H})^{\otimes l}\otimes h^{(1)} \otimes (P^{(1)})^{\otimes (k-1-l)}+\nonumber\\[-2mm]
&\hspace{3cm}+\sum_{l=0}^{k-1}(P^{(1)})^{\otimes l}\otimes I^{(1)} h^{(2)}\Pi^{(1)} \otimes (P^{(12)})^{\otimes (k-1-l)}\,.\label{eq:acthcomp}
\end{align}
One therefore needs to exercise extra care when recasting coalgebraic expressions in terms of ordinary products on $\mathcal{H}$. Observe that the action \eqref{eq:acthcomp} of the composite propagator $\mathbf{h}^{(12)}$ on the tensor space $T\mathcal{H}$ gives us an alternative possibility (besides the rule \eqref{eq:huplift}) of uplifting the map $h^{(1)}+I^{(1)}h^{(2)}\Pi^{(1)}$ from $\mathcal{H}$ to $T\mathcal{H}$, such that the respective map on $T\mathcal{H}$ satisfies the tensor version \eqref{eq:HK12} of the Hodge-Kodaira decomposition with projector $\mathbf{P}^{(12)}$.

\subsubsection{Perturbing horizontally composed SDR}

By now we should recognize that there are in principle two ways of integrating out the degrees of freedom specified by the projectors $\bar{\mathbf{P}}^{(1)}$ and $\bar{\mathbf{P}}^{(2)}$: either we can (a) perform this procedure in two steps by sequentially integrating out first the $\bar{\mathbf{P}}^{(1)}$ degrees of freedom using the propagator $\mathbf{h}^{(1)}$ and only then integrating out the $\bar{\mathbf{P}}^{(2)}$ degrees of freedom using the propagator $\mathbf{h}^{(1)}$, or, we can (b) do everything in one step by using the composite propagator $\mathbf{h}^{(12)}$. We will now show that both of these procedures lead to the same effective action.

Following the sequential procedure (a), we can first integrate out the $\bar{\mathbf{P}}^{(1)}$ degrees of freedom using the propagator $\mathbf{h}^{(1)}$ so as to obtain the perturbed SDR
\begin{align}
\mathrel{\raisebox{+14.5pt}{\rotatebox{-110}{  \begin{tikzcd}[sep=12mm,
			arrow style=tikz,
			arrows=semithick,
			diagrams={>={Straight Barb}}
			]
			\ar[to path={ node[pos=.5,C] }]{}
			\end{tikzcd}
}}}\hspace{-2.7mm}
(-\tilde{\mathbf{h}}^{(1)})\,(T\mathcal{H},\mathbf{m}=\mathbf{Q}+\delta\mathbf{m})
\hspace{-3.2mm}\raisebox{-1pt}{$\begin{array}{cc} {\text{\scriptsize $\tilde{\mathbf{\Pi}}^{(1)}$}}\\[-3.0mm] \mathrel{\begin{tikzpicture}[node distance=1cm]
		\node (A) at (0, 0) {};
		\node (B) at (1.5, 0) {};
		\draw[->, to path={-> (\tikztotarget)}]
		(A) edge (B);
		\end{tikzpicture}}\\[-4mm] {\begin{tikzpicture}[node distance=1cm]
		\node (A) at (1.5, 0) {};
		\node (B) at (0, 0) {};
		\draw[->, to path={-> (\tikztotarget)}]
		(A) edge (B);
		\end{tikzpicture}}\\[-3.0mm]
	\text{\scriptsize $\tilde{\mathbf{I}}^{(1)}$} \end{array}$} \hspace{-3.4mm}
(TP^{(1)}\mathcal{H},\tilde{\mathbf{m}}^{(1)}=\mathbf{Q}^{(1)}+\delta\tilde{\mathbf{m}}^{(1)})\,,\label{eq:SDR1pert}
\end{align}
\vspace{-7mm}

\noindent where the perturbed data are given by the by-now-familiar relations
\begin{subequations}
	\label{eq:qW}
	\begin{align}
	\delta\tilde{\mathbf{m}}^{(1)} &= \mathbf{\Pi}^{(1)}\delta\mathbf{m}\frac{1}{\mathbf{1}_{T\mathcal{H}}+\mathbf{h}^{(1)}\delta\mathbf{m}}\mathbf{I}^{(1)}\,,\label{eq:dm1}\\
	\tilde{\mathbf{h}}^{(1)} &= \frac{1}{\mathbf{1}_{T\mathcal{H}}+\mathbf{h}^{(1)}\delta\mathbf{m}}\mathbf{h}^{(1)}\,,\\
	\tilde{\mathbf{I}}^{(1)} &= \frac{1}{\mathbf{1}_{T\mathcal{H}}+\mathbf{h}^{(1)}\delta\mathbf{m}}\mathbf{I}^{(1)}\,,\\
	\tilde{\mathbf{\Pi}}^{(1)} &= \mathbf{\Pi}^{(1)}\frac{1}{\mathbf{1}_{T\mathcal{H}}+\delta\mathbf{m}\mathbf{h}^{(1)}}\,.
	\end{align}
\end{subequations}
Furthermore, integrating out the $\bar{\mathbf{P}}^{(2)}$ degrees of freedom by means of the propagator $\mathbf{h}^{(2)}$, we obtain the $\delta\tilde{\mathbf{m}}^{(1)}$-perturbed version of the SDR \eqref{eq:SDR2}
\begin{align}
\mathrel{\raisebox{+14.5pt}{\rotatebox{-110}{  \begin{tikzcd}[sep=12mm,
			arrow style=tikz,
			arrows=semithick,
			diagrams={>={Straight Barb}}
			]
			\ar[to path={ node[pos=.5,C] }]{}
			\end{tikzcd}
}}}\hspace{-2.7mm}
(-\tilde{\mathbf{h}}^{(2)})\,(TP^{(1)}\mathcal{H},\tilde{\mathbf{m}}^{(1)}=\mathbf{Q}^{(1)}+\delta\tilde{\mathbf{m}}^{(1)})
\hspace{-3.2mm}\raisebox{-1pt}{$\begin{array}{cc} {\text{\scriptsize $\tilde{\mathbf{\Pi}}^{(2)}$}}\\[-3.0mm] \mathrel{\begin{tikzpicture}[node distance=1cm]
		\node (A) at (0, 0) {};
		\node (B) at (1.5, 0) {};
		\draw[->, to path={-> (\tikztotarget)}]
		(A) edge (B);
		\end{tikzpicture}}\\[-4mm] {\begin{tikzpicture}[node distance=1cm]
		\node (A) at (1.5, 0) {};
		\node (B) at (0, 0) {};
		\draw[->, to path={-> (\tikztotarget)}]
		(A) edge (B);
		\end{tikzpicture}}\\[-3.0mm]
	\text{\scriptsize $\tilde{\mathbf{I}}^{(2)}$} \end{array}$} \hspace{-3.4mm}
(TP^{(2)}P^{(1)}\mathcal{H},\tilde{\mathbf{m}}^{(2)}=\mathbf{Q}^{(2)}+\delta\tilde{\mathbf{m}}^{(2)})\,,\label{eq:SDR2pert}
\end{align}
\vspace{-7mm}

\noindent where we can immediately write down the explicit expressions
\begin{subequations}
	\begin{align}
	\delta\tilde{\mathbf{m}}^{(2)} &= \mathbf{\Pi}^{(2)}\delta\tilde{\mathbf{m}}^{(1)}\frac{1}{\mathbf{1}_{TP^{(1)}\mathcal{H}}+\mathbf{h}^{(2)}\delta\tilde{\mathbf{m}}^{(1)}}\mathbf{I}^{(2)}\,,\label{eq:dm2}\\
	\tilde{\mathbf{h}}^{(2)} &= \frac{1}{\mathbf{1}_{TP^{(1)}\mathcal{H}}+\mathbf{h}^{(2)}\delta\tilde{\mathbf{m}}^{(1)}}\mathbf{h}^{(2)}\,,\label{eq:delU}\\
	\tilde{\mathbf{I}}^{(2)} &= \frac{1}{\mathbf{1}_{TP^{(1)}\mathcal{H}}+\mathbf{h}^{(2)}\delta\tilde{\mathbf{m}}^{(1)}}\mathbf{I}^{(2)}\,,\\
	\tilde{\mathbf{\Pi}}^{(2)} &= \mathbf{\Pi}^{(2)}\frac{1}{\mathbf{1}_{TP^{(1)}\mathcal{H}}+\delta\tilde{\mathbf{m}}^{(1)}\mathbf{h}^{(2)}}\,.
	\end{align}
\end{subequations}
Using our discussion in subsection \eqref{subsubsec:compprop}, we can also write down the corresponding composite SDR
\begin{align}
\mathrel{\raisebox{+14.5pt}{\rotatebox{-110}{  \begin{tikzcd}[sep=12mm,
			arrow style=tikz,
			arrows=semithick,
			diagrams={>={Straight Barb}}
			]
			\ar[to path={ node[pos=.5,C] }]{}
			\end{tikzcd}
}}}\hspace{-2.7mm}
(-\tilde{\mathbf{h}}^{(1)}\circ \tilde{\mathbf{h}}^{(2)} )\,(T\mathcal{H},\mathbf{m}=\mathbf{Q}+\delta\mathbf{m})
\hspace{-3.2mm}\raisebox{-1pt}{$\begin{array}{cc} {\text{\scriptsize $\tilde{\mathbf{\Pi}}^{(2)}\tilde{\mathbf{\Pi}}^{(1)}$}}\\[-3.0mm] \mathrel{\begin{tikzpicture}[node distance=1cm]
		\node (A) at (0, 0) {};
		\node (B) at (2, 0) {};
		\draw[->, to path={-> (\tikztotarget)}]
		(A) edge (B);
		\end{tikzpicture}}\\[-4mm] {\begin{tikzpicture}[node distance=1cm]
		\node (A) at (2, 0) {};
		\node (B) at (0, 0) {};
		\draw[->, to path={-> (\tikztotarget)}]
		(A) edge (B);
		\end{tikzpicture}}\\[-3.0mm]
	\text{\scriptsize $\tilde{\mathbf{I}}^{(1)}\tilde{\mathbf{I}}^{(2)}$} \end{array}$} \hspace{-3.4mm}
(TP^{(2)}P^{(1)}\mathcal{H},\tilde{\mathbf{m}}^{(2)}=\mathbf{Q}^{(2)}+\delta\tilde{\mathbf{m}}^{(2)})\,,\label{eq:SDR12pert}
\end{align}
\vspace{-7mm}

\noindent where
\begin{align}
\tilde{\mathbf{h}}^{(1)}\circ \tilde{\mathbf{h}}^{(2)} \equiv \tilde{\mathbf{h}}^{(1)}+\tilde{\mathbf{I}}^{(1)}\tilde{\mathbf{h}}^{(2)}\tilde{\mathbf{\Pi}}^{(1)}
\end{align}
gives the composite propagator.

On the other hand, we can alternatively proceed to construct an effective action by pursuing method (b), that is by integrating out everything in one go by using the composite propagator $\mathbf{h}^{(12)}$. We would then directly obtain a perturbed version of the SDR \eqref{eq:SDR12}
\begin{align}
\mathrel{\raisebox{+14.5pt}{\rotatebox{-110}{  \begin{tikzcd}[sep=12mm,
			arrow style=tikz,
			arrows=semithick,
			diagrams={>={Straight Barb}}
			]
			\ar[to path={ node[pos=.5,C] }]{}
			\end{tikzcd}
}}}\hspace{-2.7mm}
(-\tilde{\mathbf{h}}^{(12)})\,(T\mathcal{H},\mathbf{m}=\mathbf{Q}+\delta\mathbf{m})
\hspace{-3.2mm}\raisebox{-1pt}{$\begin{array}{cc} {\text{\scriptsize $\tilde{\mathbf{\Pi}}^{(12)}$}}\\[-3.0mm] \mathrel{\begin{tikzpicture}[node distance=1cm]
		\node (A) at (0, 0) {};
		\node (B) at (1.5, 0) {};
		\draw[->, to path={-> (\tikztotarget)}]
		(A) edge (B);
		\end{tikzpicture}}\\[-4mm] {\begin{tikzpicture}[node distance=1cm]
		\node (A) at (1.5, 0) {};
		\node (B) at (0, 0) {};
		\draw[->, to path={-> (\tikztotarget)}]
		(A) edge (B);
		\end{tikzpicture}}\\[-3.0mm]
	\text{\scriptsize $\tilde{\mathbf{I}}^{(12)}$} \end{array}$} \hspace{-3.4mm}
(TP^{(2)} P^{(1)}\mathcal{H},\mathbf{Q}^{(2)}+\delta'\tilde{\mathbf{m}}^{(2)})\,,\label{eq:SDR12pertp}
\end{align}
\vspace{-7mm}

\noindent where the homological perturbation lemma instructs us that we should put
\begin{subequations}
	\begin{align}
	\delta'\tilde{\mathbf{m}}^{(2)} &= \mathbf{\Pi}^{(12)}\delta\mathbf{m}\frac{1}{\mathbf{1}_{T\mathcal{H}}+\mathbf{h}^{(12)}\delta \mathbf{m}}\mathbf{I}^{(12)}\,,\label{eq:dm2p}\\
	\tilde{\mathbf{h}}^{(12)}&=\frac{1}{\mathbf{1}_{T\mathcal{H}}+\mathbf{h}^{(12)}\delta \mathbf{m}}\mathbf{h}^{(12)}\,,\\
	\tilde{\mathbf{I}}^{(12)}&=\frac{1}{\mathbf{1}_{T\mathcal{H}}+\mathbf{h}^{(12)}\delta \mathbf{m}}\mathbf{I}^{(12)}\,,\\
	\tilde{\mathbf{\Pi}}^{(12)}&=\mathbf{\Pi}^{(12)}\frac{1}{\mathbf{1}_{T\mathcal{H}}+\delta \mathbf{m}\mathbf{h}^{(12)}}\,.
	\end{align}
\end{subequations}
As we have already hinted at above, it turns out that the methods (a) and (b) yield the same effective action. Put in quantitative terms, we will now show that we in fact have
\begin{subequations}
	\label{eq:reseq}
	\begin{align}
	\delta\tilde{\mathbf{m}}^{(2)}&=\delta'\tilde{\mathbf{m}}^{(2)}\,,\label{eq:res1}\\
		\tilde{\mathbf{h}}^{(12)}&=\tilde{\mathbf{h}}^{(1)}\circ \tilde{\mathbf{h}}^{(2)}\,,\label{eq:res2}\\
		\tilde{\mathbf{I}}^{(12)}&=	\tilde{\mathbf{I}}^{(1)}\tilde{\mathbf{I}}^{(2)}\,,\label{eq:relIota}\\
			\tilde{\mathbf{\Pi}}^{(12)}&=	\tilde{\mathbf{\Pi}}^{(2)}\tilde{\mathbf{\Pi}}^{(1)}\,, \label{eq:relPi}
	\end{align}
\end{subequations}
namely that the SDRs \eqref{eq:SDR12pert} and \eqref{eq:SDR12pertp} are identical. In particular the interaction vertices of the effective SFT action after integrating out both $\bar{\mathbf{P}}_1$ and $\bar{\mathbf{P}}_2$ degrees of freedom are expressed in terms of the products which are encoded in the coderivation
\begin{align}
\delta\tilde{\mathbf{m}}^{(2)} &= \mathbf{\Pi}^{(2)}\mathbf{\Pi}^{(1)}\delta\mathbf{m}\frac{1}{\mathbf{1}_{T\mathcal{H}}+(\mathbf{h}^{(1)}+\mathbf{I}^{(1)}\mathbf{h}^{(2)}\mathbf{\Pi}^{(1)})\delta\mathbf{m}}\mathbf{I}^{(1)}\mathbf{I}^{(2)}\,.\label{eq:dm2t}
\end{align}
Here we recall that we noted in subsection \ref{subsubsec:compprop} that the composite propagator $\mathbf{h}^{(12)}=\mathbf{h}^{(1)}+\mathbf{I}^{(1)}\mathbf{h}^{(2)}\mathbf{\Pi}^{(1)}$ does not act on $T\mathcal{H}$ in the way prescribed by the rule \eqref{eq:huplift}. However, as we will see explicitly in subsection \ref{subsec:EffEx1}, it is possible to show that after unpackaging the coalgebra notation, the effective products will be given in terms of the propagator $h^{(1)}+P^{(1)}h^{(2)}$ precisely in the way as if the (uplifted) composite propagator $\mathbf{h}^{(12)}$ acted on $T\mathcal{H}$ according to \eqref{eq:huplift}. It may well be possible that in order to recover the correct perturbation expansion, it is only crucial to ensure that the uplifted propagator $\mathbf{h}^{(12)}$ satisfies the Hodge-Kodaira decomposition while the precise way in which this uplift is implemented might not play any role.

Let us first show \eqref{eq:res1}. Starting with the expression \eqref{eq:dm2} for $\delta\tilde{\mathbf{m}}^{(1)}$ and substituting for $\delta\tilde{\mathbf{m}}^{(1)}$ from \eqref{eq:dm1}, we can straightforwardly obtain (it helps to expand various denominators as power series)
\begin{align}
\delta\tilde{\mathbf{m}}^{(2)} &= \mathbf{\Pi}^{(2)}\mathbf{\Pi}^{(1)}\delta\mathbf{m}\frac{1}{\mathbf{1}_{T\mathcal{H}}+\mathbf{h}^{(1)}\delta\mathbf{m}}\frac{1}{\mathbf{1}_{TP^{(1)}\mathcal{H}}+\mathbf{I}^{(1)}\mathbf{h}^{(2)}\mathbf{\Pi}^{(1)}\delta\mathbf{m}\frac{1}{\mathbf{1}_{T\mathcal{H}}+\mathbf{h}^{(1)}\delta\mathbf{m}}}\mathbf{I}^{(1)}\mathbf{I}^{(2)}\,. \label{eq:dmstep1}
\end{align}
Noting that composition of invertible maps generally obeys $(AB)^{-1}=B^{-1}A^{-1}$, we can combine the two denominators in \eqref{eq:dmstep1} so as to show that \eqref{eq:dmstep1} is equal to \eqref{eq:dm2t},
which, in turn, is clearly equal to the expression \eqref{eq:dm2p} for $\delta'\tilde{\mathbf{m}}^{(2)}$. Continuing with \eqref{eq:res2} and taking the expressions \eqref{eq:qW} for $\tilde{\mathbf{h}}^{(1)}$, $\tilde{\mathbf{I}}^{(1)}$, $\tilde{\mathbf{\Pi}}^{(1)}$, as well as the expression \eqref{eq:delU} for $\tilde{\mathbf{h}}^{(2)}$ (where, inside $\tilde{\mathbf{h}}^{(2)}$, we also substitute for $\delta\tilde{\mathbf{m}}^{(1)}$ using \eqref{eq:dm1}), we can write
\begingroup\allowdisplaybreaks
\begin{align}
\tilde{\mathbf{h}}^{(1)}+\tilde{\mathbf{I}}^{(1)} \tilde{\mathbf{h}}^{(2)}\tilde{\mathbf{\Pi}}^{(1)} &=\frac{1}{\mathbf{1}_{T\mathcal{H}}+\mathbf{h}^{(1)}\delta\mathbf{m}}\mathbf{h}^{(1)}+\nonumber\\
&\hspace{-3cm}+\frac{1}{\mathbf{1}_{T\mathcal{H}}+\mathbf{h}^{(1)}\delta\mathbf{m}}\mathbf{I}^{(1)}\frac{1}{\mathbf{1}_{TP^{(1)}\mathcal{H}}+\mathbf{h}^{(2)}\mathbf{\Pi}^{(1)} \delta\mathbf{m} \frac{1}{\mathbf{1}_{T\mathcal{H}}+\mathbf{h}^{(1)}\delta\mathbf{m}}\mathbf{I}^{(1)}}\mathbf{h}^{(2)}\mathbf{\Pi}^{(1)}\frac{1}{\mathbf{1}_{T\mathcal{H}}+\delta\mathbf{m}\mathbf{h}^{(1)}}\,.
\end{align}
Using analogous manipulations as in the case of $\delta\tilde{\mathbf{m}}^{(2)}$ previously, this can be brought into the form
\begin{align}
\tilde{\mathbf{h}}^{(1)}+\tilde{\mathbf{I}}^{(1)} \tilde{\mathbf{h}}^{(2)}\tilde{\mathbf{\Pi}}^{(1)}
&=\frac{1}{\mathbf{1}_{T\mathcal{H}}+\mathbf{h}^{(1)}\delta\mathbf{m}}\mathbf{h}^{(1)}+\nonumber\\
&\hspace{-3cm}+\frac{1}{\mathbf{1}_{T\mathcal{H}}+\mathbf{h}^{(1)}\delta\mathbf{m}+\mathbf{I}^{(1)}\mathbf{h}^{(2)}\mathbf{\Pi}^{(1)} \delta\mathbf{m} }\mathbf{I}^{(1)}\mathbf{h}^{(2)}\mathbf{\Pi}^{(1)}\bigg(\mathbf{1}_{T\mathcal{H}}-\frac{1}{\mathbf{1}_{T\mathcal{H}}+\delta\mathbf{m}\mathbf{h}^{(1)}}\delta\mathbf{m}\mathbf{h}^{(1)}\bigg)\,.
\end{align}
A number of additional purely algebraic manipulations are then required to rewrite this as
\begin{align}
\tilde{\mathbf{h}}^{(1)}+\tilde{\mathbf{I}}^{(1)} \tilde{\mathbf{h}}^{(2)}\tilde{\mathbf{\Pi}}^{(1)}
&=\bigg(\mathbf{1}_{T\mathcal{H}}-\frac{1}{\mathbf{1}_{T\mathcal{H}}+\mathbf{h}^{(12)}\delta\mathbf{m}}\mathbf{I}^{(1)}\mathbf{h}^{(2)}\mathbf{\Pi}^{(1)}\delta\mathbf{m}\bigg)\frac{1}{\mathbf{1}_{T\mathcal{H}}+\mathbf{h}^{(1)}\delta\mathbf{m}}\mathbf{h}^{(1)}+\nonumber\\
&\hspace{5cm}+\frac{1}{\mathbf{1}_{T\mathcal{H}}+\mathbf{h}^{(12)}\delta\mathbf{m} }\mathbf{I}^{(1)}\mathbf{h}^{(2)}\mathbf{\Pi}^{(1)}\,.
\end{align}
Substituting then $\mathbf{I}^{(1)}\mathbf{h}^{(2)}\mathbf{\Pi}^{(1)}=\mathbf{h}^{(12)}-\mathbf{h}^{(1)}$, we eventually obtain
\begin{subequations}
	\begin{align}
	\tilde{\mathbf{h}}^{(1)}+\tilde{\mathbf{I}}^{(1)} \tilde{\mathbf{h}}^{(2)}\tilde{\mathbf{\Pi}}^{(1)}
	&=\frac{1}{\mathbf{1}_{T\mathcal{H}}+\mathbf{h}^{(12)}\delta\mathbf{m}}\big(\mathbf{1}_{T\mathcal{H}}
	+\mathbf{h}^{(1)}\delta\mathbf{m}
	\big)\frac{1}{\mathbf{1}_{T\mathcal{H}}+\mathbf{h}^{(1)}\delta\mathbf{m}}\mathbf{h}^{(1)}+\nonumber\\
	&\hspace{5cm}+\frac{1}{\mathbf{1}_{T\mathcal{H}}+\mathbf{h}^{(12)}\delta\mathbf{m} }\mathbf{I}^{(1)}\mathbf{h}^{(2)}\mathbf{\Pi}^{(1)}\\
	&=\frac{1}{\mathbf{1}_{T\mathcal{H}}+\mathbf{h}^{(12)}\delta\mathbf{m}}(\mathbf{h}^{(1)}+\mathbf{I}^{(1)}\mathbf{h}^{(2)}\mathbf{\Pi}^{(1)})\,,
	\end{align}
\end{subequations}
\endgroup
which is clearly equal to $\tilde{\mathbf{h}}^{(12)}$.
Finally, the relations \eqref{eq:relIota} and \eqref{eq:relPi} follow by performing steps, which are completely analogous to those which we have employed above to show that $\delta\tilde{\mathbf{m}}^{(2)}=\delta'\tilde{\mathbf{m}}^{(2)}$. This therefore concludes the proof of the equivalence \eqref{eq:reseq}.

\subsection{Vertical decomposition }\label{subsec:VertComp}

Thus far we have seen that in order to derive the effective physics of modes given by a projector $\mathbf{P}$ by integrating out the remaining modes using a propagator $\mathbf{h}$ such that $\mathbf{Q}\mathbf{h}+\mathbf{h}\mathbf{Q}=\mathbf{1}_{T\mathcal{H}}-\mathbf{P}$ (within the context of an interacting $A_\infty$ SFT with products $\mathbf{m}=\mathbf{Q}+\delta\mathbf{m}$), one simply needs to apply the homological perturbation lemma on the ``free'' SDR \eqref{eq:SDRtensor}, treating the interactions $\delta\mathbf{m}$ as a perturbation. Sometimes, however, it is natural to decompose the perturbation $\delta \mathbf{m}$ into two separate parts $\delta \mathbf{m}^{(1)}$ and $\delta\mathbf{m}^{(2)}$ as $\delta \mathbf{m}=\delta \mathbf{m}^{(1)}+\delta \mathbf{m}^{(2)}$. When deriving an effective action, one can then conceive of either (a) integrating out degrees of freedom in the fully-interacting theory with products $\mathbf{m}=\mathbf{Q}+\delta\mathbf{m}$, or, (b) first deriving an effective action for the theory with interactions $\delta\mathbf{m}^{(1)}$ and only then adding the interactions $\delta \mathbf{m}^{(2)}$ by once more applying the homological perturbation lemma, thus viewing the final theory (derived from a full theory with products $\mathbf{m}$) as a $\delta\mathbf{m}^{(2)}$-perturbation of the ``intermediate'' effective theory which is derived from the full theory with products $\mathbf{m}^{(1)}=\mathbf{Q}+\delta\mathbf{m}^{(1)}$. While one could expect that these two procedures may in general give two different effective theories, we will now show that they are in fact completely equivalent.\footnote{We thank Lada Peksov\'{a} for a discussion on this topic.} 

\subsubsection{Consecutive perturbations}

As we have already hinted at above, we will consider two consecutive perturbations of the BRST charge $\mathbf{Q}$ by coderivations $\delta\mathbf{m}^{(1)}$ and $\delta\mathbf{m}^{(2)}$. Namely, we first perturb $\mathbf{Q}\longrightarrow \mathbf{m}^{(1)}=\mathbf{Q}+\delta\mathbf{m}^{(1)}$, and subsequently we perturb
$\mathbf{m}^{(1)}\longrightarrow \mathbf{m}\equiv \mathbf{m}^{(2)}=\mathbf{m}^{(1)}+\delta\mathbf{m}^{(2)}$.
The perturbations are of course chosen in such a way that we have 
\begin{align}
(\mathbf{Q}+\delta\mathbf{m}^{(1)})^2=(\mathbf{m}^{(1)}+\delta\mathbf{m}^{(2)})^2=0\,.
\end{align}
Notice that in the situations where $\delta\mathbf{m}^{(2)}$ can be consistently rescaled by a continuous parameter $\mu$ as $\delta\mathbf{m}^{(2)}\to \mu \delta\mathbf{m}^{(2)}$, we need to satisfy 
\begin{align}
(\mathbf{m}^{(1)})^2 + \mu[\mathbf{m}^{(1)},\delta\mathbf{m}^{(2)}]+\mu^2 (\delta\mathbf{m}^{(2)})^2=0
\end{align}
order by order in $\mu$, so that we need separately 
\begin{align}
[\delta\mathbf{m}^{(2)},\mathbf{m}^{(1)}]=(\delta\mathbf{m}^{(2)})^2=0\,.\label{eq:obscond}
\end{align}
In particular, we learn that under such circumstances, the coderivation $\delta\mathbf{m}^{(2)}\equiv \mathbf{e}$ yields an observable of the form \eqref{eq:generalFormTensor} for the $\mathbf{m}^{(1)}$-interacting theory (recalling our discussion in subsection \ref{sec:obs}). 

Starting with the free-theory SDR
\begin{align}
\mathrel{\raisebox{+14.5pt}{\rotatebox{-110}{  \begin{tikzcd}[sep=12mm,
			arrow style=tikz,
			arrows=semithick,
			diagrams={>={Straight Barb}}
			]
			\ar[to path={ node[pos=.5,C] }]{}
			\end{tikzcd}
}}}\hspace{-2.7mm}
(-\mathbf{h})\,(T\mathcal{H},\mathbf{Q})
\hspace{-3.2mm}\raisebox{-0.3pt}{$\begin{array}{cc} {\text{\scriptsize $\mathbf{\Pi}$}}\\[-3.0mm] \mathrel{\begin{tikzpicture}[node distance=1cm]
		\node (A) at (0, 0) {};
		\node (B) at (1.5, 0) {};
		\draw[->, to path={-> (\tikztotarget)}]
		(A) edge (B);
		\end{tikzpicture}}\\[-4mm] {\begin{tikzpicture}[node distance=1cm]
		\node (A) at (1.5, 0) {};
		\node (B) at (0, 0) {};
		\draw[->, to path={-> (\tikztotarget)}]
		(A) edge (B);
		\end{tikzpicture}}\\[-3.5mm]
	\text{\scriptsize $\mathbf{I}$} \end{array}$} \hspace{-3.4mm}
(TP\mathcal{H},\mathbf{\Pi}\mathbf{Q}\mathbf{I})\label{eq:origCon}
\end{align}
\vspace{-7mm}

\noindent
we first perturb $\mathbf{Q}$ by $\delta\mathbf{m}^{(1)}$ to obtain the interacting-theory SDR
\begin{align}
\mathrel{\raisebox{+14.5pt}{\rotatebox{-110}{  \begin{tikzcd}[sep=12mm,
			arrow style=tikz,
			arrows=semithick,
			diagrams={>={Straight Barb}}
			]
			\ar[to path={ node[pos=.5,C] }]{}
			\end{tikzcd}
}}}\hspace{-2.7mm}
(-\tilde{\mathbf{h}}^{(1)})\,(T\mathcal{H},\mathbf{m}^{(1)}=\mathbf{Q}+\delta\mathbf{m}^{(1)})
\hspace{-3.2mm}\raisebox{-1.0pt}{$\begin{array}{cc} {\text{\scriptsize $\tilde{\mathbf{\Pi}}^{(1)}$}}\\[-3.0mm] \mathrel{\begin{tikzpicture}[node distance=1cm]
		\node (A) at (0, 0) {};
		\node (B) at (1.5, 0) {};
		\draw[->, to path={-> (\tikztotarget)}]
		(A) edge (B);
		\end{tikzpicture}}\\[-4mm] {\begin{tikzpicture}[node distance=1cm]
		\node (A) at (1.5, 0) {};
		\node (B) at (0, 0) {};
		\draw[->, to path={-> (\tikztotarget)}]
		(A) edge (B);
		\end{tikzpicture}}\\[-3.0mm]
	\text{\scriptsize $\tilde{\mathbf{I}}^{(1)}$} \end{array}$} \hspace{-3.4mm}
(TP\mathcal{H},\tilde{\mathbf{m}}^{(1)}=\mathbf{\Pi Q I}+\delta\tilde{\mathbf{m}}^{(1)})\,,
\end{align}
\vspace{-7mm}

\noindent
where the coderivation $\tilde{\mathbf{m}}^{(1)}$ encodes the effective products after integrating out the degrees of freedom (which are singled out by the projector $\bar{\mathbf{P}}=\mathbf{1}_{T\mathcal{H}}-\mathbf{P}$) using the interactions given by $\delta\mathbf{m}^{(1)}$. As usual, the homological perturbation lemma gives us the following prescription for the $\delta\mathbf{m}^{(1)}$-perturbed data
\begingroup\allowdisplaybreaks
\begin{subequations}
	\begin{align}
	\delta\tilde{\mathbf{m}}^{(1)}&=\mathbf{\Pi} \delta\mathbf{m}^{(1)} \frac{1}{\mathbf{1}_{T\mathcal{H}}+\mathbf{h}\delta\mathbf{m}^{(1)}}\mathbf{I}\,,\\
	\tilde{\mathbf{h}}^{(1)} &= \frac{1}{\mathbf{1}_{T\mathcal{H}}+\mathbf{h}\delta\mathbf{m}^{(1)}}\mathbf{h}\,,\label{eq:eta1}\\
	\tilde{\mathbf{I}}^{(1)} &= \frac{1}{\mathbf{1}_{T\mathcal{H}}+\mathbf{h}\delta\mathbf{m}^{(1)}}\mathbf{I}\,,\\
	\tilde{\mathbf{\Pi}}^{(1)} &= \mathbf{\Pi}\frac{1}{\mathbf{1}_{T\mathcal{H}}+\delta\mathbf{m}^{(1)}\mathbf{h}}\,.\label{eq:exppi1}
	\end{align}
\end{subequations}
\endgroup
Furthermore, let us perturb the differential products $\mathbf{m}^{(1)}$ by adding more interactions $\delta\mathbf{m}^{(2)}$ in order to obtain yet another SDR
\begin{align}
\mathrel{\raisebox{+14.5pt}{\rotatebox{-110}{  \begin{tikzcd}[sep=12mm,
			arrow style=tikz,
			arrows=semithick,
			diagrams={>={Straight Barb}}
			]
			\ar[to path={ node[pos=.5,C] }]{}
			\end{tikzcd}
}}}\hspace{-2.7mm}
(-\tilde{\mathbf{h}}^{(2)})\,(T\mathcal{H},\mathbf{m}=\mathbf{m}^{(1)}+\delta\mathbf{m}^{(2)})
\hspace{-3.2mm}\raisebox{-1.0pt}{$\begin{array}{cc} {\text{\scriptsize $\tilde{\mathbf{\Pi}}^{(2)}$}}\\[-3.0mm] \mathrel{\begin{tikzpicture}[node distance=1cm]
		\node (A) at (0, 0) {};
		\node (B) at (1.5, 0) {};
		\draw[->, to path={-> (\tikztotarget)}]
		(A) edge (B);
		\end{tikzpicture}}\\[-4mm] {\begin{tikzpicture}[node distance=1cm]
		\node (A) at (1.5, 0) {};
		\node (B) at (0, 0) {};
		\draw[->, to path={-> (\tikztotarget)}]
		(A) edge (B);
		\end{tikzpicture}}\\[-3.0mm]
	\text{\scriptsize $\tilde{\mathbf{I}}^{(2)}$} \end{array}$} \hspace{-3.4mm}
(TP\mathcal{H},\tilde{\mathbf{m}}^{(2)}=\tilde{\mathbf{m}}^{(1)}+\delta\tilde{\mathbf{m}}^{(2)})\,,\label{eq:SDR2steps}
\end{align}
\vspace{-7mm}

\noindent
where the homological perturbation lemma instructs us to take
\begingroup\allowdisplaybreaks
\begin{subequations}
	\label{eq:SDR2stepsdata}
	\begin{align}
	\delta\tilde{\mathbf{m}}^{(2)}&=\tilde{\mathbf{\Pi}}^{(1)} \delta\mathbf{m}^{(2)} \frac{1}{\mathbf{1}_{T\mathcal{H}}+\tilde{\mathbf{h}}^{(1)}\delta\mathbf{m}^{(2)}}\tilde{\mathbf{I}}^{(1)}\,,\label{eq:dm2tv}\\
	\tilde{\mathbf{h}}^{(2)} &= \frac{1}{\mathbf{1}_{T\mathcal{H}}+\tilde{\mathbf{h}}^{(1)}\delta\mathbf{m}^{(2)}}\tilde{\mathbf{h}}^{(1)}\,,\\
	\tilde{\mathbf{I}}^{(2)} &= \frac{1}{\mathbf{1}_{T\mathcal{H}}+\tilde{\mathbf{h}}^{(1)}\delta\mathbf{m}^{(2)}}\tilde{\mathbf{I}}^{(1)}\,,\\
	\tilde{\mathbf{\Pi}}^{(2)} &= \tilde{\mathbf{\Pi}}^{(1)}\frac{1}{\mathbf{1}_{T\mathcal{H}}+\delta\mathbf{m}^{(2)}\tilde{\mathbf{h}}^{(1)}}\,.\label{eq:exppi2}
	\end{align}
\end{subequations}
\endgroup
While the coderivation $\tilde{\mathbf{m}}^{(2)}$ clearly encodes products of an interacting theory with $A_\infty$ structure for modes in $TP\mathcal{H}$, strictly speaking its physical meaning should \emph{not} be entirely clear at this point, as the $\tilde{\mathbf{m}}^{(2)}$-theory was \emph{not} obtained by perturbing a free theory.

\subsubsection{Composing the perturbations}

Our aim will now be to show that the just outlined procedure of applying the homological perturbation lemma twice for two consecutive perturbations $\delta\mathbf{m}^{(1)}$, $\delta\mathbf{m}^{(2)}$ produces the same resulting SDR as if we perturbed the BRST $\mathbf{Q}$ in the original free-theory SDR \eqref{eq:origCon} directly by $\delta\mathbf{m}\equiv\delta\mathbf{m}^{(1)}+\delta\mathbf{m}^{(2)}$ so as to obtain
\begin{align}
\mathrel{\raisebox{+14.5pt}{\rotatebox{-110}{  \begin{tikzcd}[sep=12mm,
			arrow style=tikz,
			arrows=semithick,
			diagrams={>={Straight Barb}}
			]
			\ar[to path={ node[pos=.5,C] }]{}
			\end{tikzcd}
}}}\hspace{-2.7mm}
(-\tilde{\mathbf{h}})\,(T\mathcal{H},\mathbf{m}=\mathbf{Q}+\delta\mathbf{m})
\hspace{-3.2mm}\raisebox{-1.0pt}{$\begin{array}{cc} {\text{\scriptsize $\tilde{\mathbf{\Pi}}$}}\\[-3mm] \mathrel{\begin{tikzpicture}[node distance=1cm]
		\node (A) at (0, 0) {};
		\node (B) at (1.5, 0) {};
		\draw[->, to path={-> (\tikztotarget)}]
		(A) edge (B);
		\end{tikzpicture}}\\[-4mm] {\begin{tikzpicture}[node distance=1cm]
		\node (A) at (1.5, 0) {};
		\node (B) at (0, 0) {};
		\draw[->, to path={-> (\tikztotarget)}]
		(A) edge (B);
		\end{tikzpicture}}\\[-3.0mm]
	\text{\scriptsize $\tilde{\mathbf{I}}$} \end{array}$} \hspace{-3.4mm}
(TP\mathcal{H},\tilde{\mathbf{m}}=\mathbf{\Pi Q I}+\delta\tilde{\mathbf{m}})\,,\label{eq:SDR1step}
\end{align}
\vspace{-6mm}

\noindent
where
\begingroup\allowdisplaybreaks
\begin{subequations}
	\label{eq:SDR1stepdata}
	\begin{align}
	\delta\tilde{\mathbf{m}}&=\mathbf{\Pi} \big(\delta\mathbf{m}^{(1)}+\delta\mathbf{m}^{(2)}\big) \frac{1}{\mathbf{1}_{T\mathcal{H}}+\mathbf{h}\big(\delta\mathbf{m}^{(1)}+\delta\mathbf{m}^{(2)}\big)}\mathbf{I}\,,\label{eq:dW12}\\
	\tilde{\mathbf{h}} &= \frac{1}{\mathbf{1}_{T\mathcal{H}}+\mathbf{h}\big(\delta\mathbf{m}^{(1)}+\delta\mathbf{m}^{(2)}\big)}\mathbf{h}\,,\\
	\tilde{\mathbf{I}} &= \frac{1}{\mathbf{1}_{T\mathcal{H}}+\mathbf{h}\big(\delta\mathbf{m}^{(1)}+\delta\mathbf{m}^{(2)}\big)}\mathbf{I}\,,\\
	\tilde{\mathbf{\Pi}} &= \mathbf{\Pi}\frac{1}{\mathbf{1}_{T\mathcal{H}}+\big(\delta\mathbf{m}^{(1)}+\delta\mathbf{m}^{(2)}\big)\mathbf{h}}\,.\label{eq:pi2p}
	\end{align}
\end{subequations}
\endgroup
That is, we are going to show that the SDR \eqref{eq:SDR1step} with data \eqref{eq:SDR1stepdata} is identical to the SDR \eqref{eq:SDR2steps} with data \eqref{eq:SDR2stepsdata}.
Rephrasing what we just wrote in quantitative terms, we are going to show that
\begingroup\allowdisplaybreaks
\begin{subequations}
	\label{eq:EquivVert}
	\begin{align}
	\delta\tilde{\mathbf{m}}&=\delta\tilde{\mathbf{m}}^{(1)}+\delta\tilde{\mathbf{m}}^{(2)}\,,\\
	\tilde{\mathbf{h}} &= \tilde{\mathbf{h}}^{(2)}\,,\label{eq:correta2}\\
	\tilde{\mathbf{I}} &= \tilde{\mathbf{I}}^{(2)}\,,\label{eq:corriota2}\\
	\tilde{\mathbf{\Pi}} &= \tilde{\mathbf{\Pi}}^{(2)}\,.
	\end{align}
\end{subequations}
\endgroup
In particular, this means that when one wants to view the effective theory for $\mathbf{P}$-degrees of freedom derived using the full set $\delta\mathbf{m}$ of interactions as a $\delta\mathbf{m}^{(2)}$-perturbation of the effective theory derived using a partial set $\delta\mathbf{m}^{(1)}$ of interactions, one may conveniently use the expression \eqref{eq:dm2tv} to write the effective products as
\begin{align}
\tilde{\mathbf{m}} = \tilde{\mathbf{m}}^{(1)}+\tilde{\mathbf{\Pi}}^{(1)} \delta\mathbf{m}^{(2)}\tilde{\mathbf{I}}^{(1)} +\sum_{k=1}^\infty(-1)^{k} \tilde{\mathbf{\Pi}}^{(1)} \delta\mathbf{m}^{(2)}(\tilde{\mathbf{h}}^{(1)}\delta\mathbf{m}^{(2)})^k\tilde{\mathbf{I}}^{(1)}\,.\label{eq:mtexp}
\end{align}
Here we recall that in the above-described situation where the perturbation $\delta\mathbf{m}^{(2)}$ can be consistently rescaled by a continuous parameter, it follows from \eqref{eq:obscond} that $\tilde{\mathbf{\Pi}}^{(1)} \delta\mathbf{m}^{(2)}\tilde{\mathbf{I}}^{(1)}$ yields an observable for the $\tilde{\mathbf{m}}^{(1)}$ effective theory (recalling our discussion in subsection \ref{sec:obs}). Hence, while in such situations it is true that at leading order in $\delta\mathbf{m}^{(2)}$ the effective action is perturbed by an observable of the $\tilde{\mathbf{m}}^{(1)}$ effective theory, the expression \eqref{eq:mtexp} makes it manifest that in general we need to add corrections at higher orders in $\delta\mathbf{m}^{(2)}$. This must be the case because we generally have $(\tilde{\mathbf{\Pi}}^{(1)} \delta\mathbf{m}^{(2)}\tilde{\mathbf{I}}^{(1)})^2 = \tilde{\mathbf{\Pi}}^{(1)} \delta\mathbf{m}^{(2)}\tilde{\mathbf{P}}^{(1)} \delta\mathbf{m}^{(2)}\tilde{\mathbf{I}}^{(1)}\neq 0$, unless, for instance, we have $[\tilde{\mathbf{P}}^{(1)}, \delta\mathbf{m}^{(2)}]=0$ (then the sum in \eqref{eq:mtexp} clearly vanishes by virtue of the annihilation condition $\tilde{\mathbf{P}}^{(1)}\tilde{\mathbf{h}}^{(1)}=0$). 

Let us now proceed with proving the equivalence \eqref{eq:EquivVert}. Indeed, remembering that composition of invertible maps generally satisfies $(AB)^{-1}=B^{-1}A^{-1}$, we can for instance start with the expression \eqref{eq:pi2p} for $\tilde{\mathbf{\Pi}}$ and write
\begingroup\allowdisplaybreaks
\begin{subequations}
	\begin{align}
	\tilde{\mathbf{\Pi}}&=\mathbf{\Pi} \frac{1}{\mathbf{1}_{T\mathcal{H}}+\delta\mathbf{m}^{(1)}\mathbf{h}+\delta\mathbf{m}^{(2)}\mathbf{h}}\\[+1mm]
	&=\mathbf{\Pi} \frac{1}{\big(\mathbf{1}_{T\mathcal{H}}+\delta\mathbf{m}^{(2)}\mathbf{h}\frac{1}{\mathbf{1}_{T\mathcal{H}}+\delta\mathbf{m}^{(1)}\mathbf{h}}\big)(\mathbf{1}_{T\mathcal{H}}+\delta\mathbf{m}^{(1)}\mathbf{h})}\\
	&=\mathbf{\Pi} \frac{1}{\mathbf{1}_{T\mathcal{H}}+\delta\mathbf{m}^{(1)}\mathbf{h}}\frac{1}{\mathbf{1}_{T\mathcal{H}}+\delta\mathbf{m}^{(2)}\frac{1}{\mathbf{1}_{T\mathcal{H}}+\mathbf{h}\delta\mathbf{m}^{(1)}}\mathbf{h}}\\
	&=\tilde{\mathbf{\Pi}}^{(1)}\frac{1}{\mathbf{1}_{T\mathcal{H}}+\delta\mathbf{m}^{(2)}\tilde{\mathbf{h}}^{(1)}}\\[3mm]
	&=\tilde{\mathbf{\Pi}}^{(2)}\,,
	\end{align}
\end{subequations}
\endgroup
where in the third equality we have recognized the expressions \eqref{eq:eta1} and \eqref{eq:exppi1} for $\tilde{\mathbf{h}}^{(1)}$ and $\tilde{\mathbf{\Pi}}^{(1)}$, while in the fourth equality, we have finally recognized the expression \eqref{eq:exppi2} for $\tilde{\mathbf{\Pi}}^{(2)}$. The results \eqref{eq:correta2} and \eqref{eq:corriota2} then follow by performing completely analogous steps.
Finally, starting with the expression \eqref{eq:dW12}, we can first perform some straightforward algebraic manipulations to obtain
\begingroup\allowdisplaybreaks
\begin{align}
\delta\tilde{\mathbf{m}}	&=\mathbf{\Pi}\big(\delta\mathbf{m}^{(1)}+\delta\mathbf{m}^{(2)}\big)\frac{1}{\mathbf{1}_{T\mathcal{H}}+\frac{1}{\mathbf{1}_{T\mathcal{H}}+\mathbf{h}\delta\mathbf{m}^{(1)}}\mathbf{h}\delta\mathbf{m}^{(2)}}\frac{1}{\mathbf{1}_{T\mathcal{H}}+\mathbf{h}\delta\mathbf{m}^{(1)}}\mathbf{I} 
\end{align}
Next, isolating the term starting with $\mathbf{\Pi}\delta\mathbf{m}^{(1)}$ and substituting for $\mathbf{\Pi}$ in terms of $\tilde{\mathbf{\Pi}}^{(1)}$ into the remaining term, we eventually obtain 
\begin{align}
\delta\tilde{\mathbf{m}}
&=\tilde{\mathbf{\Pi}}^{(1)}\delta\mathbf{m}^{(2)}\frac{1}{\mathbf{1}_{T\mathcal{H}}+\tilde{\mathbf{h}}^{(1)}\delta\mathbf{m}^{(2)}}\tilde{\mathbf{I}}^{(1)} +\nonumber\\
&\hspace{2.6cm}+\mathbf{\Pi}\frac{1}{\mathbf{1}_{T\mathcal{H}}+\delta\mathbf{m}^{(1)}\mathbf{h}}\delta\mathbf{m}^{(1)}\mathbf{h}\delta\mathbf{m}^{(2)}\times\nonumber\\
&\hspace{4cm}\times\frac{1}{\mathbf{1}_{T\mathcal{H}}+\frac{1}{\mathbf{1}_{T\mathcal{H}}+\mathbf{h}\delta\mathbf{m}^{(1)}}\mathbf{h}\delta\mathbf{m}^{(2)}}\frac{1}{\mathbf{1}_{T\mathcal{H}}+\mathbf{h}\delta\mathbf{m}^{(1)}}\mathbf{I} +\nonumber\\
&\hspace{2.6cm}+\mathbf{\Pi}\delta\mathbf{m}^{(1)}\frac{1}{\mathbf{1}_{T\mathcal{H}}+\frac{1}{\mathbf{1}_{T\mathcal{H}}+\mathbf{h}\delta\mathbf{m}^{(1)}}\mathbf{h}\delta\mathbf{m}^{(2)}}\frac{1}{\mathbf{1}_{T\mathcal{H}}+\mathbf{h}\delta\mathbf{m}^{(1)}}\mathbf{I}\,.\label{eq:intstep}
\end{align}
Since the prefactor-part of second term in \eqref{eq:intstep} may be rewritten as
\begin{align}
\mathbf{\Pi}\frac{1}{\mathbf{1}_{T\mathcal{H}}+\delta\mathbf{m}^{(1)}\mathbf{h}}\delta\mathbf{m}^{(1)}\mathbf{h}\delta\mathbf{m}^{(2)}=\mathbf{\Pi}\delta\mathbf{m}^{(1)}\frac{1}{\mathbf{1}_{T\mathcal{H}}+\mathbf{h}\delta\mathbf{m}^{(1)}}\mathbf{h}\delta\mathbf{m}^{(2)}
\end{align}
it can be straightforwardly combined with the last term in \eqref{eq:intstep} to yield
\begin{subequations}
	\begin{align}
\delta\tilde{\mathbf{m}}
	&=\tilde{\mathbf{\Pi}}^{(1)}\delta\mathbf{m}^{(2)}\frac{1}{\mathbf{1}_{T\mathcal{H}}+\tilde{\mathbf{h}}^{(1)}\delta\mathbf{m}^{(2)}}\tilde{\mathbf{I}}^{(1)} +\nonumber\\
	&	\hspace{1cm}+\mathbf{\Pi}\delta\mathbf{m}^{(1)}\bigg(\mathbf{1}_{T\mathcal{H}}+\frac{1}{\mathbf{1}_{T\mathcal{H}}+\mathbf{h}\delta\mathbf{m}^{(1)}}\mathbf{h}\delta\mathbf{m}^{(2)}\bigg)\times\nonumber\\
	&\hspace{2cm}\times\frac{1}{\mathbf{1}_{T\mathcal{H}}+\frac{1}{\mathbf{1}_{T\mathcal{H}}+\mathbf{h}\delta\mathbf{m}^{(1)}}\mathbf{h}\delta\mathbf{m}^{(2)}}\frac{1}{\mathbf{1}_{T\mathcal{H}}+\mathbf{h}\delta\mathbf{m}^{(1)}}\mathbf{I} \\
	&=\tilde{\mathbf{\Pi}}^{(1)}\delta\mathbf{m}^{(2)}\frac{1}{\mathbf{1}_{T\mathcal{H}}+\tilde{\mathbf{h}}^{(1)}\delta\mathbf{m}^{(2)}}\tilde{\mathbf{I}}^{(1)} 	+\mathbf{\Pi}\delta\mathbf{m}^{(1)}\frac{1}{\mathbf{1}_{T\mathcal{H}}+\mathbf{h}\delta\mathbf{m}^{(1)}}\mathbf{I}\,,
	\end{align}
\end{subequations}
\endgroup
which is clearly equal to $\delta\tilde{\mathbf{m}}^{(1)}+\delta\tilde{\mathbf{m}}^{(2)}$.
\subsection{Summary}\label{subsec:summary}
Before analyzing in detail the specific example of Witten bosonic OSFT, let us summarize the main results of this section, which are valid for all theories based on $A_\infty$ (or $L_\infty$, see appendix \ref{A:Linfty}) structures and which can be taken as general instructions to build tree-level effective actions.
\begin{itemize}
\item Given a projector $P$ projecting on the set of fields that we want to retain, we should identify a (BPZ even) propagator $h$ which provides an Hodge-Kodaira decomposition
\be
[Q,h]=1-P,
\ee
in such a way that $hP=Ph=h^2=0$. Then the Feynman diagrams which give the effective vertices (and thus the effective action) for the fields in the image of $P$ are directly obtained by running the homological perturbation lemma \eqref{eff-products} which automatizes the process of solving the equations of motion for the fields in  $\text{ker}\,P$ and plugging back the result into the original action. If we start with cyclic $A_\infty$ (or $L_\infty$) vertices we end up with cyclic $A_\infty$ (or $L_\infty$) vertices in the effective theory. Moreover, solutions to the equation of motion of the effective theory automatically uplift to solutions of the full microscopic theory.
\item A general class of observables in the UV theory can be constructed \eqref{eq:generalForm}. The homological perturbation lemma tells us what these observables become in the IR (\ref{eq:EffObs}, \ref{eq:etexpl}, \ref{eq:Esubst}) and guarantees that they will be gauge-invariant with respect to the gauge transformations of the effective theory.
\item When we integrate out in different successive steps we should in principle run the homological perturbation lemma with the given propagator and projector at every step. But equivalently we can run it just once, simply considering the sum of the involved propagators and the product of the projectors \eqref{eq:dm2t}. This is the horizontal composition.
\item When we deform the UV theory with a new consistent interaction (which is often provided by an observable), the effective theory will be accordingly deformed. The homological perturbation lemma allows to cleanly identify the new deformed structures in the effective theory which
will be given by the homotopy transfer of the deforming observable plus an infinite set of non-linear corrections \eqref{eq:mtexp}. This is the vertical decomposition.
\item It doesn't matter in which order we do horizontal composition and vertical decompositions because the two processes commute as one can easily verify by simple algebraic manipulations at the level of coalgebra operators. 

\end{itemize}

\section{Application to Witten OSFT}\label{sec:witten}
In this section we will apply our formalism to  Witten bosonic OSFT which is based on an $A_\infty$ algebra with just two multi-string products $m_1$ and $m_2$, corresponding respectively to the BRST charge and Witten's star product.
The action is given by
\begin{align}
S(\Psi) = \sum_{k=1}^2 \frac{1}{k+1}\omega(\Psi,m_k(\Psi^{\otimes k}))=\frac12\omega(\Psi, Q\Psi)+\frac13\omega\left(\Psi,m_2(\Psi,\Psi)\right)\,,
\end{align}
where $\Psi$ is a degree-even, ghost number one state of a bosonic matter/ghost factorized  BCFT$_0$. The 2-product $m_2$ is related to Witten star product as
\be
m_2(\Psi_1,\Psi_2)=(-1)^{d(\Psi_1)}\Psi_1*\Psi_2,
\ee
where the degree $d(\Psi)$ is given by the ghost number augmented by one (mod 2). Similarly the symplectic form $\omega$ is related to BPZ inner product $\langle\cdot,\cdot\rangle$ as
\be
\omega(\Psi_1,\Psi_2)=\langle \omega|\Psi_1\otimes\Psi_2=-(-1)^{d(\Psi_1)}\langle \Psi_1,\Psi_2\rangle.
\ee
The full coderivation giving rise to propagation and interactions is given by
\be
{\bf m}={\bf Q}+{\bf m}_2
\ee
and it is nilpotent
\be
{\bf m}^2=0,
\ee
which means that $Q$ is nilpotent and it is a derivation of Witten product  which is in turn  associative. The coderivations ${\bf m}_k$ are also cyclic with respect to the symplectic form
\be
\langle \omega|\pi_2{\bf m}_k=0.
\ee
To make contact with the previous general discussion, we can also write the action in WZW form as
\be
S(\Psi)=\int_0^1 dt\, \langle\omega|\pi_1 \bs{\p}_t \frac{1}{1-\Psi(t)}\otimes \pi_1{\bf{m}} \frac{1}{1-\Psi(t)},
\ee
where we have chosen a standard interpolation such that $\Psi(0)=0$ and $\Psi(1)=\Psi$.\footnote{Later on, when we will be dealing with the Ellwood invariant, we will (equivalently) choose a different value for $\Psi(0)$.}

\subsection{Effective action for massless fields}
We would like now to obtain the effective action for the massless open string states of  Witten theory. 
We will consider a generic class of open string backgrounds (BCFTs) which can be decomposed as a direct product of a $(D+1)$-dimensional non-compact and flat worldvolume with momentum $k$ (described by a standard external free-field $\text{BCFT}^\text{ext}$ with $c=D+1$ and spacetime indices $\mu,\nu,\ldots=0,\ldots , D$) times an internal  unitary $\text{BCFT}^\text{int}$ with $c=25-D$. Under such conditions, the zero-mode of the total stress-energy tensor (including the ghost sector) can be decomposed as 
\begin{align}
L_0 = \alpha' k^2 +\widehat{L}_0,
\end{align}
where $[k^2 , \widehat{L}_0]=0$. 
Our first aim will be to get an effective action for the fields in the kernel of $\widehat{L}_0$. 

To start with we will specialize to a projector $P=\widehat{P}_0$, projecting on
$\text{ker}\, \widehat{L}_0 \cup \text{ker}\, L_0$, where the cohomology of $Q$ is fully contained.
By construction $\widehat {P}_0$ projects on all massless states which may be away from the mass-shell and also on all, generally massive, states on the mass-shell. Such projector clearly commutes with $Q$ and also satisfies (manifestly) that $\text{ker}\, L_0 \subset \text{im}\widehat {P}_0$, so that the propagator
\begin{align}
h_0=\frac{b_0}{L_0}(1-\widehat {P}_0)\label{eq:propagAlg}
\end{align}
is well-defined. In practice however, we will assume the presence of a gap in the spectrum of $\widehat{L}_0$. That is, we will assume that there exists a value $h_\text{min}\neq 0$ such that
\begin{align}
|\widehat{L}_0|<|h_\text{min}|\implies \widehat{L}_0 =0\,.
\end{align}
We will then be interested in determining the effective action for the off-shell fields in $\text{ker}\,\widehat{L}_0$ at momenta $k$ well below the cut-off $h_\text{min}/{\alpha'}$, that is for $\alpha' |k^2|\ll |h_\text{min}|$. Under such circumstances we can safely ignore the presence of the massive fields in $\text{im}\,\widehat{P}_0$, thus considering $\widehat {P}_0\sim\widehat{P}'_0$ where $\widehat{P}'_0$ projects on just $\text{ker}\,\widehat{L}_0$ (excluding therefore the massive cohomology). Under this cut-off we will be effectively dealing with the propagator $b_0/{L_0}(1-\widehat{P}'_0)$ and to see to which extent is this propagator well defined, we can expand
\be
\frac1{L_0}(1-\widehat{P}'_0)=\frac1{\alpha' k^2+\widehat{L}_0}(1-\widehat{P}'_0)=\frac1{\widehat{L}_0}\,\sum_{n=0}^\infty\left(-\frac{\alpha' k^2}{\widehat{L}_0}\right)^n(1-\widehat{P}'_0).
\ee
In this expression the inverse of $\widehat{L}_0$  always appears protected by the corresponding projector $(1-\widehat{P}'_0)$ and therefore the only possible concern is that the infinite sum could not converge. The sum is in fact an expansion in $\alpha'$ and this expansion converges  precisely when $\alpha' |k|^2< |h_\text{min}|$.
This is essentially saying that every propagator will give rise to an $\alpha'$ expansion which converges whenever $\alpha' |k|^2<| h_\text{min}|$, which is just our working hypothesis of effective field theory. Part of the $\alpha'$-expansion associated to derivative couplings of the effective action will be due to this mechanism.

With these  remarks in mind, in the following we will nevertheless use the projector $\widehat{P}_0$ so that  the structure of the SDR will be cohomologically consistent, with the understanding that we will only be interested in $\alpha' |k|^2< |h_\text{min}|$.\footnote{We can  interpret the $\widehat{P}_0$ projection as giving rise to the effective action of the massless fields, together with the ``almost'' minimal model (i.e. on the mass-shell but not necessarily on the cohomology)  for the massive fields which are invisible by our cut-off. We thank Ted Erler for offering us this picture.}


The propagator \eqref{eq:propagAlg} and the projector $\widehat{P}_0$ clearly satisfy the annihilation conditions $h_0^2 =h_0\widehat{P}_0 =\widehat{P}_0 h_0=0$ as well as the Hodge-Kodaira decomposition
\begin{align}
h_0Q+Qh_0 = 1-\widehat{P}_0\,,
\end{align}
so that defining the canonical projection $\Pi_0: \mathcal{H}\longrightarrow \widehat{P}_0\mathcal{H}$ and inclusion $I_0: \widehat{P}_0\mathcal{H}\longrightarrow \mathcal{H}$ such that $I_0\Pi_0=\widehat{P}_0$ and $\Pi_0 I_0=1_{\widehat{P}_0 \mathcal{H}}$, we have the SDR
\begin{align}
\mathrel{\raisebox{+14.5pt}{\rotatebox{-110}{  \begin{tikzcd}[sep=12mm,
			arrow style=tikz,
			arrows=semithick,
			diagrams={>={Straight Barb}}
			]
			\ar[to path={ node[pos=.5,C] }]{}
			\end{tikzcd}
}}}\hspace{-2.7mm}
(-h_0)\,(\mathcal{H},Q)
\hspace{-3.2mm}\raisebox{-0.2pt}{$\begin{array}{cc} {\text{\scriptsize $\Pi_0$}}\\[-3.0mm] \mathrel{\begin{tikzpicture}[node distance=1cm]
		\node (A) at (0, 0) {};
		\node (B) at (1.5, 0) {};
		\draw[->, to path={-> (\tikztotarget)}]
		(A) edge (B);
		\end{tikzpicture}}\\[-4mm] {\begin{tikzpicture}[node distance=1cm]
		\node (A) at (1.5, 0) {};
		\node (B) at (0, 0) {};
		\draw[->, to path={-> (\tikztotarget)}]
		(A) edge (B);
		\end{tikzpicture}}\\[-3.5mm]
	\text{\scriptsize $I_0$} \end{array}$} \hspace{-3.4mm}
(\widehat{P}_0\mathcal{H},\Pi_0 QI_0)\,.\label{eq:SDRalg1}
\end{align}
\vspace{-6mm}

\noindent
Promoting all these maps on the vector spaces $\mathcal{H}$ and $\widehat{P}_0\mathcal{H}$ to the corresponding maps on the tensor coalgebras $T\mathcal{H}$ and $T\widehat{P}_0\mathcal{H}$ in the way specified in section 2, we obtain the tensor coalgebra version of the SDR \eqref{eq:SDRalg1}
\begin{align}
\mathrel{\raisebox{+14.5pt}{\rotatebox{-110}{  \begin{tikzcd}[sep=12mm,
			arrow style=tikz,
			arrows=semithick,
			diagrams={>={Straight Barb}}
			]
			\ar[to path={ node[pos=.5,C] }]{}
			\end{tikzcd}
}}}\hspace{-2.7mm}
(-\mathbf{h}_0)\,(T\mathcal{H},\mathbf{Q})
\hspace{-3.2mm}\raisebox{-0.2pt}{$\begin{array}{cc} {\text{\scriptsize $\mathbf{\Pi}_0$}}\\[-3.0mm] \mathrel{\begin{tikzpicture}[node distance=1cm]
		\node (A) at (0, 0) {};
		\node (B) at (1.5, 0) {};
		\draw[->, to path={-> (\tikztotarget)}]
		(A) edge (B);
		\end{tikzpicture}}\\[-4mm] {\begin{tikzpicture}[node distance=1cm]
		\node (A) at (1.5, 0) {};
		\node (B) at (0, 0) {};
		\draw[->, to path={-> (\tikztotarget)}]
		(A) edge (B);
		\end{tikzpicture}}\\[-3.5mm]
	\text{\scriptsize $\mathbf{I}_0$} \end{array}$} \hspace{-3.4mm}
(T\widehat{P}_0\mathcal{H},\mathbf{\Pi}_0 \mathbf{Q}\mathbf{I}_0)\,,\label{eq:SDRalg1tens}
\end{align}
where we recall from section 2 that we have defined
\begin{align}
\mathbf{P}_0\pi_k &=\mathbf{I}_0 \mathbf{\Pi}_0\pi_k=\underbrace{\widehat{P}_0\otimes\ldots \otimes \widehat{P}_0 }_{\text{$k$ times}}\,,\\
\mathbf{h}_0\pi_k &= \sum_{l=0}^{k-1}(1_\mathcal{H})^{\otimes l}\otimes h_0\otimes \widehat{P}_0^{\otimes(k-1-l)}.\label{hlift}
\end{align}
\noindent
Perturbing then $\mathbf{m}_1\equiv\mathbf{Q}\to \mathbf{m}=\mathbf{Q}+\mathbf{m}_2$ by adding the cubic interaction and applying the homological perturbation lemma, we obtain a new (perturbed) SDR 
\begin{align}
\mathrel{\raisebox{+14.5pt}{\rotatebox{-110}{  \begin{tikzcd}[sep=12mm,
			arrow style=tikz,
			arrows=semithick,
			diagrams={>={Straight Barb}}
			]
			\ar[to path={ node[pos=.5,C] }]{}
			\end{tikzcd}
}}}\hspace{-2.7mm}
(-\tilde{\mathbf{h}}_0)\,(T\mathcal{H},\mathbf{m})
\hspace{-3.2mm}\raisebox{-0.2pt}{$\begin{array}{cc} {\text{\scriptsize $\tilde{\mathbf{\Pi}}_0$}}\\[-3.0mm] \mathrel{\begin{tikzpicture}[node distance=1cm]
		\node (A) at (0, 0) {};
		\node (B) at (1.5, 0) {};
		\draw[->, to path={-> (\tikztotarget)}]
		(A) edge (B);
		\end{tikzpicture}}\\[-4mm] {\begin{tikzpicture}[node distance=1cm]
		\node (A) at (1.5, 0) {};
		\node (B) at (0, 0) {};
		\draw[->, to path={-> (\tikztotarget)}]
		(A) edge (B);
		\end{tikzpicture}}\\[-3.5mm]
	\text{\scriptsize $\tilde{\mathbf{I}}_0$} \end{array}$} \hspace{-3.4mm}
(T\widehat{P}_0\mathcal{H},\tilde{\mathbf{m}})\,,\label{eq:SDRalg1tensef}
\end{align}
\vspace{-6mm}

\noindent
where the perturbed structures are 
\begin{align}
\tilde{\mathbf{I}}_0 &= \frac{1}{\mathbf{1}_{T\mathcal{H}}+\mathbf{h}_0\mathbf{m}_2}\mathbf{I}_0\,,\label{eq:defIt1}\\
\tilde{\mathbf{\Pi}}_0&=\mathbf{\Pi}_0\frac{1}{\mathbf{1}_{T\mathcal{H}}+\mathbf{m}_2\mathbf{h}_0}\,,\label{eq:defPit1}\\
\tilde{\mathbf{P}}_0 &= \tilde{\mathbf{I}}_0\tilde{\mathbf{\Pi}}_0
	=\frac{1}{\mathbf{1}_{T\mathcal{H}}+\mathbf{h}_0\mathbf{m}_2}\mathbf{P}_0\frac{1}{\mathbf{1}_{T\mathcal{H}}+\mathbf{m}_2\mathbf{h}_0}\,,\\
\tilde{\mathbf{h}}_0
&=\frac{1}{\mathbf{1}_{T\mathcal{H}}+\mathbf{h}_0\mathbf{m}_2}\mathbf{h}_0=\mathbf{h}_0\frac{1}{\mathbf{1}_{T\mathcal{H}}+\mathbf{m}_2\mathbf{h}_0}\,
\end{align}
and finally the effective products are given by
\begin{subequations}
	\label{eq:defmt1}
	\begin{align}
	\tilde{\mathbf{m}} 
	&=\tilde{\mathbf{\Pi}}_0\mathbf{m}\tilde{\mathbf{I}}_0={\mathbf{\Pi}}_0\mathbf{m}\tilde{\mathbf{I}}_0=\tilde{\mathbf{\Pi}}_0\mathbf{m}{\mathbf{I}}_0\\
	&=\mathbf{\Pi}_0\mathbf{Q}\mathbf{I}_0+\mathbf{\Pi}_0\mathbf{m}_2\frac{1}{\mathbf{1}_{T\mathcal{H}}+\mathbf{h}_0\mathbf{m}_2}\mathbf{I}_0\,,
	\end{align}
\end{subequations}
from which, carefully using \eqref{hlift}, the products $\tilde{m}_k=\pi_1\tilde{\bf m}\pi_k$  can be extracted
\begin{subequations}
	\label{eq:effprodAlg}
	\begin{align}
	\tilde{m}_1(\psi)	 &= \widehat{P}_0 m_1(\psi)\,,\\
	\tilde{m}_2(\psi_1,\psi_2)	 &= \widehat{P}_0 m_2(\psi_1,\psi_2)\,,\\
	\tilde{m}_3(\psi_1,\psi_2,\psi_3)	 &= -\widehat{P}_0\left( m_2(h_0m_2(\psi_1,\psi_2),\psi_3)+ m_2(\psi_1,h_0m_2(\psi_2,\psi_3))\right)\,\\
	\tilde{m}_4(\psi_1,\psi_2,\psi_3,\psi_4)&=p\widehat{P}_0{\Big(}m_2(h_0m_2(h_0m_2(\psi_1,\psi_2),\psi_3),\psi_4)+m_2(h_0m_2(\psi_1,\psi_2),(h_0m_2,\psi_3,\psi_4))\0\\
	&\hspace{0.9cm}+m_2(h_0m_2(\psi_1,h_0m_2(\psi_2,\psi_3),)\psi_4)+m_2(\psi_1,h_0m_2(h_0m_2(\psi_2,\psi_3),\psi_4))\0\\
	&\hspace{0.9cm}+m_2(\psi_1,h_0m_2(\psi_2,h_0m_2(\psi_3,\psi_4))){\Big)}\,\\
	&\hspace{0.2cm}\vdots \quad\quad.\nonumber
	\end{align}
\end{subequations}
Since $h_0$ is BPZ even, it follows that the coderivation $\tilde{\mathbf{m}}$ and the cohomomorphism $\tilde{\mathbf{I}}_0$ are cyclic with respect to the symplectic form $$\langle \tilde{\omega}| \pi_2 \equiv \langle \omega|\pi_2 \mathbf{I}_0.$$The  effective action for $\psi\in\text{ker}\,\widehat{L}_0$ therefore reads
\begin{align}
\tilde{S}(\psi) = \int_0^1 dt\,\langle \tilde{\omega} | \pi_1 \bs{\p}_t \frac{1}{1-\psi(t)} \otimes \pi_1\tilde{\mathbf{m}}\frac{1}{1-\psi(t)}=\sum_{k=1}^\infty \frac1{k+1}\omega\left(\psi,\tilde m_k\left(\psi^{\otimes k}\right)\right)\,,
\end{align}
for some interpolation $\psi(t)$ such that $\psi(0)=0$ and $\psi(1)=\psi$.

\subsection{The Nakanishi-Lautrup field and horizontal composition}
\label{subsec:EffEx1}

Inside ker $\widehat{L}_0$ there is more than just the physical fields $c\mathbb{V}_1 e^{ik\cdot X}$ (for $\mathbb{V}_1$ a matter primary field with $h=1$). We also find the auxiliary Nakanishi-Lautrup field $\p c\, e^{ik\cdot X}$. 

Denoting by $\mathbb{V}_1^i$ a generic $h=1$ primary matter fields and $j^\mu = i\sqrt{\frac2{\alpha'}}\p X^\mu$, we can write 
\begin{align}
\mathbb{V}_1(k)=\phi_i(k)\mathbb{V}_1^{{i}}+	A_\mu(k) j^\mu,
\end{align}
with $\phi_i(k)\mathbb{V}_1^{{i}}\in\text{BCFT}^\text{int}$ an internal matter primary and $A_\mu(k) j^\mu \in\text{BCFT}^\text{ext}$. We can generally allow for both $\phi_i(k)$ and $A_\mu(k)$ to carry Chan-Paton factors. The only states residing in $\text{ker}\,\widehat{L}_0$ at ghost number 1 are then\footnote{The products of fields are understood to be normal-ordered.}
\begin{subequations}
	\begin{align}
	\psi_1(k) &=c\mathbb{V}_1(k)e^{ik\cdot X}\,,\\
	\hat{\psi}_1(k) &= B(k)\p ce^{ik\cdot X}\,,
	\end{align}
\end{subequations}
while at ghost number 2 we have
\begin{subequations}
	\begin{align}
	\psi_2(k) &=c\p c\tilde{\mathbb{V}}_1(k)e^{ik\cdot X}\,,\\
	\hat{\psi}_2(k) &= \tilde{B}(k)c\p^2 ce^{ik\cdot X}\,.
	\end{align}
\end{subequations}
Finally, at ghost number $0$ and ghost number 3 we find respectively \be{\psi}_0(k)& =& D(k)e^{ik\cdot X}\0\\{\psi}_3(k) &=& \tilde{D}(k)c\p c\p^2 ce^{ik\cdot X}.\0\ee
There are no states at other ghost numbers. Also note that $\psi_{1,2}(k)$ are primaries for $\mathbb{V}_1(k)=\phi_i(k)\mathbb{V}_1^{{i}}$ but generally non-primary for $\mathbb{V}_1(k)= A_\mu(k)j^\mu$. 
The full classical string field in $\text{ker}\,\widehat{L}_0$ can be expressed as
\begin{align}
\psi&= \int_{\alpha'|k^2|\ll h_\text{min}}\frac{d^{D+1}k}{(2\pi)^{D+1}}e^{ik\cdot X}\big(c\mathbb{V}_1(k)+B(k)\p c\big) \,,\label{eq:kerL0hat1}
\end{align}
and we will usually write $\psi(k)$
for the integrand
of \eqref{eq:kerL0hat1}.

\subsubsection{Algebraic propagator}

The effective products \eqref{eq:effprodAlg} yield an effective action in terms of both the physical modes $\phi_i(k)$, $A_\mu(k)$ and the unphysical modes $B(k)$. So we would like to get rid of $B(k)$. In \cite{Erbin:2019spp}, where a similar problem was solved to the first few orders in Heterotic String Field Theory, it was observed that setting the analogue of the $B$ field to zero by imposing Siegel gauge would leave out-of-Siegel gauge equations which would not be accounted for by the remaining equations of motion (as instead it happens by fixing Siegel gauge for the massive fields). Therefore to get rid of the unphysical field  $B(k)$ we have to integrate it out. If we can do this we  end up  with a gauge invariant action for $\phi_i$ and $A_\mu$ only.  But since we are now in ker $\widehat{L}_0$ we cannot use the usual  propagator to do that. Luckily a new structure comes to rescue. To this end, we recall the decomposition of the BRST charge
\begin{align}
Q=c_0 L_0 +b_0 M^++\widehat{Q}\,,\label{eq:decompQ}
\end{align}
where we have introduced the zero-mode-free operator
\begin{align}
\widehat{Q}=\sum_{n\neq 0} c_{-n}L_n^{\text{m}}-\frac{1}{2}\sum_{\substack{m,n\neq 0\\ m+n\neq 0}}(m-n):c_{-m}c_{-n}b_{m+n}:\,,\label{eq:Qhat}
\end{align}
as well as $M^+$, together with the generators
\begin{subequations}
		\label{eq:su11gen1}
	\begin{align}
	M^+ &= -\sum_{n>0}2nc_{-n}c_n\,,\\
	M^- &= -\sum_{n>0}\frac{1}{2n}b_{-n}b_n\,,\\
	M_z &= \frac{1}{2}\sum_{m>0}(c_{-m}b_m-b_{-m}c_m)\,,
	\end{align}
\end{subequations}
which satisfy the $SU(1,1)$ algebra
\begin{subequations}
	\label{eq:su111}
	\begin{align}
		[M^+,M^-]&=2M_z\,,\\
		[M_z, M^+]&=+M^+\,,\\
		[M_z, M^-]&=-M^-\,.
	\end{align}
\end{subequations}

\noindent
 Let us now consider the following operator
\begin{align}
g = c_0 M^-\widehat{P}_0\,,\label{eq:AlgProp}
\end{align}
where $M^-$ is one of the generators \eqref{eq:su11gen1} which satisfy the $SU(1,1)$  algebra \eqref{eq:su111}. In fact, it is not difficult to see that since the $M^-$ inside $g$ only acts on the states in $\text{ker}\,\widehat{L}_0$, it can always be replaced  by $(1/2)b_1b_{-1}$ so that  we can  write
\begin{align}
g =\frac{1}{2}c_0 b_1 b_{-1} \widehat{P}_0\,.\label{eq:AlgPropShort}
\end{align}
For the reasons which shall become clear below, we will call $g$ the \emph{algebraic propagator}.
Note that we have $[c_0,\widehat{P}_0]=[M^-,\widehat{P}_0]=[Q,\widehat{P}_0]=0$ as well as $g^2 =0=[g,\widehat{P}_0]$. Recalling that the BRST charge $Q$ may be decomposed according to \eqref{eq:decompQ}, it is not hard to show that the zero-mode free part $\widehat{Q}$ \eqref{eq:Qhat} of $Q$ satisfies
\begin{align}
 [M^-,\widehat{Q}]&= -\sum_{m\neq 0}\frac{1}{2m}b_{-m}L_m^{\text{m}}+ \sum_{\substack {{m,n\neq 0}\\
		{m+n\neq 0}}}(m-n)\frac{1}{2n} c_{-m}b_{-n}b_{m+n}\equiv W \,,
\end{align}
so that we also  have $[W,\widehat{P}_0]=[W,c_0]=0$. We can then define the operator
\begin{align}
p = \left\{\begin{array}{ll}
(b_0+W) c_0\widehat{P}_0  & \text{at ghost number 0, 1}\\
c_0(b_0-W)\widehat{P}_0	  & \text{at ghost number 2, 3}
\end{array}\right.\label{eq:pdef}
\end{align}
and we can show that inside ker $\widehat{L}_0$ we have
\begin{align}
\tilde{m}_1 g+g\tilde{m}_1 = 1-p\,,\label{eq:HKalg}
\end{align}
 simply by testing \eqref{eq:HKalg} on all states in $\text{ker}\,\widehat{L}_0$ at ghost numbers 0,1,2,3, as listed above. Combining \eqref{eq:HKalg} with the super-Jacobi identity then yields $[g,p]=[\tilde{m}_1,p]=0$. Furthermore, we can show (again by testing on all states in $\text{ker}\,\widehat{L}_0$) that $gp=0$ and therefore also $pg=0$. This finally enables us to get
\begin{subequations}
	\begin{align}
	(1-p)^2 &= (1-p)(\tilde{m}_1 g+g\tilde{m}_1)\\
	&=\tilde{m}_1 g+g\tilde{m}_1\\
	&=1-p\,,
	\end{align}
\end{subequations}
so that $p$ is a projector and \eqref{eq:HKalg} is therefore a Hodge-Kodaira decomposition with the algebraic propagator $g$ playing the role of a contracting homotopy for $\tilde{m}_1$. Denoting $\bar{p}=1-p$, we can then decompose ker $\widehat{L}_0$ as
\begin{align}
\widehat{P}_0 \mathcal{H} =  p\widehat{P}_0\mathcal{H}\oplus \bar{p}\widehat{P}_0\mathcal{H}\,.
\end{align}
Thinking about $p$ as a map $p:\widehat{P}_0\mathcal{H}\longrightarrow \widehat{P}_0\mathcal{H}$, let us define the associated canonical projection and inclusion
\begin{subequations}
	\begin{align}
	\pi : \widehat{P}_0 \mathcal{H}\longrightarrow p\widehat{P}_0 \mathcal{H}\,,\\
	\iota:p\widehat{P}_0 \mathcal{H}\longrightarrow \widehat{P}_0\mathcal{H}\,,
	\end{align}
\end{subequations}
so that we have $\iota \pi=p$ and $\pi\iota = 1_{p\widehat{P}_0 \mathcal{H}} $ as well as the annihilation conditions
$g\iota =\pi g =0=g^2$. This finally establishes the SDR
\begin{align}
\mathrel{\raisebox{+14.5pt}{\rotatebox{-110}{  \begin{tikzcd}[sep=12mm,
			arrow style=tikz,
			arrows=semithick,
			diagrams={>={Straight Barb}}
			]
			\ar[to path={ node[pos=.5,C] }]{}
			\end{tikzcd}
}}}\hspace{-2.7mm}
(-g)\,(\widehat{P}_0\mathcal{H},\tilde{m}_1)
\hspace{-3.2mm}\raisebox{1.3pt}{$\begin{array}{cc} {\text{\scriptsize $\pi$}}\\[-3.0mm] \mathrel{\begin{tikzpicture}[node distance=1cm]
		\node (A) at (0, 0) {};
		\node (B) at (1.5, 0) {};
		\draw[->, to path={-> (\tikztotarget)}]
		(A) edge (B);
		\end{tikzpicture}}\\[-4mm] {\begin{tikzpicture}[node distance=1cm]
		\node (A) at (1.5, 0) {};
		\node (B) at (0, 0) {};
		\draw[->, to path={-> (\tikztotarget)}]
		(A) edge (B);
		\end{tikzpicture}}\\[-4.5mm]
	\text{\scriptsize $\iota$} \end{array}$} \hspace{-3.4mm}
(p\widehat{P}_0\mathcal{H},\pi\tilde{m}_1\iota)\,.\label{eq:SDRalg}
\end{align}
\vspace{-6mm}

\noindent
The projector $p$ can also be shown to be BPZ self-conjugate: for instance, this can be easily seen by considering the Hodge-Kodaira decomposition \eqref{eq:HKalg} and the known BPZ properties of $\tilde{m}_1$ and $g$. 

\subsubsection{Algebraic reduction of \texorpdfstring{$\mathrm{ker}\, \widehat{L}_0$}{ker L0hat}}
It is interesting to explicitly look at the subspace where the projector $p$ projects. Acting with $p$ on the states in $\text{ker}\,\widehat{L}_0$, we can explicitly compute
\begin{subequations}
	\begin{align}
	\varphi(k)\equiv p\psi(k)&=\bigg[c\phi_i(k)\mathbb{V}_1^i+A(k)\cdot\bigg(c  j +\sqrt\frac{\alpha'}{2}k\p c\bigg)\bigg]e^{ikX}\,,\\
	r(k)\equiv \bar{p}\psi(k)&=\bigg(B(k)-\sqrt\frac{\alpha'}{2}k\cdot A(k)\bigg) \p c e^{ikX}\,.\label{eq:rk}
	\end{align}
\end{subequations}
Integrating out $r(k)$ will therefore yield an effective action for the physical polarizations $\phi_ i(k)$, $A_ \mu(k)$ only, which was our goal from the start. 
Interestingly we can also explicitly verify that $$L_1(p\psi(k))=L_2(p\psi(k))=0,$$ so that $\varphi(k)=p\psi(k)$ is in fact a primary of the full mater-ghost CFT. This is a good news for the effective action which will greatly simplify the computation of the associated off-shell amplitudes.
 Another explicit computation yields the useful result
\begin{align}
Q(p\psi(k))
&=c\p c\Big[-\alpha'k^2 \phi_i(k)\mathbb{V}_1^i+\alpha'\big(k_\mu k_\nu -k^2  g_{\mu\nu}\big) A^\nu (k)j^\mu \Big]e^{ikX}\,,
\end{align}
which gives the expected gauge-invariant kinetic terms for Yang-Mills and the scalar. 

\noindent
Now we can promote $g$  to a map $\mathbf{g}:T\widehat{P}_0\mathcal{H}\longrightarrow T\widehat{P}_0\mathcal{H}$ on the tensor coalgebra $T\widehat{P}_0\mathcal{H}$ such that it acts as
\begin{align}
\mathbf{g}\pi_k &= \sum_{l=0}^{k-1}(1_{\widehat{P}_0\mathcal{H}})^{\otimes l}\otimes g\otimes p^{\otimes(k-1-l)}
\end{align}
and satisfies $\mathbf{g}^2 =0$. In this way  we end up with the tensor coalgebra version of the Hodge-Kodaira decomposition \eqref{eq:HKalg}
\begin{align}
\mathbf{g}\tilde{\mathbf{m}}_1 +\tilde{\mathbf{m}}_1 \mathbf{g} =\mathbf{1}_{T\widehat{P}_0\mathcal{H}} -\mathbf{p}\,,
\end{align}
where we have defined the cohomomorphism $\mathbf{p}$ corresponding to $p$ acting as
$\mathbf{p}\pi_k =p^{\otimes k}$.
Analogously we define the coalgebra extensions $\boldsymbol{\pi}$ and $\boldsymbol{\iota}$ of the projection $\pi$ and inclusion $\iota$ satisfying the annihilation conditions $$\mathbf{g}\bs{\iota}=\bs{\pi}\mathbf{g}=0=\mathbf{g}^2.$$ This establishes the tensor coalgebra version of the SDR \eqref{eq:SDRalg}
\begin{align}
\mathrel{\raisebox{+14.5pt}{\rotatebox{-110}{  \begin{tikzcd}[sep=12mm,
			arrow style=tikz,
			arrows=semithick,
			diagrams={>={Straight Barb}}
			]
			\ar[to path={ node[pos=.5,C] }]{}
			\end{tikzcd}
}}}\hspace{-2.7mm}
(-\mathbf{g})\,(T\widehat{P}_0\mathcal{H},\tilde{\mathbf{m}}_1)
\hspace{-3.2mm}\raisebox{1.3pt}{$\begin{array}{cc} {\text{\scriptsize $\bs{\pi}$}}\\[-3.0mm] \mathrel{\begin{tikzpicture}[node distance=1cm]
		\node (A) at (0, 0) {};
		\node (B) at (1.5, 0) {};
		\draw[->, to path={-> (\tikztotarget)}]
		(A) edge (B);
		\end{tikzpicture}}\\[-4mm] {\begin{tikzpicture}[node distance=1cm]
		\node (A) at (1.5, 0) {};
		\node (B) at (0, 0) {};
		\draw[->, to path={-> (\tikztotarget)}]
		(A) edge (B);
		\end{tikzpicture}}\\[-4.5mm]
	\text{\scriptsize $\bs{\iota}$} \end{array}$} \hspace{-3.4mm}
(Tp\widehat{P}_0\mathcal{H},\bs{\pi}\tilde{\mathbf{m}}_1\bs{\iota})\,.\label{eq:SDRalg2tens}
\end{align}
\vspace{-6mm}

\noindent
Adding interactions by perturbing $\tilde{\mathbf{m}}_1\to\tilde{\mathbf{m}}=\tilde{\mathbf{m}}_1+\delta\tilde{\mathbf{m}}$ (where  $\tilde{\mathbf{m}}$ is expressed in terms of the original microscopic products as $\tilde{\mathbf{m}}=\tilde{\mathbf{\Pi}}_0\mathbf{m}\tilde{\mathbf{I}}_0$) we can apply the homological perturbation lemma to obtain a new SDR
\begin{align}
\mathrel{\raisebox{+14.5pt}{\rotatebox{-110}{  \begin{tikzcd}[sep=12mm,
			arrow style=tikz,
			arrows=semithick,
			diagrams={>={Straight Barb}}
			]
			\ar[to path={ node[pos=.5,C] }]{}
			\end{tikzcd}
}}}\hspace{-2.7mm}
(-\tilde{\mathbf{g}})\,(T\widehat{P}_0\mathcal{H},\tilde{\mathbf{m}})
\hspace{-3.2mm}\raisebox{0.5pt}{$\begin{array}{cc} {\text{\scriptsize $\tilde{\bs{\pi}}$}}\\[-3.0mm] \mathrel{\begin{tikzpicture}[node distance=1cm]
		\node (A) at (0, 0) {};
		\node (B) at (1.5, 0) {};
		\draw[->, to path={-> (\tikztotarget)}]
		(A) edge (B);
		\end{tikzpicture}}\\[-4mm] {\begin{tikzpicture}[node distance=1cm]
		\node (A) at (1.5, 0) {};
		\node (B) at (0, 0) {};
		\draw[->, to path={-> (\tikztotarget)}]
		(A) edge (B);
		\end{tikzpicture}}\\[-4.0mm]
	\text{\scriptsize $\tilde{\bs{\iota}}$} \end{array}$} \hspace{-3.4mm}
(Tp\widehat{P}_0\mathcal{H},\mathbf{M})\,.
\end{align}
\vspace{-6mm}

\noindent
In particular, this gives us the effective products
	\begin{align}
	{\mathbf{N}}&=\boldsymbol{\pi}\tilde{\mathbf{m}}\frac{1}{\mathbf{1}_{T\widehat{P}_0\mathcal{H}}+\mathbf{g}\delta\tilde{\mathbf{m}}}\boldsymbol{\iota}\,,\label{N-products}
	\end{align}
which are cyclic with respect to the symplectic form 
\begin{align}
\langle \Omega | \pi_2 \equiv \langle \tilde{\omega}|\pi_2 \bs{\iota}=\langle {\omega}|\pi_2\mathbf{I}_0 \bs{\iota}\,,
\end{align}
as is the cohomomorphism $\tilde{\bs{\iota}}$. The products $N_k = \pi_1 \mathbf{N}\pi_k$
then determine the vertices of the effective action 
\begin{align}
\tilde{S}_\text{p}(\varphi) = \int_0^1 dt\,\langle \Omega |\pi_1 \bs{\p}_t \frac{1}{1-\varphi(t)} \otimes \pi_1 \mathbf{N}\frac{1}{1-\varphi(t)}\,,\label{eq:SeffPhys}
\end{align}
which now contains only the physical modes $\phi_i(k)$, $A_\mu(k)$ and by construction has an $A_\infty$ gauge symmetry.

\subsubsection{Horizontal composition}

Let us summarize our construction. In order to obtain the  effective action \eqref{eq:SeffPhys} for the physical massless modes $\varphi(k)$, we have first used the propagator $h_0=(b_0/L_0)(1-\widehat{P}_0)$ to integrate out fields which were outside of $\text{ker}\,\widehat{L}_0$ and subsequently employed the algebraic propagator $g=c_0 M^-\widehat{P}_0$ to integrate out the unphysical fields inside $\text{ker}\,\widehat{L}_0$. It is then natural to ask if these two procedures can be combined by introducing a new propagator which would take care of both steps in one go -- this would clearly streamline the explicit evaluation of the vertices of the effective action \eqref{eq:SeffPhys}. We shall now see that the answer to this question turns out to be positive: it is not difficult to note that the two SDRs \eqref{eq:SDRalg1tens} and \eqref{eq:SDRalg2tens} can be horizontally concatenated as
\begin{align}
\mathrel{\raisebox{+14.5pt}{\rotatebox{-110}{  \begin{tikzcd}[sep=12mm,
			arrow style=tikz,
			arrows=semithick,
			diagrams={>={Straight Barb}}
			]
			\ar[to path={ node[pos=.5,C] }]{}
			\end{tikzcd}
}}}\hspace{-2.7mm}
(-\mathbf{h}_0)\,(T\mathcal{H},\mathbf{Q})
\hspace{-3.2mm}\raisebox{-0.3pt}{$\begin{array}{cc} {\text{\scriptsize ${\mathbf{\Pi}_0}$}}\\[-3.0mm] \mathrel{\begin{tikzpicture}[node distance=1cm]
		\node (A) at (0, 0) {};
		\node (B) at (1.5, 0) {};
		\draw[->, to path={-> (\tikztotarget)}]
		(A) edge (B);
		\end{tikzpicture}}\\[-4mm] {\begin{tikzpicture}[node distance=1cm]
		\node (A) at (1.5, 0) {};
		\node (B) at (0, 0) {};
		\draw[->, to path={-> (\tikztotarget)}]
		(A) edge (B);
		\end{tikzpicture}}\\[-3.5mm]
	\text{\scriptsize ${\mathbf{I}_0}$} \end{array}$} \hspace{-7.4mm}
\mathrel{\raisebox{+14.5pt}{\rotatebox{-110}{  \begin{tikzcd}[sep=12mm,
			arrow style=tikz,
			arrows=semithick,
			diagrams={>={Straight Barb}}
			]
			\ar[to path={ node[pos=.5,C] }]{}
			\end{tikzcd}
}}}\hspace{-2.7mm}(-\mathbf{g})\,(T\widehat{P}_0\mathcal{H},\mathbf{\Pi}_0 \mathbf{Q}\mathbf{I}_0)
\hspace{-3.2mm}\raisebox{1.3pt}{$\begin{array}{cc} {\text{\scriptsize ${\boldsymbol{\pi}}$}}\\[-3.0mm] \mathrel{\begin{tikzpicture}[node distance=1cm]
		\node (A) at (0, 0) {};
		\node (B) at (1.5, 0) {};
		\draw[->, to path={-> (\tikztotarget)}]
		(A) edge (B);
		\end{tikzpicture}}\\[-4mm] {\begin{tikzpicture}[node distance=1cm]
		\node (A) at (1.5, 0) {};
		\node (B) at (0, 0) {};
		\draw[->, to path={-> (\tikztotarget)}]
		(A) edge (B);
		\end{tikzpicture}}\\[-4.5mm]
	\text{\scriptsize ${\boldsymbol{\iota}}$} \end{array}$} \hspace{-3.4mm}
(Tp\widehat{P}_0\mathcal{H},\boldsymbol{\pi}\mathbf{\Pi}_0 \mathbf{Q}\mathbf{I}_0\boldsymbol{\iota})
\,,
\end{align}
\vspace{-6mm}

\noindent
so that using our discussion in  section 2, we can establish the corresponding horizontally composed SDR
\begin{align}
\mathrel{\raisebox{+14.5pt}{\rotatebox{-110}{  \begin{tikzcd}[sep=12mm,
			arrow style=tikz,
			arrows=semithick,
			diagrams={>={Straight Barb}}
			]
			\ar[to path={ node[pos=.5,C] }]{}
			\end{tikzcd}
}}}\hspace{-2.7mm}
(-\mathbf{h}_0\circ \mathbf{g})\,(T\mathcal{H},\mathbf{Q})
\hspace{-3.2mm}\raisebox{-0.3pt}{$\begin{array}{cc} {\text{\scriptsize $\bs{\pi}\mathbf{\Pi}_0$}}\\[-3.0mm] \mathrel{\begin{tikzpicture}[node distance=1cm]
		\node (A) at (0, 0) {};
		\node (B) at (1.5, 0) {};
		\draw[->, to path={-> (\tikztotarget)}]
		(A) edge (B);
		\end{tikzpicture}}\\[-4mm] {\begin{tikzpicture}[node distance=1cm]
		\node (A) at (1.5, 0) {};
		\node (B) at (0, 0) {};
		\draw[->, to path={-> (\tikztotarget)}]
		(A) edge (B);
		\end{tikzpicture}}\\[-3.5mm]
	\text{\scriptsize $\mathbf{I}_0\bs{\iota}$} \end{array}$} \hspace{-3.4mm}
(Tp\widehat{P}_0\mathcal{H},\bs{\pi}\mathbf{\Pi}_0\mathbf{Q}\mathbf{I}_0\bs{\iota})\,,\label{eq:SDRHorComposed}
\end{align}
\vspace{-6mm}

\noindent
where (minus) the composed contracting homotopy (i.e. the propagator) $\mathbf{h}_0\circ \mathbf{g}$ is defined as
\begin{align}
\mathbf{h}_0\circ \mathbf{g} = \mathbf{h}_0+ \mathbf{I}_0\mathbf{g}\mathbf{\Pi}_0\,.
\end{align}
By construction this propagator satisfy the Hodge-Kodaira decomposition at the co-algebra level
\be
{\Big[}\bs{\pi}\mathbf{\Pi}_0\mathbf{Q}\mathbf{I}_0\bs{\iota}\,,\,\mathbf{h}_0+ \mathbf{I}_0\mathbf{g}\mathbf{\Pi}_0{\Big ]}={\bf1}_{Tp\widehat{P}_0\mathcal{H}}-{\bf p}{\bf P}_0.
\ee
The products $\mathbf{N}$ \eqref{N-products} can therefore be equivalently computed by instead perturbing the horizontally composed SDR \eqref{eq:SDRHorComposed} by $\mathbf{Q}\to \mathbf{Q}+ \mathbf{m}_2$.
Applying the homological perturbation lemma, we simply obtain
\begin{align}
{\mathbf{N}}&=\boldsymbol{\pi}{\boldsymbol{\Pi}}_0{\mathbf{m}}\frac{1}{\mathbf{1}_{T\mathcal{H}}+(\mathbf{h}_0+{\mathbf{I}}_0\mathbf{g}{\boldsymbol{\Pi}}_0){\mathbf{m}_2}}{\mathbf{I}}_0\boldsymbol{\iota}\,.
\end{align}
Order by order in $\varphi$, the effective products $N_k=\pi_1\mathbf{N}\pi_k$ can be therefore  written down. In doing this it is important to be aware that the composed 
co-algebraic propagator $ \mathbf{h}_0+{\mathbf{I}}_0 \mathbf{g}\mathbf{\Pi}_0$ acts on the tensor algebra as 
\be
( \mathbf{h}_0+ \mathbf{I}_0\mathbf{g}\mathbf{\Pi}_0)\pi_k=\sum_{l=0}^{k-1}(1_\mathcal{H})^{\otimes l}\otimes h_0\otimes \widehat P_0^{\otimes(k-1-l)}+\sum_{l=0}^{k-1} \widehat P_0^{\otimes l}\otimes g\widehat P_0\otimes \left(p\widehat P_0\right)^{\otimes(k-1-l)}.
\ee
Then one can readily verify that this gives the following  cyclic-$A_\infty$ effective  products 
\begingroup\allowdisplaybreaks
\begin{subequations}
	\begin{align}
	{N}_1(\varphi) &= p\widehat{P}_0 Q\varphi\,,\\
	{N}_2(\varphi_1,\varphi_2)&=p\widehat{P}_0m_2(\varphi_1,\varphi_2)\,,\\
	{N}_3(\varphi_1,\varphi_2,\varphi_3)&=-p\widehat{P}_0\left(m_2(h'_0m_2(\varphi_1,\varphi_2),\varphi_3)+m_2(\varphi_1,h'_0m_2(\varphi_2,\varphi_3))\right)\,\\
	{N}_4(\varphi_1,\varphi_2,\varphi_3,\varphi_4)&=p\widehat{P}_0{\Big(}m_2(h'_0m_2(h'_0m_2(\varphi_1,\varphi_2),\varphi_3),\varphi_4)+m_2(h'_0m_2(\varphi_1,\varphi_2),(h'_0m_2,\varphi_3,\varphi_4))\0\\
	&\hspace{0.6cm}+m_2(h'_0m_2(\varphi_1,h'_0m_2(\varphi_2,\varphi_3),)\varphi_4)+m_2(\varphi_1,h'_0m_2(h'_0m_2(\varphi_2,\varphi_3),\varphi_4))\0\\
	&\hspace{0.6cm}+m_2(\varphi_1,h'_0m_2(\varphi_2,h'_0m_2(\varphi_3,\varphi_4))){\Big)}\,\\
	&\hspace{0.2cm}\vdots,\nonumber
	\end{align}
\end{subequations}
\endgroup
 which reproduce the string tree-level perturbation theory, but with a modified propagator 
\be
h'_0=\frac{b_0}{L_0}(1-\widehat P_0)+\frac{1}{2}c_0 b_1 b_{-1} \widehat P_0.\label{modprop}
\ee
%
This expression for the propagator  is tightly related to Sen's prescription for computing amplitudes at zero momentum given in \cite{Sen:2020cef}: the first term gives the standard world-sheet amplitude where all the logarithmic divergences due to massless fields are removed thanks to $(1-P_0)$\footnote{Tachyon divergences are also automatically taken care of as in \cite{Larocca:2017pbo ,Sen:2019jpm}.}. The second term adds the contribution from integrating out the NL field $c_0$ without gauge fixing, but  using the classical gauge invariant action for the path integral. Notice that this second step, while obviously needed for consistency, does not have a natural world-sheet interpretation in terms of moduli space but it is essentially field-theoretical.

Notice also that at zero momentum
we have $N_1(\varphi) \equiv p{P}_0 Q\varphi=0$, so that our method gives an explicit construction of the minimal model $(pP_0\mathcal{H},\{N_k\}_{k=2}^\infty)$ for the zero-momentum cubic OSFT.  It would be interesting to extend this mechanism to the massive fields above the threshold and to give an explicit construction of the complete minimal model to get the full ``correct'' prescription to compute all tree-level amplitudes.

At generic momentum, the results derived in this subsection can be used to  explicitly compute the vertices of the $\text{ker}\,\widehat{L}_0$ effective action \eqref{eq:SeffPhys}  and to investigate the induced  $A_\infty$ gauge symmetry and the associated derivative couplings. On this regard let us  note that exploiting the primariness of $\varphi$, it is easy to verify (using the explicit expression for the Witten's star product  \cite{wedges} which is expressed in terms of a symmetric OPE) that $\widehat{P}_0 m_2(\varphi,\varphi)$ is proportional to $c\p c$, with no contamination of $c\p^2 c$. Therefore we have that $$\widehat{P}_0 g\,m_2(\varphi,\varphi)=0,\quad\quad \textrm{(Witten vertex)}$$ so that up to quartic order the algebraic propagator does not contribute into the effective action for $\varphi\in p\widehat{P}_0 \mathcal{H}$. Note however, that this would cease to be true had we used a non-twist invariant cubic vertex (see \cite{Sen:2020cef} for a related  discussion).  The algebraic propagator will anyhow give contributions at loop level even in Witten theory.

\subsection{Deformations by closed string backgrounds and vertical decomposition}\label{Ellwood}
In the context of Witten theory we now consider deforming the original action by adding the 
Ellwood invariant \cite{Hashimoto:2001sm, Gaiotto:2001ji, Ellwood:2008jh}, so that we will be dealing
with a deformed UV theory of the form
\be
S^{(\mu)}(\Psi) &=&\frac12\omega(\Psi,Q\Psi)+\frac13\omega(\Psi,m_2(\Psi,\Psi))+\mu\, \omega(\Psi,e)\0\\
&=&S^{(\mu)}(\Psi_0)+ \int_0^1 dt\, \langle\omega|\pi_1 \bs{\p}_t \frac{1}{1-\Psi(t)}\otimes \pi_1{\bf{M}}(\mu) \frac{1}{1-\Psi(t)}\,,\label{eq:Sdef}
\ee
where $\Psi(t)$ is a generic interpolation for the full string field such that $\Psi(1)=\Psi$ and $\Psi(0)=\Psi_0$. Here $\Psi_0$ is any constant open string field which will be later fixed to a convenient value. Notice that for $\mu\neq 0$ the theory has a tree-level tadpole, whose consequences will be analyzed in the next subsection.

The full coderivation describing propagation and interactions is 
\be
{\bf{M}}(\mu) = {\bf m}+\mu\, {\bf e},
\ee
which is composed of the usual coderivations  of the Witten theory ${\bf m}={\bf Q}+{\bf m}_2$ and the coderivation $\bf e$ associated to the 0-string product $e$ 
\be
\pi_1{\bf e}\pi_0=e=V(i,-i)|I\rangle,
\ee
which corresponds to the insertion of a weight zero physical closed string field  $V(z,\bar z)$ at the midpoint of the identity string field. From the on-shellness and the midpoint properties of $V$ we have $$[{\bf m},{\bf e}]=0,$$ which, together with the trivial $${\bf e}^2=0,$$ 
gives rise to a (weak) $A_\infty$ algebra.  
In this section, for simplicity, we will be interested in the zero momentum sector and therefore we will consider the projector on the kernel of $L_0$, $P_0$, which gives rise to the following SDR for the free theory 
\begin{align}
\mathrel{\raisebox{+14.5pt}{\rotatebox{-110}{  \begin{tikzcd}[sep=12mm,
			arrow style=tikz,
			arrows=semithick,
			diagrams={>={Straight Barb}}
			]
			\ar[to path={ node[pos=.5,C] }]{}
			\end{tikzcd}
}}}\hspace{-2.7mm}
(-\mathbf{h}_0)\,(T\mathcal{H},\mathbf{Q})
\hspace{-3.2mm}\raisebox{-0.3pt}{$\begin{array}{cc} {\text{\scriptsize $\mathbf{\Pi}_0$}}\\[-3.0mm] \mathrel{\begin{tikzpicture}[node distance=1cm]
		\node (A) at (0, 0) {};
		\node (B) at (1.5, 0) {};
		\draw[->, to path={-> (\tikztotarget)}]
		(A) edge (B);
		\end{tikzpicture}}\\[-4mm] {\begin{tikzpicture}[node distance=1cm]
		\node (A) at (1.5, 0) {};
		\node (B) at (0, 0) {};
		\draw[->, to path={-> (\tikztotarget)}]
		(A) edge (B);
		\end{tikzpicture}}\\[-3.5mm]
	\text{\scriptsize $\mathbf{I}_0$} \end{array}$} \hspace{-3.4mm}
(T{P}_0\mathcal{H},\mathbf{\Pi}_0\mathbf{Q}\mathbf{I}_0)\,.\label{eq:SDRVerTens}
\end{align}
To compute the effective action we  run the homotopy transfer triggered by 
\begin{align}
\mathbf{Q}\to \bf{M}(\mu)= \mathbf{Q}+\delta\bf{M}(\mu)
\end{align}
to get the deformed SDR
\begin{align}
\mathrel{\raisebox{+14.5pt}{\rotatebox{-110}{  \begin{tikzcd}[sep=12mm,
			arrow style=tikz,
			arrows=semithick,
			diagrams={>={Straight Barb}}
			]
			\ar[to path={ node[pos=.5,C] }]{}
			\end{tikzcd}
}}}\hspace{-2.7mm}
(-\tilde{\bs{\mathfrak{h}}}_0(\mu))\,(T\mathcal{H},{\bf M}(\mu))
\hspace{-3.2mm}\raisebox{0.5pt}{$\begin{array}{cc} {\text{\scriptsize $\tilde{\bs{\mathfrak{P}}}_0(\mu)$}}\\[-2.5mm] \mathrel{\begin{tikzpicture}[node distance=1cm]
		\node (A) at (0, 0) {};
		\node (B) at (1.5, 0) {};
		\draw[->, to path={-> (\tikztotarget)}]
		(A) edge (B);
		\end{tikzpicture}}\\[-4mm] {\begin{tikzpicture}[node distance=1cm]
		\node (A) at (1.5, 0) {};
		\node (B) at (0, 0) {};
		\draw[->, to path={-> (\tikztotarget)}]
		(A) edge (B);
		\end{tikzpicture}}\\[-3.5mm]
	\text{\scriptsize $\tilde{\bs{\mathfrak{I}}}_0(\mu)$} \end{array}$} \hspace{-3.4mm}
(T{P}_0\mathcal{H},\tilde{\bf M}(\mu))\,.\label{eq:SDRVerTensDef}
\end{align}
The deformed inclusion and projection can be expressed using the homological perturbation lemma as
\begin{subequations}
\begin{align}
\tilde{\bs{\mathfrak{I}}}_0(\mu) &=\frac{1}{\mathbf{1}_{T\mathcal{H}}+\mathbf{h}_0\delta\bf{M}(\mu)}\mathbf{I}_0\,,\label{eq:cohIdef}\\
\tilde{\bs{\mathfrak{P}}}_0(\mu) &=\mathbf{\Pi}_0\frac{1}{\mathbf{1}_{T\mathcal{H}}+\delta\bf{M}(\mu)\mathbf{h}_0}\,.
\end{align}
\end{subequations}
From this we can express the total field $\Psi$ as a function of the massless one $\psi$ as
\begin{align}
\Psi(\mu; \psi)  = \pi_1 \tilde{\bs{\mathfrak{I}}}_0(\mu) \frac{1}{1-\psi}\,.\label{eq:interpVer}
\end{align}
It should be stressed that, due to the presence of the zero string product ${\bf e}$ in \eqref{eq:interpVer}, the total field $\Psi(\mu; \psi)$ does not vanish when the massless field is set to zero, but it is given by a constant value
\be
\Psi(\mu; 0)&=&\pi_1 \tilde{\bs{\mathfrak{I}}}_0(\mu) \,1_{T\mathcal{H}}=\pi_1\frac1{\mathbf{1}_{T\mathcal{H}}+\frac{1}{\mathbf{1}_{T\mathcal{H}}+{\bf h}_0{\bf m}_2}\mu\,{\bf e}}\,1_{T\mathcal{H}}\0\\
&=&-\mu h_0\,e+\mu^2\,h_0\,m_2\left(h_0\,e,h_0\,e\right)\label{const-shift}\\
&&-\mu^3\,h_0\,\left(m_2\left(h_0e,m_2\left(h_0e,h_0e\right)\right)+(m_2\left(m_2\left(h_0e,h_0e\right),h_0e\right)\right)+ O(\mu^4)\0.
\ee
The effective products in the infrared will be given by
\begin{align}
\tilde{\bf M }(\mu) = \mathbf{\Pi}_0{\bf M }(\mu)\frac{1}{\mathbf{1}_{T\mathcal{H}}+\mathbf{h}_0(\mu\mathbf{e}+\mathbf{m}_2)}\mathbf{I}_0\,,
\end{align} 
but a more intelligible form is given by applying the vertical (de)composition discussed in section 2 and write them as 
\begin{subequations}
\begin{align}
\tilde{\bf M}(\mu)&=\tilde{\mathbf{m}}+\mu\tilde{\mathbf{\Pi}}_0 \mathbf{e}\frac{1}{\mathbf{1}_{T\mathcal{H}}+\mu\tilde{\mathbf{h}}_0\mathbf{e}}\tilde{\mathbf{I}}_0\\
&=\tilde{\mathbf{m}}+\mu\tilde{\mathbf{e}}-\sum_{\alpha=1}^\infty(-\mu)^{\alpha+1}\tilde{\mathbf{\Pi}}_0\mathbf{e}(\tilde{\mathbf{h}}_0\mathbf{e})^\alpha\tilde{\mathbf{I}}_0\,,\label{eq:secondline}
\end{align}
\end{subequations}
where we recall the main objects obtained by deforming the free theory with just ${\bf m}_2$ (the middle step in the vertical decomposition)
\begin{subequations}
	\begin{align}
	\tilde{\mathbf{I}}_0 &= \frac{1}{\mathbf{1}_{T\mathcal{H}}+\mathbf{h}_0 \mathbf{m}_2}{\mathbf{I}}_0\,,\\
	\tilde{\mathbf{\Pi}}_0 &= {\mathbf{\Pi}}_0\frac{1}{\mathbf{1}_{T\mathcal{H}}+ \mathbf{m}_2\mathbf{h}_0}\,,\\
	\tilde{\mathbf{h}}_0 &= \frac{1}{\mathbf{1}_{T\mathcal{H}}+\mathbf{h}_0 \mathbf{m}_2}{\mathbf{h}}_0	\,,\\
	\tilde {\bf m}&=  {\mathbf{\Pi}}_0({\bf Q}+{\bf m}_2)\frac{1}{\mathbf{1}_{T\mathcal{H}}+\mathbf{h}_0 \mathbf{m}_2}{\mathbf{I}}_0.
	\end{align}
\end{subequations}
In equation \eqref{eq:secondline} we see that the effective products in the infrared are the sum of three contributions. The first is  $\tilde{\bf m}$ which is the effective coderivation of original Witten theory without the closed string deformation. The second is $\tilde {\bf e}$, which is the homotopy transfer (not deformed by the closed string) of the UV coderivation $\bf e$
\be
\tilde {\bf e}= \tilde{\mathbf{\Pi}}_0 {\bf e} \tilde{\mathbf{I}}_0.
\ee
This coderivation can be used to construct an observable in the effective theory (not deformed by the closed string) via
\begin{align}
\tilde{\mathcal{E}}(\psi) &= \sum_{k=0}^\infty \frac{1}{k+1}\omega(\psi,\tilde e_k(\psi^{\otimes k}))\,\\
& =\int_0^1 dt\,\langle\omega| \pi_1 \bs{\p}_t \frac{1}{1-\psi(t)}\otimes \pi_1 \tilde{\mathbf{e}} \frac{1}{1-\psi(t)}\,\\
&=\omega\left(\pi_1\frac1{\mathbf{1}_{T\mathcal{H}}+\mathbf{h}_0 \mathbf{m}_2}\frac1{1-\psi}\,,\,e\right),\label{eq:generalFormTensor}
\end{align}
which indeed coincides with the original Ellwood invariant  when the full string field $\Psi$ is expressed in terms of the massless field $\psi$  $without$ the closed string deformation. Notice that this quantity computes S-matrix elements between massless open strings and a single physical  closed string.\footnote{These amplitudes are relevant for constructing the boundary state associated to a perturbative solution of 
the massless equations, see \cite{Larocca:2017pbo} for a fully computable example, although not in Siegel gauge.} This is an observable of the effective theory because, from the general construction of section 2 we have
\be
[\tilde{\bf  m}, \tilde{\bf e}]=0.
\ee
However it is easy to check that ${\tilde{\bf e}}^2\neq 0$ and therefore $\tilde{\bf m}+\mu \tilde{\bf e}$ is not a nilpotent coderivation. The $A_\infty$ structure in the infrared is saved thanks to the third term in \eqref{eq:secondline} which couples an arbitrary number of open strings to at least two closed strings. Therefore we see that even if in the UV theory the closed string couples linearly to the open string field, in the infrared an infinite number of non-linear couplings between open and closed strings is generated.

Also note that the algebraic properties of $\tilde{\bf M}(\mu)$ (namely that it is a cyclic coderivation) need to by satisfied order by order in $\mu$. Hence, it follows from \eqref{eq:secondline} that $\tilde{\mathbf{\Pi}}_0\mathbf{e}(\tilde{\mathbf{h}}_0\mathbf{e})^\alpha \tilde{\mathbf{I}}_0$ are cyclic coderivations for all $\alpha \geq 0$. In general we can write \eqref{eq:secondline} as a double expansion
\begin{align}
\tilde{\bf M}(\mu) =\sum_{k=0}^\infty  \tilde{\bf M}_k(\mu)=\sum_{k=0}^\infty \sum_{\alpha=0}^\infty \mu^\alpha \mathbf{n}_{k\alpha}\,,
\end{align}
where we have introduced cyclic coderivations $\mathbf{n}_{k\alpha}$ with $k$ counting the open string inputs and $\alpha$ counting the closed string insertions. We can therefore write the effective products as
\begin{align}
\tilde M_k^{(\mu)}\equiv \pi_1\tilde{\bf M}_k(\mu)\pi_k = \sum_{\alpha=0}^\infty \mu^\alpha n_{k\alpha}\,,
\end{align}
where we can express
\begin{align}
n_{k\alpha}=\frac{1}{\alpha!}\frac{d^\alpha}{d\mu^\alpha}\pi_1\tilde{\bf M}(\mu)\pi_k\big|_{\mu=0}\,.
\end{align}
In more detail, we can explicitly  write 
\begin{subequations}\label{eq:scoop1}
	\begin{align}
	n_{k0} &= \tilde{m}_k\,,\\
	n_{k1} &= \tilde{e}_k\,,\\
		n_{01}&=P_0 e\,,\\
	n_{k\alpha}(\psi^{\otimes k}) &=\hspace{-2mm} \sum_{\substack{{l_1,\ldots,l_\alpha\geq 0}\\ \sum_{i=1}^{\alpha+1} l_i = k}} (-1)^\alpha\tilde{m}_{k+\alpha}(\psi^{\otimes l_1},h_0e,\psi^{\otimes l_2},h_0e,\ldots,\psi^{\otimes l_\alpha},h_0e,\psi^{\otimes l_{\alpha+1}})\,,\label{eq:scoop2}
		\end{align}
\end{subequations}
 where the last line is valid for $(k,\alpha)\neq(0,1)$. 
 
 We can now write down the effective action. We can do it directly by substituting \eqref{eq:interpVer} into \eqref{eq:Sdef}. In order to do this we choose the natural interpolation
 \be
 \Psi(t)=\Psi\left(\mu;\psi(t)\right)=\pi_1 \tilde{\bs{\mathfrak{I}}}_0(\mu) \frac{1}{1-\psi(t)},\,
 \ee
 where $\psi(0)=0$ and $\psi(1)=\psi$. In this interpolation we have from \eqref{const-shift}
 \be
 \Psi(0)=\pi_1 \tilde{\bs{\mathfrak{I}}}_0(\mu) \,1_{TH}\equiv \Psi_0,
 \ee
which fixes $\Psi_0$ in \eqref{eq:Sdef}.
Then we explicitly get
\be
\tilde S^{(\mu)}(\psi)&=&S^{(\mu)}\left(\Psi(\mu;\psi)\right)=S^{(\mu)}\left(\Psi_0\right)+\int_0^1 dt\, \langle \tilde{\omega}|\pi_1 \bs{\p}_t \frac{1}{1-\psi(t)}\otimes \pi_1\tilde{\bf M}(\mu) \frac{1}{1-\psi(t)}\0\\
&=&S^{(\mu)}\left(\Psi_0\right)+\sum_{k=0}^\infty\frac1{n+1}\omega\left(\psi, \tilde M^{(\mu)}_k\left(\psi^{\otimes k}\right)\right)\0\\
&=&S^{(\mu)}\left(\Psi_0\right)+\sum_{k=0}^\infty\sum_{\alpha=0}^{\infty}\frac{\mu^\alpha}{n+1}\omega\left(\psi,  n_{k\alpha}\left(\psi^{\otimes k}\right)\right).\label{seffell}
\ee
 Notice that the constant term $S^{(\mu)}\left(\Psi_0\right)$ contributes to the vacuum energy but not to the effective equation of motion and the (tree-level) dynamics of the massless fields $\psi$  is  governed by the effective open-closed couplings $n_{k\alpha}$ which have been defined in \eqref{eq:scoop1}.
\subsubsection{Tadpole removal and bulk-induced boundary flows}
Keeping on the interpretation that the Ellwood invariant is a gauge invariant deformation of the original action,  we have now to address the fact that the deformed action \eqref{eq:Sdef}  contains a tadpole, which means that $\Psi=0$ is not a vacuum anymore. In order to remove the tadpole, we need to shift the vacuum of the theory by a classical solution $\Psi_\mathrm{v}(\mu)$ to the  equation of motion of the deformed theory 
\begin{align}
\mu e+Q\Psi+m_2(\Psi,\Psi)=0.
\end{align}
With the assumption that $\lim_{\mu\to0}\Psi_\mathrm{v}(\mu)=0$ we can search for the solution perturbatively\footnote{In this discussion we are limiting ourselves to solutions which are analytic at $\mu=0$}
\begin{align}
\Psi_\mathrm{v}(\mu) = \sum_{\alpha=1}^\infty \mu^\alpha\Psi_\alpha\,,
\end{align}
and, order by order in $\mu$, we obtain the following equations for $\Psi_\alpha$
\begin{subequations}
	\begin{align}
	0&=Q\Psi_1+e\,,\\
	0&=Q\Psi_2+m_2(\Psi_1,\Psi_1)\,,\\
	0&=Q\Psi_3+m_2(\Psi_1,\Psi_2)+m_2(\Psi_2,\Psi_1),\\
	&\hspace{0.2cm}\vdots\nonumber
	\end{align}
\end{subequations}
If we use  Siegel gauge to invert $Q$, the solution can be expressed as (denoting $h_0 = (b_0/L_0)\bar{P}_0$)
\begin{subequations}
	\label{eq:PsiVacSol}
\begin{align}
\Psi_1 &= -h_0e+\psi_1\,,\\
\Psi_2 &= h_0m_2(h_0 e-\psi_1,h_0 e-\psi_1) +\psi_2\,,\\
&\hspace{0.2cm}\vdots\nonumber
\end{align}
\end{subequations}
where $\psi_1, \psi_2, \cdots \psi_\alpha, \cdots$ are in the kernel of $L_0$ (since the component in the complementary space is already accounted for by the part of the state with $h_0$ in front)
\be
P_0\psi_\alpha=\psi_\alpha.
\ee
 Analogously to the discussion in \cite{Vosmera:2019mzw, Mattiello:2019gxc}, we will have a solution provided 
 that the following obstructions (obtained by hitting  \eqref{eq:PsiVacSol} with $Q$, in order to verify the equations of motion)
\begin{subequations}
	\label{eq:PsiVacObst}
\begin{align}
O_1 &= P_0 e+ Q\psi_1\,,\\
O_2 &= P_0 m_2(-h_0 e+\psi_1,-h_0 e+\psi_1) + Q\psi_2\,,\\
&\hspace{0.2cm}\vdots\nonumber
\end{align}
\end{subequations}
all vanish. As we can see, setting to zero the obstructions means to impose dynamical equations for $\psi_\alpha\in {\rm ker}L_0$.
And in fact these obstructions are nothing but the equations of motion of the effective action \eqref{seffell} for the massless fields
\be
\pi_1\tilde {\bf M}(\mu)\frac1{1-\psi}=0,
\ee
when we perturbatively expand $\psi$ in powers of $\mu$
\be
\psi=\sum_{\alpha=1}^\infty \mu^\alpha \psi_\alpha.
\ee
To see this  consider that a solution for the vacuum shift in the full theory $\Psi_\mathrm{v}(\mu)$ can be obtained from a solution of the effective theory $\psi_\mathrm{v}(\mu)$ via \eqref{eq:interpVer}
\begin{align}
\Psi_\mathrm{v}(\mu) = \pi_1 \tilde{\bs{\mathfrak{I}}}_0(\mu) \frac{1}{1-\psi_\mathrm{v}(\mu)}\,.
\end{align}
 In order to verify whether $\Psi_\mathrm{v}(\mu)$  solves the equation of motion derived from the action \eqref{eq:Sdef}, we compute
\begin{subequations}
	\label{eq:compEOMp}
\begin{align}
\pi_1 {\bf M} (\mu)\frac{1}{1-\Psi_\mathrm{v}(\mu)}&=\pi_1 {\bf M} (\mu) \tilde{\bs{\mathfrak{I}}}_0(\mu) \frac{1}{1-\psi_\mathrm{v}(\mu)}\\
&=\pi_1 \tilde{\bs{\mathfrak{I}}}_0(\mu)  \tilde{\bf M} (\mu) \frac{1}{1-\psi_\mathrm{v}(\mu)}\,,
\end{align}
\end{subequations}
where we have used that $\tilde{\bs{\mathfrak{I}}}_0(\mu)$ is an $A_\infty$-morphism intertwining between the (weak) $A_\infty$ structures ${\bf M}(\mu)$ and $\tilde{\bf M}(\mu) $
\begin{align}
{\bf M} (\mu)\tilde{\bs{\mathfrak{I}}}_0(\mu)=\tilde{\bs{\mathfrak{I}}}_0(\mu)\tilde{\bf M} (\mu)\,.\label{eq:intert}
\end{align}
Therefore if $\psi_\mathrm{v}(\mu)$ solves the equation of motion of the effective theory then $\Psi_\mathrm{v}(\mu)$ will solve the equation of motion of the full theory.
Finally writing $$\psi_\mathrm{v}(\mu) = \sum_{\alpha=1}^\infty \mu^\alpha \psi_\alpha$$
and expanding 
 order by order in $\mu$, we can verify that the obstructions \eqref{eq:PsiVacObst} are just the coefficients of the power series expansion of the equation of motion
\begin{subequations}
\begin{align}
\pi_1\tilde{\bf M} (\mu) \frac{1}{1-\psi_\mathrm{v}(\mu)} &=\mu \big(P_0  e+Q\psi_1\big)+\nonumber\\
& +\mu^2 \big(P_0 m_2(h_0 e+\psi_1,h_0 e+\psi_1)+
Q\psi_2\big)+\mathcal{O}(\mu^3)\label{eq:EffVacEOM}\\
&=\sum_{k=1}^\infty \mu^\alpha O_\alpha\,.
\end{align}
\end{subequations}
Just as it happens that putative solutions for marginal deformations may fail to solve the massless equations of motion and are thus obstructed  (which corresponds to the fact that the current used to define the solution is not exactly marginal), here also it is not guaranteed that these massless equations (and therefore the full equations) will have a (perturbative) solution. However if a perturbative solution of the effective theory $\psi_\mathrm{v}(\mu)$ is found, then we will be able to expand around it and by construction the theory around the vacuum shift  will have a proper $A_\infty$ structure with no zero-product, i.e. with no tadpole. And thanks to the homotopy transfer that we have discussed the same will be true for the full theory. This new $A_\infty$ structure will vary continuously with $\mu$ and therefore the cohomology of its 1-product will be a $\mu$-deformation of the cohomology of the original $Q$. This describes how the physical spectrum of the D-brane changes as we  change the closed string background, when the starting boundary conditions are compatible with the exactly marginal bulk deformation parametrized by $\mu$. Examples of this will be reported in \cite{MV}.

If, on the other hand, there is no (perturbative) solution to the massless equation of motion, this means that there is not going to be a stationary point (at least not one $\mu$-parametrically close  to $\Psi=0$).  From a physical point of view this can happen for a couple of reasons. The closed string insertion in the Ellwood invariant can still be an exactly marginal deformation of the closed string background but the boundary conditions of the starting OSFT D-brane are unable to adapt to the bulk deformation.\footnote{ The simplest example of this is given by a generic $SU(2)$ boundary condition for a free boson at the self-dual radius \cite{Gaberdiel:2001xm}, under a change in the radius, which will only be compatible with Neumann or Dirichlet boundary conditions.} Then the D-brane will decay presumably towards some other D-branes whose boundary conditions can adapt to the bulk deformation \cite{Fredenhagen:2006dn}, or simply to the tachyon vacuum. In this case indeed  we would expect that other non-perturbative solutions (e.g. \cite{ Erler:2019fye, Erler:2014eqa}) will admit a consistent $\mu$-deformation and therefore will survive, corresponding to the D-branes that are compatible with the bulk marginal deformation. In particular, if the closed string is exactly marginal, we would always expect to find the (properly deformed) tachyon vacuum solution \cite{Schnabl:2005gv}, at least  for reasonable small-but-finite values of $\mu$. It would be interesting to see this explicitly.
The story is expected to be different however when the closed string insertion is not exactly marginal. In this case the physical picture suggests that all the existing solutions at $\mu=0$ (including the tachyon vacuum) should just cease to exist  and no vacuum will be found at all (we don't expect to have  a consistent OSFT when the bulk is not conformal).  This is also a quite interesting area to investigate.

\section{Conclusions and outlook}\label{concl}
In this paper we have analyzed several aspects of string field theory effective actions whose gauge invariance is encoded in
homotopy structures of $A_\infty$ or $L_\infty$ type.  The associated co-algebra description allows to efficiently package the whole perturbation theory and closed-form expressions for the whole tree-level perturbative series become easily accessible to all orders. 
In the particular case of $A_\infty$ we have defined a new class of observables and we have studied their fate in the effective theory. 
We have moreover discussed two variations on the process of integrating out which we have called horizontal composition and vertical decomposition. 
These two operations allow, respectively, to  extract the effective action after two subsequent processes of integration out, and to systematically obtain the corrections in the effective action after a consistent deformation of the initial UV  theory.
After having discussed these general structures, we have then considered Witten bosonic open string field theory as a simple theoretical lab to test them at work. 
The horizontal composition has been used to efficiently integrate out the Nakanishi-Lautrup field from the set of level zero fields, resulting in corrections to the usual scattering amplitudes of physical massless fields due to an extra algebraic propagator. This is related to 
the discussions in \cite{Sen:2020cef} concerning the role of the extra ghost zero mode $c_0$ in the massless sector and it would be interesting to further explore this relation. The vertical decomposition has been used to account for the effective open-closed couplings that are generated in the infrared by deforming the original theory with the Ellwood Invariant, which acts as a tadpole. 
Depending on the nature of the on-shell closed string insertion the tadpole can be removed by a vacuum shift in the open string field. This vacuum shift physically describes how the original D-brane adapts itself to the new background given by the closed string deformation. If the tadpole can be removed in the effective theory then the same will be true for the full theory. The interplay between closed-strings deformations and change in open-string boundary conditions will be further discussed and developed in \cite{MV} also in the context of superstring theories.

In the next future we would like to further investigate the structure of the low-energy effective field theory for the massless fields (for example the gauge field on a stack of D-branes) and compare it with more conventional approaches to this problem. By construction our effective action has an $A_\infty$-gauge symmetry inherited from the UV and it is interesting to understand how this structure will relate to the $\alpha'$-expansion.

Our analysis has been purely classical (tree-level) but it should be possible to generalize the horizontal composition and vertical decomposition to include loop corrections and thus to work at the level of the full perturbative path-integral, following the general construction of \cite{Sen:2016qap} which reduces to ours in the leading  saddle-point approximation around the perturbative vacuum. 

Continuing in this direction one could  also address the possibility of computing non-perturbative corrections to the effective action, for example due to D-brane instantons, in situations  where the bosonic string makes sense at the quantum level (see e.g. \cite{Klebanov:1991qa, Balthazar:2019ypi, Balthazar:2019rnh, Sen:2019qqg, Sen:2020cef}) and where exact OSFT solutions describing any D-brane system are analytically known in closed form \cite{Schnabl:2005gv, Erler:2019fye, Erler:2014eqa}.

 With an eye towards a superstring generalization, we hope that the effective string field theory  approach that is being developed will be instrumental for a better understanding of string theory at the perturbative as well as at the non-perturbative level.

\section*{Acknowledgments}
We thank Ted Erler, Bra\v{n}o Jur\v{c}o, Martin Markl, Hiroaki Matsunaga, Lada Peksov\'{a} and Ashoke Sen for interesting discussions and Yuji Okawa for correspondence on related results on open-closed couplings.
We thank the organizers of ``Fundamental Aspects of String Theory'', Sao Paolo 1-12 June 2020,  and in particular Nathan Berkovits for giving us the opportunity to present our results.
HE and CM thanks CEICO and the Czech Academy of Science for hospitality during part of this work. JV also thanks INFN Turin for their hospitality during the initial stages of this work. MS thanks the Department of Physics of UC Davis for their hospitality.  
The research of JV has been supported by the Czech Science Foundation - GA\v{C}R, project 19-06342Y.
The research of MS was supported by the GAČR project 18-07776S and RVO: 67985840, as well as by ESIF and M\v{S}MT (Project CoGraDS -CZ.02.1.01/0.0/0.0/15\_ 003/0000437). The work of HE and CM is partially supported by the MIUR PRIN
Contract 2015MP2CX4 ``Non-perturbative Aspects Of Gauge Theories And Strings''.


\appendix 
\section{Homological perturbation lemma}\label{A:hpl}
The goal of this appendix will be  to review the homological perturbation lemma in as simple terms as possible. In other words, we will only aim for a bare minimum, which will enable us to understand how to make use of this powerful concept in writing down compact expressions for tree-level effective SFT actions. As we explain in section \ref{sec:eff}, the main virtue of the lemma lies in its ability to automatically provide the Feynman diagram expansions for the tree-level effective interactions. For a more mathematically minded exposition, the reader should consult \cite{crainic2004perturbation,Pulmann:thesis}, as well as \cite{Costello:2007ei,Gwilliam:thesis} where the applications in the BV formalism are detailed. See also \cite{Konopka:2015tta,Erler:2016rxg,Matsunaga:2019fnc,Masuda:2020tfa} for recent applications of the lemma in string field theory. 

We will specialize on perturbing a particular type of homotopy equivalence data (the strong deformation retract), which we will recognize in section \ref{sec:eff} as naturally fitting into the context of tree-level effective SFT actions.

\subsection{Strong deformation retract}

Consider two $\mathbb{Z}$-graded vector spaces $V$ and $W$ together with maps
$\pi: V\longrightarrow W$,
$\iota: W\longrightarrow V$.
We will assume that the $\iota$-image of $W$ inside $V$ is a \emph{retract} of $V$, namely that 
\begin{align}
\pi  \iota ={1}_W\,.\label{eq:retract}
\end{align}
Let us pause here for a while and think about some implications of this definition: defining further the map $p:V\longrightarrow V$ by $p=\iota \pi$, we learn that $p^2 =\iota \pi\iota \pi = \iota 1_W \pi = p $, as well as $p\iota W = \iota \pi \iota W = \iota W$, so that $p$ is a projector onto $\iota W\subset V$ (that is $\iota W = p V$). Also, we can use the properties recorded so far to write
$\pi V= \pi\iota\pi V=\pi p V = \pi \iota W = W$,
so that the map $\pi$ is necessarily onto. 

We will further assume that the vector spaces $V,W$ are equipped with degree-odd nilpotent differentials $d_V:V\longrightarrow V$, $d_W:W\longrightarrow W$
(that is $(d_V)^2=0$ and $(d_W)^2=0$) such that we have the \emph{chain-map properties}
\begin{subequations}
\begin{align}
d_W \pi &=\pi d_V\,,\label{eq:inter1}\\
\iota  d_W &= d_V \iota\,,\label{eq:inter2}
\end{align}
\end{subequations}
together with $\pi$ and $\iota$ being quasi-isomorphisms (meaning that they induce isomorphisms on the respective cohomologies).
Note that given the retract property \eqref{eq:retract}, it is then possible to show that $d_W$ is the ``pull-back'' of $d_V$ on $W$, that is $\pi d_V\iota= d_W\pi\iota = d_W$. We will also assume that we have a degree-odd map $\eta:V\longrightarrow V$ (called the contracting homotopy) which, together with $d_V,\iota$ and $\pi$, satisfy the \emph{Hodge-Kodaira decomposition}
\begin{align}
\iota \pi - {1}_V = d_V \eta + \eta d_V\,.\label{eq:HodgeKodaira}
\end{align}
The map $\eta$ is therefore a chain homotopy between ${1}_V$ and $\iota \pi=p$. 
We will also assume the \emph{annihilation conditions}
\begin{align}
\eta^2=\eta \iota = \pi  \eta =0
\end{align}
(so that we also have $p\eta = \eta p =0$).
Since the composition of maps on $V$ forms a (graded) associative algebra, the super-Jacobi identity can be used to show
that $[p,d_V]=0$. This finally gives us the chain-map properties.

Altogether, the structure just presented is usually schematized as
\begin{align}
\mathrel{\raisebox{+14.5pt}{\rotatebox{-110}{  \begin{tikzcd}[sep=12mm,
			arrow style=tikz,
			arrows=semithick,
			diagrams={>={Straight Barb}}
			]
			\ar[to path={ node[pos=.5,C] }]{}
			\end{tikzcd}
}}}\hspace{-2.7mm}
\eta\,(V,d_V)
\hspace{-3.2mm}\raisebox{+1.3pt}{$\begin{array}{cc} {\text{\scriptsize $\pi$}}\\[-3.0mm] \mathrel{\begin{tikzpicture}[node distance=1cm]
		\node (A) at (0, 0) {};
		\node (B) at (1.5, 0) {};
		\draw[->, to path={-> (\tikztotarget)}]
		(A) edge (B);
		\end{tikzpicture}}\\[-3.75mm] {\begin{tikzpicture}[node distance=1cm]
		\node (A) at (1.5, 0) {};
		\node (B) at (0, 0) {};
		\draw[->, to path={-> (\tikztotarget)}]
		(A) edge (B);
		\end{tikzpicture}}\\[-4.0mm]
	\text{\scriptsize $\iota$} \end{array}$} \hspace{-3.4mm}
(W,d_W)\,,\label{eq:SDR}
\end{align}
\vspace{-7mm}

\noindent 
and is called a \emph{strong deformation retract} (SDR) or a \emph{contraction} (sometimes also called \emph{special} deformation retract). This is the structure we will find most relevant for our applications in computing tree-level effective actions in string field theory. Note, however, that the homological perturbation lemma (which we are about to state) can be similarly formulated for ordinary deformation retracts (that is, without assuming the annihilation conditions $\eta^2=\eta \iota = \pi  \eta =0$) or even the so-called standard situations (with general homotopy equivalence data, i.e.\ only assuming the chain-map properties $d_W\pi=\pi d_V$ and $\iota d_W = d_V\iota$, as well as the decomposition \eqref{eq:HodgeKodaira} and \emph{not} assuming that $\pi\iota=1_W$). In the simple DR case, however, the retract property $\pi\iota=1_W$ does not turn out to be preserved by the perturbation: see \cite{crainic2004perturbation} and below for more details. Finally, note that \cite{crainic2004perturbation} gives a method of dressing $\eta$, which can be used to turn any deformation retract into a strong deformation retract.

\subsection{Homological perturbation lemma: statement}

Consider now a strong deformation retract of the form \eqref{eq:SDR} and a perturbation $\delta_V:V\longrightarrow V$ of the differential $d_V$ such that the perturbed map $\tilde{d}_V=d_V+\delta_V$ satisfies $(d_V+\delta_V)^2=0$. The homological perturbation lemma then states that for every such perturbation $(V,\tilde{d}_V)=(V,d_V+\delta_V)$ of $(V,d_V)$, we can define the perturbed data
\begin{subequations}
	\label{eq:HPLdefs}
	\begin{align}
	\delta_W &= \pi\delta_V \frac{1}{{1}_V- \eta \delta_V}  \iota\,,\\
	\tilde{\iota}&=\iota+\eta \delta_V\frac{1}{{1}_V-\eta \delta_V}\iota\,,\\
	\tilde{\pi}&=\pi+\pi \delta_V\frac{1}{{1}_V-\eta\delta_V}\eta\,,\\
	\tilde{\eta}&=\eta+\eta\delta_V\frac{1}{{1}_V-\eta\delta_V} \eta\,,
	\end{align}
\end{subequations}
so that upon setting $\tilde{d}_W = d_W + \delta_W$, we obtain a new SDR
\begin{align}
\mathrel{\raisebox{+14.5pt}{\rotatebox{-110}{  \begin{tikzcd}[sep=12mm,
			arrow style=tikz,
			arrows=semithick,
			diagrams={>={Straight Barb}}
			]
			\ar[to path={ node[pos=.5,C] }]{}
			\end{tikzcd}
}}}\hspace{-2.7mm}
\tilde{\eta}\,(V,\tilde{d}_V)
\hspace{-3.2mm}\raisebox{+0.5pt}{$\begin{array}{cc} {\text{\scriptsize $\tilde{\pi}$}}\\[-3.0mm] \mathrel{\begin{tikzpicture}[node distance=1cm]
		\node (A) at (0, 0) {};
		\node (B) at (1.5, 0) {};
		\draw[->, to path={-> (\tikztotarget)}]
		(A) edge (B);
		\end{tikzpicture}}\\[-4mm] {\begin{tikzpicture}[node distance=1cm]
		\node (A) at (1.5, 0) {};
		\node (B) at (0, 0) {};
		\draw[->, to path={-> (\tikztotarget)}]
		(A) edge (B);
		\end{tikzpicture}}\\[-4.0mm]
	\text{\scriptsize $\tilde{\iota}$} \end{array}$} \hspace{-3.4mm}
(W,\tilde{d}_W)\,.
\end{align}
\vspace{-6mm}

\noindent
Note that one can also show that the maps $\tilde{\iota}$, $\tilde{\pi}$ are quasi-isomorphisms (see \cite{crainic2004perturbation} for a proof, which will not be presented here). Altogether we can schematize the statement of the lemma as
\begin{align}
\begin{array}{l}
\mathrel{\raisebox{+14.5pt}{\rotatebox{-110}{  \begin{tikzcd}[sep=12mm,
			arrow style=tikz,
			arrows=semithick,
			diagrams={>={Straight Barb}}
			]
			\ar[to path={ node[pos=.5,C] }]{}
			\end{tikzcd}
}}}\hspace{-2.7mm}
\eta\,(V,d_V)
\hspace{-3.2mm}\raisebox{+1.3pt}{$\begin{array}{cc} {\text{\scriptsize $\pi$}}\\[-3.0mm] \mathrel{\begin{tikzpicture}[node distance=1cm]
		\node (A) at (0, 0) {};
		\node (B) at (1.5, 0) {};
		\draw[->, to path={-> (\tikztotarget)}]
		(A) edge (B);
		\end{tikzpicture}}\\[-4mm] {\begin{tikzpicture}[node distance=1cm]
		\node (A) at (1.5, 0) {};
		\node (B) at (0, 0) {};
		\draw[->, to path={-> (\tikztotarget)}]
		(A) edge (B);
		\end{tikzpicture}}\\[-4.5mm]
	\text{\scriptsize $\iota$} \end{array}$} \hspace{-3.4mm}
(W,d_W)\\[-1.1cm]
\hspace{1.7cm}\mathrel{\rotatebox{90}{$\begin{array}{c}
		\text{\scriptsize $\delta_V$}\\[-2.5mm]
		\mathrel{\begin{tikzpicture}[node distance=1cm]
			\node (A) at (1.5, 0) {};
			\node (B) at (0, 0) {};
			\draw[->, to path={-> (\tikztotarget)}]
			(A) edge (B);
			\end{tikzpicture}}
		\end{array}$}}
\hspace{2.05cm}\mathrel{\rotatebox{90}{$\begin{array}{c}
		\text{\scriptsize $\delta_W$}\\[-2.5mm]
		\mathrel{\begin{tikzpicture}[node distance=1cm]
			\node (A) at (1.5, 0) {};
			\node (B) at (0, 0) {};
			\draw[->, to path={-> (\tikztotarget)}]
			(A) edge (B);
			\end{tikzpicture}}
		\end{array}$}}\\[-0.7cm]
\mathrel{\raisebox{+14.5pt}{\rotatebox{-110}{  \begin{tikzcd}[sep=12mm,
			arrow style=tikz,
			arrows=semithick,
			diagrams={>={Straight Barb}}
			]
			\ar[to path={ node[pos=.5,C] }]{}
			\end{tikzcd}
}}}\hspace{-2.7mm}
\tilde{\eta}\,(V,\tilde{d}_V)
\hspace{-3.2mm}\raisebox{+0.5pt}{$\begin{array}{cc} {\text{\scriptsize $\tilde{\pi}$}}\\[-3.0mm] \mathrel{\begin{tikzpicture}[node distance=1cm]
		\node (A) at (0, 0) {};
		\node (B) at (1.5, 0) {};
		\draw[->, to path={-> (\tikztotarget)}]
		(A) edge (B);
		\end{tikzpicture}}\\[-4mm] {\begin{tikzpicture}[node distance=1cm]
		\node (A) at (1.5, 0) {};
		\node (B) at (0, 0) {};
		\draw[->, to path={-> (\tikztotarget)}]
		(A) edge (B);
		\end{tikzpicture}}\\[-4.0mm]
	\text{\scriptsize $\tilde{\iota}$} \end{array}$} \hspace{-3.4mm}
(W,\tilde{d}_W)
\end{array}
\end{align}
\vspace{-6mm}

\noindent 
with $\delta_W$, $\tilde{\iota}$, $\tilde{\pi}$, $\tilde{\eta}$ as in \eqref{eq:HPLdefs}.

\subsection{Homological perturbation lemma: proof}

In order to prove the lemma, we will find it convenient to define
\begin{align}
a = \delta_V  \frac{1}{1_V-\eta\delta_V}\,,\label{eq:adef}
\end{align}
which allows us to write simply
\begin{subequations}
	\label{eq:simpledef}
	\begin{align}
	\delta_W &= \pi a \iota\,,\\
	\tilde{\iota}&=\iota+\eta a\iota\,,\\
	\tilde{\pi}&=\pi+\pi  a\eta\,,\\
	\tilde{\eta}&=\eta+\eta a \eta\,.
	\end{align}
\end{subequations}
It is then easy to see that we have expressions
\begin{subequations}
	\begin{align}
	1_V+\eta a &=\frac{1}{1_V-\eta\delta_V}\,,\label{eq:1v1}\\ 
	1_V+a\eta  &=\frac{1}{1_V-\delta_V\eta}\,.\label{eq:1v2}
	\end{align}
\end{subequations}
Let us first establish the identity
\begin{align}
a\iota\pi a+a d_V+d_V a=0\,.\label{eq:aiota}
\end{align}
Indeed, substituting first the Hodge-Kodaira decomposition \eqref{eq:HodgeKodaira} for $\iota\pi$ and then using \eqref{eq:1v2}, \eqref{eq:1v2}, we obtain
\begingroup\allowdisplaybreaks
\begin{subequations}
	\begin{align}
	&a\iota\pi a+a d_V+d_V a= \\[+4mm]
	&\hspace{1cm}= a(1_V+\eta d_V+d_V\eta) a+a d_V+d_V a\\[+4mm]
	&\hspace{1cm}=a a +a d_V(1_V+\eta a)+(1_V+a \eta) d_V a\\[+2mm]
	&\hspace{1cm}=a a +a d_V\frac{1}{1_V-\eta\delta_V}+\frac{1}{1_V-\delta_V\eta} d_V a\,.
	\end{align}
\end{subequations}
Introducing further manipulations, using the definition \eqref{eq:adef} and finally using the fact that $(d_V)^2=0$, we find
\begin{subequations}
	\begin{align}
	&a\iota\pi a+a d_V+d_V a= \\[+2mm]
	&\hspace{1cm}=\frac{1}{1_V-\delta_V\eta}\bigg\{
	(1_V-\delta_V\eta) a a(1_V-\eta\delta_V)+\nonumber\\
	&\hspace{2cm}+(1_V-\delta_V\eta) a d_V+d_V a (1_V-\eta\delta_V)
	\bigg\} \frac{1}{1_V-\eta\delta_V}\\
	&\hspace{1cm}=\frac{1}{1_V-\delta_V\eta}\bigg\{
	(\delta_V)^2+\delta_V d_V+d_V \delta_V
	\bigg\} \frac{1}{1_V-\eta\delta_V}\\
	&\hspace{1cm}=\frac{1}{1_V-\delta_V\eta}\bigg\{
	(\delta_V)^2+\delta_V d_V+d_V \delta_V
	+(d_V)^2\bigg\} \frac{1}{1_V-\eta\delta_V}\\
	&\hspace{1cm}=\frac{1}{1_V-\delta_V\eta}(d_V+\delta_V)^2\frac{1}{1_V-\eta\delta_V}\\[+2mm]
	&\hspace{1cm}=0\,,
	\end{align}
\end{subequations}
\endgroup
where the last equality holds as per our assumption that $(\tilde{d}_V)^2=(d_V+\delta_V)^2=0$. We can then use this result to show that
\begingroup\allowdisplaybreaks
\begin{subequations}
	\begin{align}
	(d_W+\delta_W)^2 &=(d_W+\pi a\iota)^2\\
	&=(d_W)^2 + d_W\pi a\iota+\pi a\iota d_W+\pi a\iota \pi a\iota\\
	&=\pi d_V a\iota+\pi a d_V\iota+\pi a\iota \pi a\iota\\
	&=\pi (d_V a+a d_V+a\iota \pi a)\iota\\
	&=0\,,
	\end{align}
\end{subequations}
\endgroup
where in the third equality we have used the chain-map properties \eqref{eq:inter1} and \eqref{eq:inter2}. This shows that given the definitions \eqref{eq:HPLdefs}, the perturbed differential $\tilde{d}_W = d_W+\delta_W$ is indeed nilpotent.
Substituting for the perturbed data from \eqref{eq:simpledef} and using the chain-map property \eqref{eq:inter1}, as well as the Hodge-Kodaira decomposition \eqref{eq:HodgeKodaira}, we also have
\begingroup\allowdisplaybreaks
\begin{subequations}
	\begin{align}
	\tilde{d}_W\tilde{\pi}-\tilde{\pi}\tilde{d}_V &=(d_W+\pi a\iota)(\pi+\pi a\eta)-(\pi+\pi a\eta)(d_V+\delta_V)\\
	&=\pi (a\iota\pi a+  d_V a+ a d_V)\eta+\nonumber\\
	&\hspace{3cm}-\pi \delta_V+\pi a (1_V-\eta \delta_V)\,,
	\end{align}	
\end{subequations}
\endgroup
so that substituting the definition \eqref{eq:adef} of $a$, as well as the identity \eqref{eq:aiota}, gives the perturbed chain-map relation $\tilde{d}_W\tilde{\pi}=\tilde{\pi}\tilde{d}_V$.
Similarly, we have
\begingroup\allowdisplaybreaks
\begin{subequations}
	\begin{align}
	\tilde{\iota}\tilde{d}_W-\tilde{d}_V\tilde{\iota}
	&=(\iota +\eta a \iota)(d_W+\pi a \iota)-({d}_V+\delta_V)(\iota +\eta a \iota)\\
	&=\iota d_W +\eta a \iota d_W
	+\iota \pi a \iota +\eta a \iota \pi a \iota+\nonumber\\
	&\hspace{2cm}-{d}_V\iota-\delta_V\iota
	-{d}_V \eta a \iota-\delta_V\eta a \iota\\
	&= d_V \iota +\eta a  d_V \iota
	+\iota \pi a \iota +\eta a \iota \pi a \iota+\nonumber\\
	&\hspace{2cm}-{d}_V\iota-\delta_V\iota
	-(-\eta{d}_V+\iota\pi-1_V ) a \iota-\delta_V\eta a \iota\\
	&= 
	\eta (a \iota \pi a+a  d_V+d_V a) \iota+\nonumber\\
	&\hspace{2cm}
	+ a \iota-\delta_V(1+\eta a) \iota\\
	&=  a \iota-a \iota\\[1mm]
	&=0\,,
	\end{align}
\end{subequations}
\endgroup
where in the third equality we have used the Hodge-Kodaira decomposition together with the chain-map property \eqref{eq:inter2}, while in the fourth equality we have made use of the identity \eqref{eq:aiota}, as well as of the definition \eqref{eq:adef}. 
In order to show that the perturbed Hodge-Kodaira decomposition holds, we first write
\begingroup\allowdisplaybreaks
\begin{subequations}
	\begin{align}
	\tilde{\eta}\tilde{d}_V+\tilde{d}_V\tilde{\eta} &= (\eta+\eta a \eta)(d_V+\delta_V)+(d_V+\delta_V)(\eta+\eta a \eta)\\
	&= \eta d_V+\eta a \eta d_V
	+\eta \delta_V+\eta a \eta \delta_V+\nonumber\\
	&\hspace{1cm}
	+d_V\eta+\delta_V\eta
	+d_V\eta a \eta+\delta_V\eta a \eta\\
	&= (\eta d_V+d_V\eta)+\eta a (-d_V \eta+\iota\pi-1_V )
	+\eta \delta_V+\eta a \eta \delta_V+\nonumber\\
	&\hspace{1cm}
	+\delta_V\eta
	+(-\eta d_V+\iota\pi-1_V) a \eta+\delta_V\eta a \eta\\
	&= \iota\pi-1_V+\eta(-ad_V-d_V a)\eta+\iota\pi a \eta+\eta a i\pi +\nonumber\\
	&\hspace{1cm}
	+\delta_V\eta
	- a \eta+\delta_V\eta a \eta-\eta a
	+\eta \delta_V+\eta a \eta \delta_V
	\end{align}
\end{subequations}
where we have used the unperturbed Hodge-Kodaira decomposition in the third equality. Substituting now the identity \eqref{eq:aiota} and using the definition \eqref{eq:adef} we have
\begin{subequations}
	\begin{align}
	\tilde{\eta}\tilde{d}_V+\tilde{d}_V\tilde{\eta} &= \iota\pi-1_V+\eta a\iota\pi a\eta+\iota\pi a \eta+\eta a \iota\pi +\nonumber\\
	&\hspace{1cm}
	+\delta_V\eta
	-(1_V-\delta_V \eta) a \eta-\eta a(1_V-\eta \delta_V)
	+\eta \delta_V\\
	&= (\iota+\eta a\iota)(\pi+\pi a\eta)-1_V +\nonumber\\
	&\hspace{1cm}
	+\delta_V\eta
	-\delta_V \eta-\eta\delta_V
	+\eta \delta_V\\
	&=\tilde{\iota}\tilde{\pi}-1_V\,,
	\end{align}
\end{subequations}
\endgroup
where in the last line we have made use of the definition \eqref{eq:simpledef} of the perturbed maps $\tilde{\iota}$ and $\tilde{\pi}$. Notice that up to this point, we have only been using the homotopy-equivalence properties of the unperturbed data in our proofs (that is, the chain-map properties \eqref{eq:inter1}, \eqref{eq:inter2}, as well as the decomposition \eqref{eq:HodgeKodaira}). We can therefore conclude that the perturbed data are again homotopy equivalence data, even without assuming DR, or even full SDR properties of the unperturbed data.
On the other hand, note that we have
\begin{subequations}
	\begin{align}
	\tilde{\pi}\tilde{\iota} &= \pi\frac{1}{1_V-\delta_V \eta}\frac{1}{1_V-\eta\delta_V }\iota\\
	&=\pi\iota\\[2mm]
	&=1_W\,,
	\end{align}
\end{subequations}
where the second line follows from the unperturbed annihilation conditions $\eta^2=\pi\eta=\eta\iota=0$. Therefore, in order to show that the perturbed data are a deformation retract, we need to assume that the unperturbed data we start with are a strong deformation retract. 
Finally, in order to establish the perturbed annihilation conditions, we may write
\begingroup
\allowdisplaybreaks
\begin{subequations}
	\begin{align}
	\tilde{\eta}^2 &= (\eta+\eta a \eta)^2\\
	&= (1_V+\eta a )\eta\eta(1_V+ a \eta)\\
	&=0\,,
	\end{align}
\end{subequations}
\endgroup
because $\eta^2=0$, as well as
\begingroup
\allowdisplaybreaks
\begin{subequations}
	\begin{align}
	\tilde{\eta}\tilde{\iota} &= (\eta+\eta a \eta)(\iota+\eta a \iota)\\
	&= (1_V+\eta a )\eta(\iota+\eta a \iota)\\
	&= (1_V+\eta a )(\eta\iota+\eta\eta a \iota)\\
	&=0\,,
	\end{align}
\end{subequations}
\endgroup
because $\eta^2=\eta\iota =0$, together with
\begingroup
\allowdisplaybreaks
\begin{subequations}
	\begin{align}
	\tilde{\pi}\tilde{\eta} &= (\pi +\pi a \eta)(\eta+\eta a \eta)\\
	&=(\pi +\pi a \eta)\eta(1_V+ a \eta)\\
	&=(\pi\eta +\pi a \eta\eta)(1_V+ a \eta)\\
	&=0\,,
	\end{align}
\end{subequations}
\endgroup
because $\eta^2=\pi\eta=0$. We also introduce the perturbed projector
\begin{align}
\tilde{p}
&=\frac{1}{1_V-\eta \delta_V}p\frac{1}{1_V-\delta_V \eta}\,,
\end{align}
so that the Hodge-Kodaira decomposition and the super-Jacobi identity together imply $[\tilde{p},\tilde{d}_V]=0$.
We therefore also have the property
\begingroup\allowdisplaybreaks
\begin{subequations}
	\begin{align}
	\tilde{\pi}\tilde{d}_V\tilde{\iota}&=\tilde{d}_W \tilde{\pi}\tilde{\iota}\\
	&=\tilde{d}_W\,,
	\end{align}
\end{subequations}
\endgroup
namely that $\tilde{d}_W$ may be thought of as the ``pull-back'' of $\tilde{d}_V$ by the perturbed maps $\tilde{\pi}$, $\tilde{\iota}$.
That is, we may also write
\begin{align}
\tilde{d}_W = \pi \frac{1}{1_V-\delta_V\eta}(d_V+\delta_V)\frac{1}{1_V-\eta\delta_V}\iota\,.
\end{align}
Having verified all the properties which enter the definition of a strong deformation retract, we have therefore successfully verified the homological perturbation lemma.

\section{Effective physics of $L_\infty$ theories}\label{A:Linfty}

Having outlined the construction of effective actions for string field theories based on $A_\infty$ structures at some length in section \ref{sec:eff}, we will now briefly turn to discuss the main points of the corresponding story for  string field theories based on $L_\infty$ structures, whose paradigm is Zwiebach's closed string field theory \cite{Zwiebach:1992ie}. As opposed to the $A_\infty$ case,  $L_\infty$ SFTs are most naturally formulated on symmetrized tensor coalgebras (see in particular \cite{Erler:2014eba,Goto:2015pqv} for an overview). As we will see,  the main obstacle to overcome in order to be able to use the homological perturbation lemma, is to define a suitable uplift of the propagator to a map on the symmetrized tensor coalgebra in such a way that it satisfies an Hodge-Kodaira decomposition.

\subsection{Product notation}

Let us first introduce the framework for string field theories based on cyclic $L_\infty$ structures using the simple notation of graded-symmetrized products on a vector space of states $\mathcal{H}$.

\subsubsection{Basic definitions}

Starting with a degree-graded vector space of states $\mathcal{H}$, let us consider the graded-symmetrized spaces $\mathcal{H}^{\wedge k}$ which consist of the linear combinations of states of the form
\begin{align}
A_1\wedge\ldots \wedge A_k = \sum_{\sigma \in S_k}(-1)^{\epsilon(\sigma)}A_{\sigma(1)}\otimes \ldots \otimes A_{\sigma(k)}\,,\label{eq:wedge}
\end{align}
where $(-1)^{\epsilon(\sigma)}$ are the signs picked up by moving the entries past each other in the manner prescribed by the permutation $\sigma$ (here $S_k$ denotes the symmetric group on $k$ elements). Let us now consider the graded-symmetric multi-linear products $l_k : \mathcal{H}^{\wedge k}\longrightarrow \mathcal{H}$.
Note that we will often simply write $l_k(A_1\wedge\ldots \wedge A_k) = l_k(A_1,\ldots, A_k)$.
Given now two such products $c_k : \mathcal{H}^{\wedge k}\longrightarrow \mathcal{H}$,
$d_l : \mathcal{H}^{\wedge l}\longrightarrow \mathcal{H}$,
let us define a new product
\begin{align}
c_k d_l : \mathcal{H}^{\wedge k+l-1}\longrightarrow	\mathcal{H}
\end{align}
by requiring 
\begin{align}
c_k d_l (A_1,\ldots ,A_{k+l-1}) &=\sum_{\sigma\in S_{k+l-1}}\frac{(-1)^{\epsilon(\sigma)} }{l!(k-1)!}\times\nonumber\\
&\hspace{-0.4cm}\times c_k(d_l(A_{\sigma(1)},\ldots,A_{\sigma(l)}),A_{\sigma(l+1)},\ldots,A_{\sigma(k+l-1)})\,.\label{eq:cd}
\end{align}
As in the non-symmetrized case, this can be rewritten more succinctly. First, given any two multi-linear maps 
$\alpha : \mathcal{H}^{\wedge k}\longrightarrow \mathcal{H}^{\wedge l}$,
$\beta : \mathcal{H}^{\wedge m}\longrightarrow \mathcal{H}^{\wedge n}$,
we can define their wedge product
\begin{align}
\alpha \wedge \beta: \mathcal{H}^{\wedge k+m}\longrightarrow \mathcal{H}^{\wedge k+n}
\end{align}
by writing
\begin{align}
\alpha \wedge \beta(A_1,\ldots,A_{k+m}) &=\!\!\!\sum_{\sigma\in S_{k+m}}\!\!\!\frac{(-1)^{\epsilon(\sigma)}}{k!m!} \alpha(A_{\sigma(1)},\ldots,A_{\sigma{(k)}})\wedge\nonumber\\
&\hspace{3cm}\wedge \beta(A_{\sigma(k+1)},\ldots ,A_{\sigma(k+m)})\,.
\end{align}
The definition \eqref{eq:cd} is then equivalent to writing
\begin{align}
c_k d_l = c_k(d_l\wedge 1_{\mathcal{H}^{\wedge k-1}})\,,
\end{align}
where 
\begin{align}
1_{\mathcal{H}^{\wedge k}} &= \frac{1}{k!}(1_\mathcal{H})^{\wedge k}=(1_\mathcal{H})^{\otimes k}
\end{align}
is the identity operator on $\mathcal{H}^{\wedge k}$. 
We can also define the graded commutator of $c_k$ and $d_l$ by writing
\begin{align}
[c_k,d_l] = c_k d_l - (-1)^{d(c_k)d(d_l)} d_l c_k\,.
\end{align}
Considering a vector space $\mathcal{H}$ equipped with a collection of graded-symmetric degree-odd products $l_k$, we will say that the pair $(\mathcal{H},\{l_k\}_{k\geq 1})$ forms an $L_\infty$ algebra provided that we have
\begin{align}
\sum_{l=1}^{k}  l_l l_{k+1-l} = \frac{1}{2}\sum_{l=1}^{k}  [l_l, l_{k+1-l}]=0\,,\label{eq:Linfrel}
\end{align}
for each $k\geq 1$. In an analogy to the $A_\infty$ case, if the sequence $\{l_k\}_{k\geq 1}$ of products truncates at some $k=N<\infty$, we will call the algebra $L_N$. For instance, for $k=1,2,3,\ldots$,  these relations can be explicitly listed as
\begingroup\allowdisplaybreaks
\begin{subequations}
	\begin{align}
	0&= l_1(l_1(A_1))\,,\label{eq:Linfrel1}\\
	0&= l_1(l_2(A_1,A_2))+l_2(l_1(A_1),A_2)+(-1)^{d(A_1)d(A_2)}l_2(l_1(A_2),A_1)\,,\label{eq:Linfrel2}\\
	0&=l_1(l_3(A_1,A_2,A_3))+l_2(l_2(A_1,A_2),A_3)+\nonumber\\
	&\hspace{1cm}+(-1)^{d(A_1)(d(A_2)+d(A_3))}l_2(l_2(A_2,A_3),A_1)+\nonumber\\
	&\hspace{2cm}+(-1)^{d(A_3)(d(A_1)+d(A_2))}l_2(l_2(A_3,A_1),A_2)+\nonumber\\
	&\hspace{1cm}+l_3(l_1(A_1),A_2,A_3)+(-1)^{d(A_1)}l_3(A_1,l_1(A_2),A_3)+\nonumber\\
	&\hspace{5cm}+(-1)^{d(A_1)+d(A_2)}l_3(A_1,A_2,l_1(A_3))\,,\label{eq:Linfrel3}\\
	&\hspace{0.2cm}\vdots\nonumber
	\end{align}
\end{subequations}
\endgroup
for all $A_1,A_2,A_3,\ldots \in \mathcal{H}$. The relation \eqref{eq:Linfrel1} tells us that the map $l_1$ is nilpotent, the relation \eqref{eq:Linfrel2} says that $l_1$ is a derivation of $l_2$ while the relation \eqref{eq:Linfrel3} says that the failure of $l_1$ to be a derivation of $l_3$ is exactly compensated by the failure of $l_2$ to satisfy the super-Jacobi identity. We will further say that the products are cyclic with respect to a symplectic form $\omega$ provided that we have
\begin{align}
\omega(A_1,l_k(A_2,\ldots,A_{k+1})) = -(-1)^{d(A_1)}\omega(l_k(A_1,\ldots,A_k),A_{k+1})\,.
\end{align}
If this is satisfied, the triple $(\mathcal{H},\{l_k\}_{k\geq 1},\omega)$ will then be called a cyclic $L_\infty$ algebra.

\subsubsection{$L_\infty$ SFT action and symmetrization of $A_\infty$ structures}\label{Asubsub:symm}

Similarly to the $A_\infty$ case, requiring the degree of the dynamical string field to be even, the action of an $L_\infty$ SFT then takes the form
\begin{align}
S(\Psi) = \sum_{k=1}^\infty\frac{1}{(k+1)!}\omega(\Psi,l_k(\Psi^{\wedge k}))\,.\label{eq:SLinf}
\end{align}
Alternatively, introducing an arbitrary smooth interpolation $\Psi(t)$ for $0\leq t \leq 1$ with $\Psi(0)=0$ and $\Psi(1)=\Psi$, the cyclic property of the products $l_k$ with respect to $\omega$ allows us to rewrite \eqref{eq:SLinf} as
\begin{align}
S(\Psi) = \int_0^1 dt\,\sum_{k=1}^\infty\frac{1}{k!}\omega(\dot{A}(t),l_k(\Psi(t)^{\wedge k}))\,.
\end{align}
Varying this action with respect to $\Psi$ and using cyclicity of the products $l_k$ with respect to $\omega$, we obtain the equation of motion (Maurer-Cartan equation)
\begin{align}
\text{EOM}(\Psi)=\sum_{k=1}^\infty \frac{1}{k!}l_k(\Psi^{\wedge k}) = Q\Psi+ \mathcal{J}(\Psi)\,,
\end{align}
where we have separated interactions
\begin{align}
\mathcal{J}(\Psi) = \sum_{k=2}^\infty\frac{1}{k!}l_k(\Psi^k)\,.
\end{align}
Also note that the action \eqref{eq:SLinf} is invariant under the linearized gauge transformation
\begin{align}
\delta_\Lambda\Psi
&=\sum_{k=1}^\infty \frac{1}{(k-1)!}l_k(\Lambda\wedge \Psi^{\wedge k-1})\,,
\end{align}
where $\Lambda\in\mathcal{H}$ is a degree-odd gauge parameter (the corresponding calculation is very similar to what we did in \eqref{eq:AinfGauge}).

Finally, let us consider an $A_\infty$ algebra $(\mathcal{H},\{m_k\}_{k\geq 1})$ and define the graded-symmetrized products
\begin{align}
l_k (A_1,\ldots,A_k) &= \sum_{\sigma\in S_k}(-1)^{\epsilon(\sigma)} m_k (A_{\sigma(1)},\ldots,A_{\sigma(k)})\,,
\end{align}
where $(-1)^{\varepsilon(\sigma)}$ is the obvious sign obtained by moving $A_1,\ldots,A_k$ past each other. In particular, we have
\begin{subequations}
	\begin{align}
	l_1(A_1) &= m_1(A_1)\,,\\
	l_2(A_1,A_2) &= m_2(A_1,A_2)+(-1)^{d(A_1)d(A_2)}m_2(A_2,A_1)\,,\\
	&\hspace{0.2cm}\vdots\nonumber
	\end{align}
\end{subequations}
It is then straightforward to show that the products $l_k$ satisfy the relations of an $L_\infty$ algebra. We therefore observe that given any $A_\infty$ algebra, one may always construct an $L_\infty$ algebra by symmetrizing. In this sense, the notion of an $A_\infty$ algebra appears to be somewhat stronger than that of an $L_\infty$ algebra. Moreover, given a symplectic form $\omega$ on $\mathcal{H}$, such that the products $m_k$ are cyclic with respect to $\omega$, one can also show that the corresponding products $l_k$ are also cyclic with respect to $\omega$.
Noting that we have
\begin{align}
l_k(A ^{\wedge k}) = k!\, m_k(A^{\otimes k})\,,
\end{align}
these facts allow us to rewrite any $A_\infty$ SFT action in an $L_\infty$ form.

\subsubsection{Integrating out unwanted degrees of freedom}

Similarly to what we did in the $A_\infty$ case, let us split the string field as $\Psi=\psi+\mathcal{R}$, where $\psi=P\Psi$ and $\mathcal{R}=(1-P)\Psi\equiv \bar{P}\Psi$. Here $P$ is a BPZ even projector which is such that $\mathcal{R}$ can be integrated out (upon fixing the gauge $h\mathcal{R}=0$) using a propagator $h$ satisfying the Hodge-Kodaira decomposition $Qh+hQ = 1-P$, as well as the annihilation conditions $Ph=hP=h^2=0$ (so that we can write an SDR of the form \eqref{eq:SDRH}).
The equations of motion for $\psi$ and $\mathcal{R}$ then read
\begin{subequations}
	\begin{align}
	\text{EOM}_\psi(\psi,\mathcal{R}) &= P\,\text{EOM}(\psi+\mathcal{R})= Q\psi + P\mathcal{J}(\psi+\mathcal{R})\,,\label{eq:EOMpsiL}\\
	\text{EOM}_\mathcal{R}(\psi,\mathcal{R}) &= \bar{P}\,\text{EOM}(\psi+\mathcal{R})= Q\mathcal{R} + \bar{P}\mathcal{J}(\psi+\mathcal{R})\,.\label{eq:EOMRL}
	\end{align}
\end{subequations}
In a completely parallel way to what we did in the $A_\infty$ case, we will now solve \eqref{eq:EOMRL} (fixing the gauge $h\mathcal{R}=0$) to obtain the in-gauge component $R$ of $\mathcal{R}$ as a function of $\psi$. Denoting $\mathcal{G} = -h\mathcal{J}$, the solution for $\Psi(\psi)\equiv \psi+R(\psi)$ again reads
\begin{align}
\Psi(\psi) = \psi+\mathcal{G}(\psi+\mathcal{G}(\psi+\mathcal{G}(\psi+\ldots)))\,,
\end{align}
or, explicitly up to quartic order in $\psi$,  
\begin{align}
\Psi(\psi) &=\psi -\frac{1}{2!}hl_2(\psi,\psi)-\frac{1}{3!}hl_3(\psi,\psi,\psi)+\frac{2}{(2!)^2}hl_2(hl_2(\psi,\psi),\psi)+\nonumber\\
&\hspace{0.5cm}-\frac{1}{4!}hl_4(\psi,\psi,\psi,\psi)+\frac{2}{2!3!}hl_2(hl_3(\psi,\psi,\psi),\psi)+\frac{3}{2!3!}hl_3(hl_2(\psi,\psi),\psi,\psi)+\nonumber\\
&\hspace{0.5cm}-\frac{1}{(2!)^3}hl_2(hl_2(\psi,\psi),hl_2(\psi,\psi))-\frac{2^2}{(2!)^3}hl_2(hl_2(hl_2(\psi,\psi),\psi),\psi)+\mathcal{O}(\psi^5)\,.\label{eq:Psi(psi)L}
\end{align}
It is then again possible to show that the resulting out-of-gauge constraints are trivialized whenever $\psi$ solves \eqref{eq:EOMpsiL} (the proof is completely parallel to the $A_\infty$ case so that we will not reproduce it here). Substituting \eqref{eq:Psi(psi)L} into the equation of motion \eqref{eq:EOMpsiL} for $\psi$, we obtain the effective equation of motion
\begin{align}
\text{eom}(\psi) = \sum_{k=1}\frac{1}{k!}\tilde{l}_k (\Psi^{\wedge k})\,,\label{eq:eomL}
\end{align}
where the effective products $\tilde{l}_k$ can be expressed as, for $A_k\in\mathcal{H}$,
\begin{subequations}
	\label{eq:effprodL}
	\begin{align}
	\tilde{l}_1(A_1)&=PQ\,,\\
	\tilde{l}_2(A_1,A_2)&=Pl_2(A_1,A_2)\,,\\
	\tilde{l}_3(A_1,A_2,A_3) &= Pl_3(A_1,A_2,A_3)-Pl_2(A_1,hl_2(A_2,A_3))+\nonumber\\
	&\hspace{2cm}-(-1)^{d(A_1)(d(A_2)+d(A_3))}Pl_2(A_2,hl_2(A_3,A_1))+\nonumber\\
	&\hspace{3cm}-(-1)^{d(A_3)(d(A_1)+d(A_2))}Pl_2(A_3,hl_2(A_1,A_2))\,.\label{eq:l3t}\\
	&\hspace{0.2cm}\vdots\nonumber
	\end{align}
\end{subequations}
Using the symmetrized tensor coalgebra language (which is to be introduced below in subsection \ref{Asub:symm}), we can prove that the products $\tilde{l}_k$ satisfy $L_\infty$ relations. Moreover, assuming the BPZ property \eqref{eq:BPZproph}, we can also show order by order that the products $\tilde{l}_k$ are cyclic with respect to the symplectic form $\tilde{\omega}$ on $P\mathcal{H}$ (defined identically as in the $A_\infty$ case). It then follows that the effective action for $\psi$ can be written as
\begin{align}
\tilde{S}(\psi)=S(\Psi(\psi)) = \sum_{k=1}^\infty \frac{1}{(k+1)!}\omega(\psi,\tilde{l}_k(\psi^{\wedge k}))\,.
\end{align}
As a consequence of the fact that the out-of-gauge constraints vanish at classical solutions of \eqref{eq:eomL}, we can say that $\tilde{S}(\psi)$ completely captures the dynamics of $\psi$.

Finally, in the cases when the full SFT products $l_k$ are given by a symmetrization of $A_\infty$ products $m_k$ (see our discussion in subsection \eqref{Asubsub:symm} above), the reader can easily convince herself that the effective products $\tilde{l}_k$, as given by \eqref{eq:effprodL}, can be obtained by symmetrizing the effective products $\tilde{m}_k$, as given by \eqref{eq:effprod}. The property that a particular $L_\infty$ structure is obtained by symmetrizing an $A_\infty$ structure is therefore (classically) preserved by going to IR.

\subsection{Symmetrized tensor coalgebra}\label{Asub:symm}

We will now explain how to derive closed-form expressions for the effective products $\tilde{l}_k$ using the formalism of symmetrized tensor coalgebras and homological perturbation theory. We will observe that once we manage to establish a suitable uplift for the propagator $h$ from $\mathcal{H}$ to $S\mathcal{H}$, the discussion will become identical to the $A_\infty$ case which was dealt with in quite some detail in section \ref{sec:eff}.

\subsubsection{Basic definitions and $L_\infty$ SFT action}

As in the non-symmetric $A_\infty$ case considered in section \ref{sec:eff}, we will now explore the possibility of using tensor constructions to package various structures in the symmetric $L_\infty$ case (see \cite{Erler:2014eba,Goto:2015pqv} for some details on symmetrized tensor coalgebras). First, note that the symmetrized spaces $\mathcal{H}^{\wedge k}$ can be conveniently combined into the symmetrized tensor product space
\begin{align}
S\mathcal{H} = \mathcal{H}^{\wedge 0}\oplus\mathcal{H}^{\wedge 1}\oplus\mathcal{H}^{\wedge 2}\oplus\ldots\label{eq:symmetrized}
\end{align}
We can introduce a coproduct on $S\mathcal{H}$ (recall \eqref{eq:wedge} and \eqref{eq:symmetrized} for the definition of $S\mathcal{H}$) as a linear map
$\Delta_{S\mathcal{H}}: S\mathcal{H}\longrightarrow S\mathcal{H}\otimes' S\mathcal{H}$
satisfying (see e.g.\ \cite{Goto:2015pqv,Pulmann:thesis})
\begin{align}
\Delta_{S\mathcal{H}}(A_1\wedge \ldots \wedge A_k) &=\sum_{\substack {l_1,l_2\\ l_1+l_2=k}}\sum_{\sigma\in S_{l_1+l_2}}\!\!\!\frac{(-1)^{\epsilon(\sigma)}}{l_1!l_2!} (A_{\sigma(1)}\wedge\ldots\wedge A_{\sigma{(l_1)}})\otimes'\nonumber\\[-2mm]
&\hspace{4.0cm}\otimes' (A_{\sigma(l_1+1)}\wedge \ldots \wedge A_{\sigma(l_1+l_2)})\,.\label{eq:coprS}
\end{align}
Clearly one can replace the sum over $\sigma \in S_{l_1+l_2}$ in \eqref{eq:coprS} with the sum over all $(l_1,l_2)$-unshuffles so that there would be no need for the $(l_1!l_2!)^{-1}$ prefactor which is currently compensating for overcounting.
It is easy to see that the definition \eqref{eq:coprS} can be induced from our previous definition \eqref{eq:coprT} of the coassociative coproduct on $T\mathcal{H}$ by using the correspondence \eqref{eq:wedge}. For any multi-linear symmetric $k$-string product $c_k :\mathcal{H}^{\wedge k}\longrightarrow \mathcal{H}$, let us define the coderivations $\mathbf{c}_k : S\mathcal{H}\longrightarrow S\mathcal{H}$
derived from $c_k$ by requiring that for $N\geq k$, these act on $\mathcal{H}^{\wedge N}$ as
\begin{align}
\mathbf{c}_k\pi_N = \big[c_k\wedge 1_{\mathcal{H}^{\wedge N-k}}\big]\pi_N\label{eq:correspL}
\end{align}
and that they vanish on $\mathcal{H}^{\wedge N}$ for $N< k$. These may be straightforwardly shown to satisfy the co-Leibniz rule with respect to the coproduct \eqref{eq:coprS}. In an obvious way, we can also define the graded commutator $[\mathbf{c}_k,\mathbf{d}_l]$ and we may show it to agree with the coderivation derived (using the relation \eqref{eq:correspL}) from the commutator $[c_k,d_l]$. Defining a cohomomorphism $\mathbf{F}$ as a linear map $\mathbf{F}:S\mathcal{H}\longrightarrow S\mathcal{H}'$ satisfying the property $\Delta_{S\mathcal{H}'} \mathbf{F}=(\mathbf{F}\otimes' \mathbf{F})\Delta_{S\mathcal{H}}$ with respect to the coproduct \eqref{eq:coprS}, we may alternatively express $\mathbf{F}$ as a collection of degree zero multilinear maps (for each $k\geq 0$)
$F_k:\mathcal{H}^{\wedge k}\longrightarrow \mathcal{H}'$
by writing, for each $j,k\geq 0$
\begin{align}
\pi_j\mathbf{F}\pi_k = \sum_{l_1+\ldots+l_j=k} \frac{1}{j!}\big[F_{l_1}\wedge \ldots \wedge F_{l_j}\big]\pi_k\,.
\end{align}
We will often find it convenient to work with (for a degree-even $A\in\mathcal{H}$) a group-like element
\begin{align}
e^{\wedge A} = 1_{T\mathcal{H}} + A +\frac{1}{2!}A\wedge A + \frac{1}{3!}A\wedge A\wedge A+\ldots\,.
\end{align} 
Acting with the coproduct \eqref{eq:coprS} and using the identity
\begin{align}
\frac{1}{k!}\Delta_{S\mathcal{H}}(A^{\wedge k}) = \sum_{\substack{l_1,l_2\\ l_1+l_2=k}}\frac{1}{l_1!l_2!}A^{\wedge l_1}\otimes' A^{\wedge l_2}\,,
\end{align}
it is then easy to see that we indeed have the property
\begin{align}
\Delta_{S\mathcal{H}}(e^{\wedge A}) = e^{\wedge A}\otimes' e^{\wedge A}\,.
\end{align}
It therefore follows that cohomomorphisms map group-like elements to group-like elements. In particular, we can write
\begin{align}
\mathbf{F}(e^{\wedge A}) = e^{\wedge \pi_1 \mathbf{F}(e^{\wedge A})}\,,
\end{align}
as well as
\begin{align}
\mathbf{d}(e^{\wedge A}) = e^{\wedge A}\wedge\pi_1 \mathbf{d}(e^{\wedge A})\,.
\end{align}
for any cohomomorphism $\mathbf{F}$ and coderivation $\mathbf{d}$.

We are now prepared to consider an $L_\infty$ structure $(\mathcal{H},\{l_k\}_{k\geq 1})$ and use \eqref{eq:correspL} to define the corresponding coderivations $\boldsymbol{l}_k$, together with the total coderivation
\begin{align}
\boldsymbol{l}=\sum_{k=1}^\infty\boldsymbol{l}_k\,.
\end{align}
The $L_\infty$ relations can then be succinctly packaged as
\begin{align}
\boldsymbol{l}^2 = \frac{1}{2}[\boldsymbol{l},\boldsymbol{l}]=0\,.
\end{align}
Using the above-collected properties, it follows that the action \eqref{eq:SLinf} may be rewritten in a more compact way as
\begin{align}
S(\Psi) = \int_0^1 dt\,\langle \omega|\pi_1 \boldsymbol{\p}_t\, e^{\wedge \Psi(t)}\otimes\pi_1 \boldsymbol{l}\, e^{\wedge \Psi(t)}\,.\label{eq:actL}
\end{align}
Again, we should keep in mind that the construction of $S(\Psi)$ ensures that it only depends on the endpoint $\Psi(1)=\Psi$ of the interpolation $\Psi(t)$. Varying the action \eqref{eq:actL}  with respect to $\Psi$, we would obtain the equation of motion
\begin{align}
\text{EOM}(\Psi) = \pi_1 \boldsymbol{l}\, e^{\wedge \Psi}\,.
\end{align}
Furthermore, the action \eqref{eq:actL} is invariant under the gauge transformation
\begin{align}
\delta\Psi = \pi_1[\boldsymbol{l},\mathbf{\Lambda}]\,e^{\wedge \Psi}\,,
\end{align}
where we have introduced a cyclic degree-odd coderivation $\mathbf{\Lambda}$ which acts as a gauge parameter. Again, unless we have $\Lambda_k\equiv \pi_1\mathbf{\Lambda}\pi_k=0$ for $k>0$, the gauge transformation will contain trivial pieces which vanish on-shell.

\subsubsection{Uplifting the propagator to $S\mathcal{H}$}\label{subsub:uplift}

In order to continue paralleling the discussion of $A_\infty$ effective actions in the coalgebra language, we now need to define suitable uplifts of $h$ and $P$ from $\mathcal{H}$ to $S\mathcal{H}$ (this of course also needs to be done for the canonical inclusion $I:P\mathcal{H}\longrightarrow H$ and canonical projection $\Pi:\mathcal{H}\longrightarrow P\mathcal{H}$). This is very easy in the cases of $P$, $I$ and $\Psi$, where we define the associated cohomomorphisms $\mathbf{P}:S\mathcal{H}\longrightarrow S\mathcal{H}$, $\mathbf{I}:SP\mathcal{H}\longrightarrow S\mathcal{H}$ and $\mathbf{\Pi}:S\mathcal{H}\longrightarrow SP\mathcal{H}$ to simply act as
\begin{subequations}
	\begin{align}
	\mathbf{P}\pi_k &= \frac{1}{k!} P^{\wedge k}\pi_k\,,\\
	\mathbf{I}\pi_k &= \frac{1}{k!} I^{\wedge k}\pi_k\,,\\
	\mathbf{\Pi}\pi_k &= \frac{1}{k!} \Pi^{\wedge k}\pi_k\,.
	\end{align}
\end{subequations}
In the case of the propagator $h$, we would now like to define a map $\mathbf{h}:S\mathcal{H}\longrightarrow S\mathcal{H}$ which would satisfy the Hodge-Kodaira decomposition\footnote{Here $\mathbf{1}_{S\mathcal{H}}$ denotes the identity cohomomorphism on $S\mathcal{H}$ which acts as
	\begin{align}
	\mathbf{1}_{S\mathcal{H}}\pi_k = \frac{1}{k!} (1_{\mathcal{H}})^{\wedge k}\pi_k\,.
	\end{align}
}
\begin{align}
\mathbf{Qh}+\mathbf{hQ} = \mathbf{1}_{S\mathcal{H}}-\mathbf{P}\,,\label{eq:HKL}
\end{align}
as well as the annihilation conditions $\mathbf{hI}=\mathbf{h}^2 = \mathbf{\Pi h}=0$. Here $\mathbf{Q}\equiv \boldsymbol{l}_1:S\mathcal{H}\longrightarrow S\mathcal{H}$ is of course the coderivation corresponding to $Q$ which acts as
\begin{align}
\mathbf{Q}\pi_k
&=\frac{1}{(k-1)!}\big[Q\wedge(1_\mathcal{H})^{\wedge(k-1)}\big]\pi_k\,.
\end{align}
We propose to define $\mathbf{h}$ so that it acts as\footnote{We have learned that in parallel to our work, the same result has been obtained by H.\ Kunitomo \cite{kunitomo-talk}.}
\begin{align}
\mathbf{h}\pi_k 
&=\frac{1}{k!}\sum_{j=0}^{k-1}\big[h\wedge (P)^{\wedge j} \wedge (1_\mathcal{H})^{\wedge (k-1-j)}\big]\pi_k\,.\label{eq:hupliftL}
\end{align}
Indeed, it is straightforward to compute that 
\begingroup\allowdisplaybreaks
\begin{subequations}
	\begin{align}
	\mathbf{Qh}\pi_k &=\frac{1}{k!}\sum_{j=0}^{k-1}\bigg\{Qh\wedge (P)^{\wedge j} \wedge (1_\mathcal{H})^{\wedge (k-1-j)}+\nonumber\\[-2mm]
	&\hspace{3cm}-jh\wedge QP \wedge (P)^{\wedge (j-1)}\wedge (1_\mathcal{H})^{\wedge (k-1-j)}+\nonumber\\
	&\hspace{3cm}-(k-1-j)h\wedge (P)^{\wedge j} \wedge Q\wedge  (1_\mathcal{H})^{\wedge (k-2-j)}
	\bigg\}\pi_k\,,\\
	\mathbf{hQ}\pi_k
	&=\frac{1}{k!}\sum_{j=0}^{k-1}\bigg\{hQ\wedge (P)^{\wedge j} \wedge (1_\mathcal{H})^{\wedge (k-1-j)}+\nonumber\\[-2mm]
	&\hspace{3cm}+jh\wedge QP \wedge ( P)^{\wedge (j-1)} \wedge (1_\mathcal{H})^{\wedge (k-1-j)}+\nonumber\\
	&\hspace{3cm}+(k-1-j)h\wedge (P)^{\wedge j} \wedge Q\wedge  (1_\mathcal{H})^{\wedge (k-2-j)}
	\bigg\}\pi_k\,,
	\end{align}
\end{subequations}
\endgroup
so that we eventually obtain
\begingroup\allowdisplaybreaks
\begin{subequations}
	\begin{align}
	(\mathbf{hQ}+\mathbf{Qh}) \pi_k 
	&=\frac{1}{k!}\sum_{j=0}^{k-1}\big[(hQ+Qh)\wedge (P)^{\wedge j} \wedge (1_\mathcal{H})^{\wedge (k-1-j)}\big]\pi_k\\
	&=\frac{1}{k!}\sum_{j=0}^{k-1}\big[(1_\mathcal{H}-P)\wedge (P)^{\wedge j} \wedge (1_{\mathcal{H}})^{\wedge (k-1-j)}\big]\pi_k\\
	&=\frac{1}{k!}\bigg\{
	(1_\mathcal{H}-P)\wedge (1_\mathcal{H})^{\wedge (k-1)}+
	(1_\mathcal{H}-P)\wedge P\wedge (1_\mathcal{H})^{\wedge (k-2)}+\ldots\nonumber\\
	&\hspace{6.2cm}\ldots +(1_\mathcal{H}-P)\wedge (P)^{\wedge (k-1)}
	\bigg\}\pi_k\\
	&=\frac{1}{k!}\big[(1_\mathcal{H})^{\wedge k}-P^{\wedge k}\big]\pi_k\\[+2mm]
	&=(\mathbf{1}_{S\mathcal{H}}-\mathbf{P})\pi_k\,.
	\end{align}
\end{subequations}
\endgroup
This means that given that we define the uplift of $h$ from $\mathcal{H}$ to $S\mathcal{H}$ using \eqref{eq:hupliftL}, the resulting map $\mathbf{h}$ satisfies the Hodge-Kodaira decomposition \eqref{eq:HKL}. We also clearly have the annihilation conditions $\mathbf{h I} = \mathbf{Ph}=\mathbf{h}^2=0$. We can therefore write down the following free-theory SDR
\begin{align}
\mathrel{\raisebox{+14.5pt}{\rotatebox{-110}{  \begin{tikzcd}[sep=12mm,
			arrow style=tikz,
			arrows=semithick,
			diagrams={>={Straight Barb}}
			]
			\ar[to path={ node[pos=.5,C] }]{}
			\end{tikzcd}
}}}\hspace{-2.7mm}
(-\mathbf{h})\,(S\mathcal{H},\mathbf{Q})
\hspace{-3.2mm}\raisebox{-0.2pt}{$\begin{array}{cc} {\text{\scriptsize $\mathbf{\Pi}$}}\\[-3.0mm] \mathrel{\begin{tikzpicture}[node distance=1cm]
		\node (A) at (0, 0) {};
		\node (B) at (1.5, 0) {};
		\draw[->, to path={-> (\tikztotarget)}]
		(A) edge (B);
		\end{tikzpicture}}\\[-4mm] {\begin{tikzpicture}[node distance=1cm]
		\node (A) at (1.5, 0) {};
		\node (B) at (0, 0) {};
		\draw[->, to path={-> (\tikztotarget)}]
		(A) edge (B);
		\end{tikzpicture}}\\[-3.5mm]
	\text{\scriptsize $\mathbf{I}$} \end{array}$} \hspace{-3.4mm}
(SP\mathcal{H},\mathbf{\Pi Q I})\,.\label{eq:SDRtensorL}
\end{align}
\vspace{-9mm}

\noindent It would be interesting to investigate in more detail the properties of $\mathbf{h}$ in relation to the coproduct $\Delta_{S\mathcal{H}}$: in contrast to the non-symmetrized case, it seems to be non-trivial to write in a closed form the corresponding map on $S\mathcal{H}\otimes'S\mathcal{H}$ which arises from the action of $\Delta_{S\mathcal{H}}$ on $\mathbf{h}$.

\subsubsection{Homotopy transfer of $L_\infty$ structures}

In parallel to the $A_\infty$ case, we can proceed to show that the Feynman diagrams describing the effective interactions of the action for $\psi\in P\mathcal{H}$ can be neatly organized using the homological perturbation lemma. Introducing a perturbation $\delta\boldsymbol{l} = \sum_{k>1}\boldsymbol{l}_k$ to the free-theory SDR \eqref{eq:SDRtensorL}, we obtain the interacting SDR
\begin{align}
\mathrel{\raisebox{+14.5pt}{\rotatebox{-110}{  \begin{tikzcd}[sep=12mm,
			arrow style=tikz,
			arrows=semithick,
			diagrams={>={Straight Barb}}
			]
			\ar[to path={ node[pos=.5,C] }]{}
			\end{tikzcd}
}}}\hspace{-2.7mm}
(-\tilde{\mathbf{h}})\,(S\mathcal{H},\boldsymbol{l})
\hspace{-3.2mm}\raisebox{-1pt}{$\begin{array}{cc} {\text{\scriptsize $\tilde{\mathbf{\Pi}}$}}\\[-3.0mm] \mathrel{\begin{tikzpicture}[node distance=1cm]
		\node (A) at (0, 0) {};
		\node (B) at (1.5, 0) {};
		\draw[->, to path={-> (\tikztotarget)}]
		(A) edge (B);
		\end{tikzpicture}}\\[-4mm] {\begin{tikzpicture}[node distance=1cm]
		\node (A) at (1.5, 0) {};
		\node (B) at (0, 0) {};
		\draw[->, to path={-> (\tikztotarget)}]
		(A) edge (B);
		\end{tikzpicture}}\\[-3.0mm]
	\text{\scriptsize $\tilde{\mathbf{I}}$} \end{array}$} \hspace{-3.4mm}
(SP\mathcal{H},\mathbf{\Pi Q I}+\delta \tilde{\boldsymbol{l}})\,,\label{eq:SDRtensorLint}
\end{align}
\vspace{-9mm}

\noindent where the perturbed data can be expressed as usual
\begin{subequations}
	\begin{align}
	\delta\tilde{\boldsymbol{l}} &= \mathbf{\Pi}\delta \boldsymbol{l}\frac{1}{\mathbf{1}_{S\mathcal{H}} +\mathbf{h}\delta\boldsymbol{l} }\mathbf{I}\,,\\
	\tilde{\mathbf{h}} &= \frac{1}{\mathbf{1}_{S\mathcal{H}} +\mathbf{h}\delta\boldsymbol{l} }\mathbf{h}\,,\\
	\tilde{\mathbf{I}} &= \frac{1}{\mathbf{1}_{S\mathcal{H}} +\mathbf{h}\delta\boldsymbol{l} }\mathbf{I}\,,\label{eq:ItL}\\
	\tilde{\mathbf{\Pi}} &= \mathbf{\Pi}\frac{1}{\mathbf{1}_{S\mathcal{H}} +\delta\boldsymbol{l}\mathbf{h} }\,.
	\end{align}
\end{subequations}
We can indeed easily check that order by order in $\psi$, the perturbed inclusion map $\tilde{\mathbf{I}}$ (as given by \eqref{eq:ItL}) indeed provides the solution $\Psi(\psi)$ (as given by \eqref{eq:Psi(psi)L}) for integrating out the modes $R$ which lie outside of $\text{im}\, P$ (fixing the gauge $hR=0$). Using \eqref{eq:correspL} and \eqref{eq:hupliftL}, we can write
\begin{subequations}
	\begin{align}
	\Psi(\psi) &= \pi_1\frac{1}{\mathbf{1}_{S\mathcal{H}} +\mathbf{h}\delta\boldsymbol{l} }\mathbf{I} e^{\wedge \psi}\\
	&=\pi_1 \mathbf{I} e^{\wedge \psi} - \pi_1 (\mathbf{h}\delta\boldsymbol{l})\mathbf{I} e^{\wedge \psi} + \pi_1 (\mathbf{h}\delta\boldsymbol{l})(\mathbf{h}\delta\boldsymbol{l})\mathbf{I} e^{\wedge \psi}+\nonumber\\
	&\hspace{5cm}
	- \pi_1 (\mathbf{h}\delta\boldsymbol{l})(\mathbf{h}\delta\boldsymbol{l})(\mathbf{h}\delta\boldsymbol{l})\mathbf{I} e^{\wedge \psi}+\mathcal{O}(\psi^{\wedge 5})\\
	&=\psi - \frac{1}{2!}\pi_1 (\mathbf{h}\boldsymbol{l}_2)\mathbf{I} (\psi\wedge \psi)- \frac{1}{3!}\pi_1 (\mathbf{h}\boldsymbol{l}_3)\mathbf{I} (\psi\wedge \psi\wedge \psi)+\nonumber\\
	&\hspace{6cm}- \frac{1}{4!}\pi_1 (\mathbf{h}\boldsymbol{l}_4)\mathbf{I} (\psi\wedge \psi\wedge \psi \wedge \psi)\\
	&\hspace{0.5cm}
	+ \frac{1}{3!}\pi_1 (\mathbf{h}\boldsymbol{l}_2)(\mathbf{h}\boldsymbol{l}_2)\mathbf{I} (\psi\wedge\psi\wedge \psi) + \frac{1}{4!}\pi_1 (\mathbf{h}\boldsymbol{l}_2)(\mathbf{h}\boldsymbol{l}_3)\mathbf{I} (\psi\wedge\psi\wedge\psi\wedge \psi)
	+\nonumber\\
	&\hspace{6cm} + \frac{1}{4!}\pi_1 (\mathbf{h}\boldsymbol{l}_3)(\mathbf{h}\boldsymbol{l}_2)\mathbf{I} (\psi\wedge\psi\wedge\psi\wedge \psi)+\nonumber\\
	&\hspace{0.5cm}
	- \frac{1}{4!}\pi_1 (\mathbf{h}\boldsymbol{l}_2)(\mathbf{h}\boldsymbol{l}_2)(\mathbf{h}\boldsymbol{l}_2)\mathbf{I}(\psi\wedge\psi\wedge\psi\wedge \psi)+\mathcal{O}(\psi^{\wedge 5})\\
	&=\psi -\frac{1}{2!}hl_2(\psi,\psi)-\frac{1}{3!}hl_3(\psi,\psi,\psi)-\frac{1}{4!}hl_4(\psi,\psi,\psi,\psi)+\nonumber\\
	&\hspace{0.5cm}+\frac{1}{2} hl_2(hl_2(\psi,\psi), \psi)+ \frac{1}{6}hl_2(hl_3(\psi,\psi,\psi),\psi)+ \frac{1}{4}
	hl_3(hl_2(\psi,\psi), \psi, \psi) +\nonumber\\
	&\hspace{0.5cm}-\frac{1}{8}hl_2(hl_2(\psi,\psi),hl_2(\psi,\psi))-\frac{1}{2}hl_2(hl_2(hl_2(\psi,\psi),\psi),\psi)+\mathcal{O}(\psi^{\wedge 5})\,,
	\end{align}
\end{subequations}
so that we see that we have recovered the quartic-order result \eqref{eq:Psi(psi)L}, including all symmetry factors. Also, while it is trivial to establish that $\tilde{l}_1 = Pl_1$, $\tilde{l}_2 = Pl_2$, we have, for instance,
\begin{subequations}
	\begin{align}
	\tilde{l}_3(\psi_1,\psi_2,\psi_3) &= \pi_1\mathbf{\Pi} \boldsymbol{l}_3 \mathbf{I}(\psi_1\wedge \psi_2 \wedge \psi_3) -\pi_1\mathbf{\Pi} \boldsymbol{l}_2\mathbf{h}\boldsymbol{l}_2 \mathbf{I}(\psi_1\wedge \psi_2 \wedge \psi_3)\\
	&=Pl_3(\psi_1,\psi_2,\psi_3)-Pl_2(hl_2(\psi_1,\psi_2),\psi_3)+\nonumber\\
	&\hspace{2cm}-(-1)^{d(\psi_1)(d(\psi_2)+d(\psi_3))}Pl_2(hl_2(\psi_2,\psi_3),\psi_1)+\nonumber\\
	&\hspace{2cm}-(-1)^{d(\psi_3)(d(\psi_1)+d(\psi_2))}Pl_2(hl_2(\psi_1,\psi_2),\psi_3)
	\end{align}
\end{subequations}
which clearly agrees with \eqref{eq:l3t} which we derived using the product notation. Similarly for higher effective products. 

As we have already mentioned, it is possible to show order by order in $\psi$ that the effective products $\tilde{l}_k$ are cyclic with respect to $\tilde{\omega}$ (assuming suitable BPZ properties of $h$). It would be  interesting to generalize the all-order proof of cyclicity of $A_\infty$ effective products from subsection \ref{subsub:cyc} to the $L_\infty$ case. Here the main obstacle  is the very definition of the action of $\langle \tilde{\omega}|$ on $S\mathcal{H}$: acting with a graded-antisymmetric form on a graded-symmetrized space gives automatically zero so some new structure seems to be needed. 

It would be interesting to generalize the proof \cite{Konopka:2015tta} that $\tilde{\mathbf{I}}$ and $\tilde{\mathbf{\Pi}}$ are cohomomorphisms from the $A_\infty$ to the $L_\infty$ case (the fact that $\delta\tilde{\bs{l}}$ is a coderivation would then follow immediately from a computation analogous to \eqref{eq:prev}). However, in order to do this, we would need to have at our disposal a closed-form expression for the action $\Delta_{S\mathcal{H}}$ on $\mathbf{h}$, which, as we have commented at the end of subsection \ref{subsub:uplift}, is missing at the moment. In fact, it is easy to see that $\tilde{\mathbf{I}}$ is a cohomomorphism provided that order by order in $\delta\bs{l}$, we satisfy
\begin{align}
\Delta_{S\mathcal{H}} (\mathbf{h}\delta\bs{l})^k\mathbf{I} = \sum_{l=0}^{k} \big[(\mathbf{h}\delta\bs{l})^l\mathbf{I}\otimes' (\mathbf{h}\delta\bs{l})^{k-l}\mathbf{I}\big]\Delta_{SP\mathcal{H}}\,.\label{eq:cohomopert}
\end{align}
Let us now perform an explicit check of \eqref{eq:cohomopert} for $k\leqslant 2$. The case $k=0$ trivially follows from the fact that $\mathbf{I}$ is a cohomomorphism. After a little algebra (making use of the annihilation condition $hI=0$), one can show that 
\begin{align}
(\mathbf{h}\delta \bs{l})\mathbf{I}=\sum_{k=2}^\infty \sum_{l=k}^\infty \big[hl_k (I^{\otimes k}) \wedge I^{\otimes (l-k)}\big]\pi_l\,,
\end{align}
so that we clearly have
\begin{align}
\Delta_{S\mathcal{H}}(\mathbf{h}\delta \bs{l})\mathbf{I} = \big[\mathbf{I}\otimes' (\mathbf{h}\delta \mathbf{l})\mathbf{I}+(\mathbf{h}\delta \mathbf{l})\mathbf{I}\otimes' \mathbf{I}\big]\Delta_{SP\mathcal{H}}\,,
\end{align}
which verifies \eqref{eq:cohomopert} for $k=1$.
Going through somewhat more algebra, we have
\begin{subequations}
	\begin{align}
	\delta\bs{l}(\mathbf{h}\delta \bs{l})\mathbf{I}
	&=\sum_{k=2}^\infty \sum_{l=k}^\infty\sum_{m=2}^{l-k+1} \big[l_m(hl_k(I^{\otimes k})\wedge I^{\otimes (m-1)})\wedge I^{\otimes l-k-m+1}\big]\pi_l+\nonumber\\
	&\hspace{2cm}+\sum_{k=2}^\infty \sum_{l=k}^\infty\sum_{m=2}^{l-k} \big[l_m(I^{\otimes m})\wedge hl_k(I^{\otimes k})\wedge I^{\otimes l-k-m}\big]\pi_l\,,
	\end{align}
\end{subequations}
so that using the annihilation conditions $hI=\Pi h = h^2 =0$, we eventually obtain
\begin{subequations}
	\begin{align}
	(\mathbf{h}\delta \bs{l})^2\mathbf{I}
	&=\sum_{k=2}^\infty \sum_{l=k}^\infty\sum_{m=2}^{l-k+1} \big[hl_m(hl_k(I^{\otimes k})\wedge I^{\otimes (m-1)})\wedge I^{\otimes l-k-m+1}\big]\pi_l+\nonumber\\
	&\hspace{2cm}+\sum_{k=2}^\infty \sum_{l=k}^\infty\sum_{m=2}^{l-k+1}\sum_{j=0}^{l-k-m+1} \frac{(l-k-m-j+1)}{(l-k-m+2)(l-k-m+1)}\times\nonumber\\
	&\hspace{6cm}\times   \big[hl_m(I^{\otimes m})\wedge hl_k(I^{\otimes k})\wedge I^{\otimes l-k-m}\big]\pi_l\\
	&=\sum_{k=2}^\infty \sum_{l=k}^\infty\sum_{m=2}^{l-k+1} \big[hl_m(hl_k(I^{\otimes k})\wedge I^{\otimes (m-1)})\wedge I^{\otimes l-k-m+1}\big]\pi_l+\nonumber\\
	&\hspace{2cm}+\frac{1}{2}\sum_{k=2}^\infty \sum_{l=k}^\infty\sum_{m=2}^{l-k}   \big[hl_m(I^{\otimes m})\wedge hl_k(I^{\otimes k})\wedge I^{\otimes l-k-m}\big]\pi_l\,,\label{eq:hdl2I}
	\end{align}
\end{subequations}
where in the last step we have used the result
\begin{align}
\sum_{j=0}^{l-k-m+1} \frac{(l-k-m-j+1)}{(l-k-m+2)(l-k-m+1)} = \frac{1}{2}\,.
\end{align}
From \eqref{eq:hdl2I} it is then immediate that we can write
\begin{align}
\Delta_{S\mathcal{H}} (\mathbf{h}\delta \bs{l})^2\mathbf{I} = \big[(\mathbf{h}\delta \bs{l})^2\mathbf{I}\otimes' \mathbf{I}+\mathbf{I}\otimes' (\mathbf{h}\delta \bs{l})^2\mathbf{I} + (\mathbf{h}\delta \bs{l})\mathbf{I}\otimes' (\mathbf{h}\delta \bs{l})\mathbf{I}\big] \Delta_{SP\mathcal{H}}\,,
\end{align}
which verifies \eqref{eq:cohomopert} for $k=2$. Similarly, we can verify order by order in $\delta\bs{l}$ that $\tilde{\mathbf{\Pi}}$ is a cohomomorphism. Nevertheless, it should be worthwhile to look for an all-order method of proving these results.

\end{document}